%% file: exciting.tex
\documentclass{WileyMSP-template}
\usepackage[numbers,sort&compress]{natbib}

\usepackage{orcidlink}
\usepackage{array} 
\usepackage{amsmath}
\usepackage{mathbbol}
\usepackage{amssymb}
\usepackage{mathrsfs}
\usepackage{siunitx}
\usepackage{dutchcal}
\usepackage{tikz}
\usepackage{makecell}
\usepackage{hyperref}
\usepackage[acronym,nomain,hyperfirst=false]{glossaries}
\usepackage{braket}
\usepackage{xspace}
\usepackage{amsmath}
\usepackage[capitalise]{cleveref}
\usepackage{multirow}
\usepackage{booktabs}
\usepackage{bm}
\usepackage{float}

\usetikzlibrary{positioning,calc,arrows.meta}
\usetikzlibrary{shapes.geometric, arrows}
\usetikzlibrary{arrows.meta}

\newcommand{\bftt}[1]{{\fontfamily{lmtt}\fontseries{b}\selectfont #1}}

\DeclareSIUnit\au{a.u.}
\DeclareSIUnit\Bohr{a_0}
\DeclareSIUnit\Hartree{Ha}
\DeclareSIUnit\Rydberg{Ry}
\DeclareSIUnit\electron{e}
\DeclareSIUnit\electronmass{m_\mathrm{e}}
\DeclareSIUnit\atomictime{t_\mathrm{a}}
\DeclareSIUnit\atomicfield{E_\mathrm{a}}
\DeclareSIUnit\atomicmagfield{B_\mathrm{a}}
\DeclareSIUnit\atomicvelocity{v_\mathrm{a}}
\DeclareSIUnit\atomicforce{F_\mathrm{a}}
\DeclareSIUnit\atomicdipole{d_\mathrm{a}}
\DeclareSIUnit\atomicpolarizability{\alpha_\mathrm{a}}
\DeclareSIUnit\atomiccurrent{I_\mathrm{a}}
\DeclareSIUnit\atomicmoment{\mu_\mathrm{a}}

\newcommand{\glossaryentry}[2]{%
  \noindent
  \begin{tabular}{@{}p{3cm}p{5cm}p{10cm}@{}}
    \ensuremath{#1} & \texttt{\detokenize{#1}} & #2 \\[4pt] 
  \end{tabular}\par
}
\newcommand{\glossarycode}[2]{%
  \noindent
  \begin{tabular}{@{}p{3cm}p{5cm}p{10cm}@{}}
    #1 & \texttt{\detokenize{#1}} & #2 \\[4pt] 
  \end{tabular}\par
}

\newcommand{\ts}{\scriptscriptstyle\xspace}
\newcommand{\ie}{{\it i.e.}, }
\newcommand{\eg}{{\it e.g.}, }
\newcommand{\etal}{{\it et al.}\xspace}

\newcommand{\pos}{\ensuremath{\mathbf{r}}}
\newcommand{\unitCellVol}{\ensuremath{\Omega}}
\newcommand{\latvec}{\ensuremath{\mathbf{T}}}
\newcommand{\cellidx}{\ensuremath{{L}}}
\newcommand{\atomincell}{\ensuremath{\mathbf{S}}}
\newcommand{\atomidx}{\ensuremath{\mathrm{\alpha}}}
\newcommand{\atompos}{\ensuremath{\mathbf{R}}}
\newcommand{\atomdisp}{\ensuremath{\boldsymbol{\tau}}}
\newcommand{\ff}{\ensuremath{\mathbf{f}}}
\newcommand{\kk}{\ensuremath{\mathbf{k}}}
\newcommand{\GG}{\ensuremath{\mathbf{G}}}
\newcommand{\PP}{\ensuremath{\mathbf{P}}}
\newcommand{\qq}{\ensuremath{\mathbf{q}}}

\newcommand{\rr}{\ensuremath{\mathbf{r}}} 

\newcommand{\hKS}{\ensuremath{\hat{h}_{\ts \mathrm{KS}}}}
\newcommand{\vKS}{\ensuremath{{v}_{\ts \mathrm{KS}}}}
\newcommand{\vXC}{\ensuremath{\hat{v}_{\mathrm{xc}}}}
\newcommand{\vXCnohat}{\ensuremath{{v}_{\mathrm{xc}}}}
\newcommand{\vExt}{\ensuremath{\hat{v}_{\mathrm{ext}}}}
\newcommand{\vH}{\ensuremath{\hat{v}_{\ts \mathrm{H}}}}
\newcommand{\nKS}{\ensuremath{n_{\mathrm{\ts KS}}}} 
 
\newcommand{\occ}[2]{\ensuremath{f}_{#1#2}}
\newcommand{\kptweight}[1]{\ensuremath{w}_{#1}}
\newcommand{\mH}{\ensuremath{\hat{\mathcal{H}}}}
\newcommand{\basis}{\ensuremath{\phi}}
\newcommand{\mbasis}{\ensuremath{\mathcal{B}}}
\newcommand{\mwf}{\ensuremath{\Psi}}
\newcommand{\kswf}{\ensuremath{\psi}}
\newcommand{\eigval}{\ensuremath{\epsilon}}
\newcommand{\diel}{\ensuremath{\varepsilon}}
\newcommand{\dielmac}{\ensuremath{\varepsilon_{\mathrm{mac}}}}
\newcommand{\xckernel}{\ensuremath{f_{\mathrm{xc}}}}
\newcommand{\radial}{\ensuremath{\mathcal{u}}}
\newcommand{\pola}{\ensuremath{{\chi}^0}}
\newcommand{\reducpola}{\ensuremath{\chi}}
\newcommand{\barcoul}{\ensuremath{v_{\ts \mathrm{H}}}}
\newcommand{\scrcoul}{\ensuremath{\mathrm{W}}}
\newcommand{\cwf}{\ensuremath{c}}
\newcommand{\matching}{\ensuremath{\mathcal{A}}}
\newcommand{\locoeff}{\ensuremath{\mathcal{a}}}
\newcommand{\tel}{\ensuremath{T_{\mathrm{\ts el}}}}
\newcommand{\tph}{\ensuremath{T_{\mathrm{\ts ph}}}}
\newcommand{\freq}{\ensuremath{\omega}}
\newcommand{\phmode}{\ensuremath{\nu}}
\newcommand{\pheig}{\ensuremath{\mathbf{w}}}
\newcommand{\phdmat}{\ensuremath{\mathscr{D}}}
\newcommand{\vel}{\ensuremath{\mathbf{v}}}
\newcommand{\born}{\ensuremath{\mathbf{Z}^\ast}}
\newcommand{\green}{\ensuremath{\mathcal{G}}}
\newcommand{\timeevol}{\ensuremath{\hat{\mathcal{U}}}}
\newcommand{\order}[2]{\ensuremath{#1^{{(#2)}}}}
\newcommand{\elphmatel}{\ensuremath{\mathcal{g}}}
\newcommand{\selfenergy}{\ensuremath{\Sigma}}
\newcommand{\BZpoint}[1]{\ensuremath{\mathrm{#1}}}
\newcommand{\im}{\ensuremath{\mathrm{i}}\hspace*{0.20pt}}
\newcommand{\re}{\ensuremath{\Re}}
\newcommand{\imag}{\ensuremath{\Im}}
\newcommand{\cond}{\ensuremath{\sigma}}
\newcommand{\loss}{\ensuremath{\mathbb{L}}}
\newcommand{\speedOfLight}{\ensuremath{\mathrm{c}}}
\newcommand{\vecpot}{\ensuremath{\mathbf{A}}}
\newcommand{\efield}{\ensuremath{\mathbf{E}}}
\newcommand{\currentDensity}{\ensuremath{\mathbf{J}}}
\newcommand{\polarization}{\ensuremath{\mathbf{P}}}
\newcommand{\TimeOrdering}{\ensuremath{\mathcal{\hat{T}}}}
\newcommand{\force}{\ensuremath{\mathbf{F}}}
\newcommand{\drm}{\ensuremath{\mathrm{d}}}
\newcommand{\ee}{\ensuremath{\mathrm{e}}}
\newcommand{\gw}{$GW$\xspace}
\newcommand{\gwb}{$\mathbf{GW}$\xspace}
\newcommand{\pI}{{\phantom{I}}}
\newcommand{\bigO}{\mathcal{O}} 
\newcommand{\shup}[1]{{\hspace*{0.65pt}#1}}

\newcommand{\exciting}{\texttt{exciting}\xspace} 
\newcommand{\excitingb}{\bftt{exciting}\xspace}  
\newcommand{\elastic}{\texttt{ElaStic}\xspace}   
\newcommand{\elasticb}{\bftt{ElaStic}\xspace}    
\newcommand{\brixs}{\texttt{BRIXS}\xspace}         
\newcommand{\pybrixs}{\texttt{pyBRIXS}\xspace}     
\newcommand{\prexas}{\texttt{PreXAS-Exciton}\xspace}  
\newcommand{\python}{Python\xspace}
\newcommand{\excitingtools}{\texttt{exciting\-tools}\xspace}

\newcommand{\excitingworkflow}{\texttt{exciting\-workflow}\xspace}

\newcommand{\excitingscripts}{\texttt{exciting\-scripts}\xspace}

\newcommand{\fsvisual}{\texttt{FSvisual}\xspace}
\newcommand{\fsvisualb}{\bftt{FSvisual}\xspace}
\newcommand{\jobflow}{Jobflow\xspace}
\newcommand{\jobflowremote}{Jobflow Remote\xspace}
\newcommand{\slurm}{Slurm\xspace}
\newcommand{\abinit}{Abinit\xspace}
\newcommand{\vasp}{VASP\xspace}
\newcommand{\gpaw}{GPAW\xspace}
\newcommand{\elphbolt}{elphbolt\xspace}
\newcommand{\quantumespresso}{Quantum ESPRESSO\xspace}
\newcommand{\cell}{\texttt{CELL}\xspace}
\newcommand{\cellb}{\bftt{CELL}\xspace}

\newcommand{\matid}{MatID\xspace}

\newcommand{\orange}[1]{\color{orange}#1}
\newcommand{\gruen}[1]{\color{green}#1}

\begin{document}

\input{acronyms}

\title{An \exciting approach to theoretical spectroscopy} 

\maketitle

\author{Martí Raya-Moreno$^1$\orcidlink{0000-0001-6190-9769}, 
        Alexander Buccheri$^4$\orcidlink{0000-0001-5983-8631},
        Noah Alexy Dasch$^1$\orcidlink{0009-0007-5259-7733},      
        Nasrin Farahani$^1$\orcidlink{0000-0001-5499-8556}, 
        Ignacio Gonzalez Oliva$^1$\orcidlink{0000-0002-4229-708X},
        Andris Gulans$^3$\orcidlink{0000-0001-7304-1952},
        Seokhyun Hong$^1$\orcidlink{0009-0000-4544-5415},
        Manoar Hossain$^2$\orcidlink{0000-0002-0737-7981}, 
        Hannah Kleine$^1$\orcidlink{0000-0003-2251-8719},
        Martin Kuban$^1$\orcidlink{0000-0002-1619-2460}, 
        Sven Lubeck$^1$\orcidlink{0000-0003-4405-9656}, 
        Benedikt Maurer$^1$\orcidlink{0000-0001-9152-7390},
        Pasquale Pavone$^1$\orcidlink{0000-0001-5093-9997}, 
        Fabian Peschel$^1$\orcidlink{0000-0003-0619-6713}, 
        Daria Popova-Gorelova$^7$\orcidlink{0000-0002-3036-0467},
        Lu Qiao$^1$\orcidlink{0009-0002-0344-1082}, 
        Elias Richter$^1$\orcidlink{0009-0008-2773-1248}, 
        Santiago Rigamonti$^1$\orcidlink{0000-0002-8939-6173},
        Ronaldo Rodrigues Pela$^5$\orcidlink{0000-0002-2413-7023},
        Maximilian Schebek$^{6}$\orcidlink{0009-0003-4308-3926},
        Kshitij Sinha$^1$\orcidlink{0009-0000-6776-269X},
        Daniel T. Speckhard$^1$\orcidlink{0000-0002-9849-0022}, 
        Jan Stutz$^1$\orcidlink{0009-0009-8583-9115},
        Sebastian Tillack$^1$\orcidlink{0000-0001-6402-4157}, 
        Dmitry Tumakov$^1$\orcidlink{0000-0003-0708-2427},
        J{\=a}nis U\v{z}ulis$^3$\orcidlink{0000-0003-2754-3972},
        Mara Voiculescu$^1$\orcidlink{0000-0003-4393-8528},
        Cecilia Vona$^1$\orcidlink{0000-0001-7161-6951},
        Mao Yang$^1$\orcidlink{0000-0003-1488-4671},
        and Claudia Draxl$^{1,*}$\orcidlink{0000-0003-3523-6657}
}









\begin{affiliations}
\textsuperscript{1} Department of Physics and CSMB, Humboldt-Universität zu Berlin,
Zum Großen Windkanal 2, 12489 Berlin, Germany \\
\textsuperscript{2} Paderborn Center for Parallel Computing (PC2), Paderborn University, Warburger Stra\ss e~100, 33098 Paderborn, Germany \\
\textsuperscript{3} Department of Physics, University of Latvia,
Jelgavas iela 3, Riga, LV-1004, Latvia \\
\textsuperscript{4} Max Planck Institute for the Structure and Dynamics of Matter, Luruper Chaussee 149, 22761 Hamburg, Germany\\
\textsuperscript{5} Distributed Algorithms and Supercomputing Department, Zuse Institute Berlin (ZIB),
Takustra\ss e~7, 14195 Berlin, Germany \\
\textsuperscript{6} Department of Physics, Freie Universität Berlin, Arnimallee 14, 14195 Berlin, Germany\\
\textsuperscript{7} Institute of Physics, Brandenburg University of Technology
Cottbus--Senftenberg, Erich-Weinert-Stra\ss e~1, 03046 Cottbus, Germany \\
\end{affiliations}


\begin{abstract}
{Theoretical spectroscopy, and more generally, electronic-structure theory, are powerful concepts for describing the complex many-body interactions in materials. They comprise a variety of methods that can capture all aspects, from ground-state properties to lattice excitations to different types of light-matter interaction, including time-resolved variants. Modern electronic-structure codes implement either a few or several of these methods. Among them, \exciting is an all-electron full-potential package that has a very rich portfolio of all levels of theory, with a particular focus on excitations. It implements the \acrfull{lapwlo} basis, which is known as the gold standard for solving the Kohn-Sham equations of \acrfull{dft}. Based on this, it also offers benchmark-quality results for a wide range of excited-state methods. In this review, we provide a comprehensive overview of the features implemented in \exciting in recent years, accompanied by short summaries on the state of the art of the underlying methodologies. They comprise DFT and \acrfull{tddft}, \acrfull{dfpt} for phonons and electron-phonon coupling, and many-body perturbation theory in terms of the \gw approach and the \acrfull{bse}. Moreover, \exciting can handle \acrfull{rixs}, pump-probe spectroscopy as well as \acrfull{exph}. Finally, we cover workflows and a view on data and \acrfull{ml}. All aspects are demonstrated with examples for scientifically relevant materials.}
\end{abstract}

\newpage

\newpage

\section{Introduction}
\label{sec:Intro}
\input{sections/introduction}

\section{The \bftt{exciting} package}
\label{sec:Code} 
\input{sections/code}

\section{Ground state}
\label{sec:Groundstate}
\input{sections/groundstate}

\section{Lattice dynamics}
\label{sec:DFPT}
\input{sections/dfpt}

\section{GW}
\label{sec:GW}
\input{sections/gw}


\section{Bethe-Salpeter equation}
\label{sec:BSE}
\input{sections/bse}

\section{Time-dependent DFT}
\label{sec:TDDFT}
\input{sections/tddft}


\section{Pump-probe spectroscopy}
\label{sec:PP}
\input{sections/pumpprobe}


\section{Tools, workflows, and interfaces}
\label{sec:Tools}
\input{sections/tools}


\section{Data handling and machine learning}
\label{sec:Data}
\input{sections/data}

\section{Outlook}
\label{sec:Outlook}
\input{sections/outlook}

\section*{Acknowledgement}

Work supported by the European Union’s Horizon 2020 research and innovation program under the grant agreement No. 951786 (NOMAD CoE), the German Science Foundation (DFG) through the Collaborative Research Center HIOS (SFB 951), Project No. 182087777, the Collaborative Research Center FONDA, Project No. 414984028, the priority program SPP2196 Perovskite Semiconductors, project No. 424709454, and the NFDI consortium FAIRmat Project No. 460197019. D.~T.~S. acknowledges support by the Max-Planck Graduate Center for Quantum Materials. N.~F., D.~T. and D.~P.-G. acknowledge funding from the Volkswagen Foundation under Grant No.~96237. J.~U. acknowledges funding provided by the project “Strengthening the Research and Development Capacity of Doctoral Studies at the University of Latvia in the Fields of Smart Specialisation”, identification no. 1.1.1.8/1/24/I/003. A.G. acknowledges funding provided by the Latvian Council of Science, project No. lzp-2024/1-0202. R.~R.~P. acknowledges computing time on the high-performance computer “Lise” at the NHR center NHR@ZIB. M. H. acknowledges computing time provided on the high-performance computer Noctua~2 at the NHR Center PC$^2$. The NHR center is jointly supported by the Federal Ministry of Education and Research and the state governments participating in the NHR, \texttt{www.nhr-verein.de}. We thank Fabian Nemitz, Ben Alex, Elisa Stephan, Nora Salas-Illanes, Christian Vorwerk, and Maria Troppenz for their valuable contributions to the \exciting code.

\bibliographystyle{apsrev4-1} 
\bibliography{bibliography}

\end{document}

%% file: acronyms.tex
%
%
\newacronym{ac}{AC}{adiabatic connection}
\newacronym{ace}{ACE}{adaptively compressed exchange}
\newacronym{ai}{AI}{artificial intelligence}
\newacronym{apw}{APW}{augmented planewave}
\newacronym{apwlo}{APW+LO}{augmented planewave plus local orbital}
\newacronym{arpes}{ARPES}{angle-resolved photoemission spectroscopy}
\newacronym{ase}{ASE}{Atomic Simulation Environment}
\newacronym{asr}{ASR}{Atomic Simulation Recipes}
\newacronym{bo}{BO}{Born-Oppenheimer}
\newacronym{bse}{BSE}{Bethe-Salpeter equation}
\newacronym{bsh}{BSH}{Bethe-Salpeter Hamiltonian}
\newacronym{bte}{BTE}{Boltzmann transport equation}
\newacronym{bvk}{BvK}{Born-von Karman}
\newacronym{bz}{BZ}{Brillouin zone}
\newacronym{cb}{CB}{conduction band}
\newacronym{cbs}{CBS}{complete basis set}
\newacronym{cc}{CC}{coupled cluster}
\newacronym{cdft}{cDFT}{constrained density-functional theory}
\newacronym{ce}{CE}{cluster expansion}
\newacronym{ci}{CI}{continuous integration}
\newacronym{cvt}{CVT}{centroidal Voronoi tessalation}
\newacronym{dbsv}{DBSV}{dual basis self-validation}
\newacronym{dfpt}{DFPT}{density-functional perturbation theory}
\newacronym{dft}{DFT}{density-functional theory}
\newacronym{dha}{DHA}{double-hybrid approximation}
\newacronym{dmft}{DMFT}{dynamical mean-field theory}
\newacronym{dos}{DOS}{density of states}
\newacronym{eas}{EAS}{expansion addition screening}
\newacronym{epc}{EPC}{electron-phonon coupling}
\newacronym{exph}{EXPC}{exciton-phonon coupling}
\newacronym{exx}{EXX}{exact-exchange}
\newacronym{fs}{FS}{Fermi surface}
\newacronym{fspl}{FS}{Fermi surfaces}
\newacronym{gdft}{gDFT}{generalized DFT}
\newacronym{gga}{GGA}{generalized gradient approximation}
\newacronym{ghs}{GHs}{global hybrids}
\newacronym{gks}{gKS}{generalized Kohn-Sham}
\newacronym{gs}{GS}{ground state}
\newacronym{html}{HTML}{HyperText Markup Language}
\newacronym{hf}{HF}{Hartree-Fock}
\newacronym{ibc}{IBC}{incomplete basis set correction}
\newacronym{ibz}{IBZ}{irreducible Brillouin zone}
\newacronym{ifc}{IFC}{interatomic force constant}
\newacronym{ipa}{IPA}{independent particle approximation}
\newacronym{isdf}{ISDF}{interpolative separable density fitting}
\newacronym{ked}{KED}{kinetic-energy density}
\newacronym{ks}{KS}{Kohn-Sham}
\newacronym{lapw}{LAPW}{linearized augmented planewave}
\newacronym{llm}{LLM}{large language model}
\newacronym{lapwlo}{LAPW+LO}{linearized augmented planewave plus local orbital}
\newacronym{lda}{LDA}{local density approximation}
\newacronym{lo}{LO}{local orbital}
\newacronym{lfe}{LFE}{local-field effect}
\newacronym{lh}{LH}{local hybrid}
\newacronym{lr}{LR}{linear response}
\newacronym{lnd}{LND}{lowest non-degenerate}
\newacronym{lrtddft}{LR-TDDFT}{linear-response TDDFT}

\newacronym{mb}{MB}{manybody}
\newacronym{mc}{MC}{Monte Carlo}
\newacronym{md}{MD}{molecular dynamics}
\newacronym{ml}{ML}{machine learning}
\newacronym{mbpt}{MBPT}{manybody perturbation theory}
\newacronym{mgga}{mGGA}{meta-GGA}
\newacronym{mlwf}{MLWF}{maximally localized Wannier function}
\newacronym{moke}{MOKE}{magneto-optical Kerr effect}
\newacronym{mt}{MT}{muffin-tin}
\newacronym{oep}{OEP}{optimized effective potential}
\newacronym{pbc}{PBC}{periodic boundary conditions}
\newacronym{qp}{QP}{quasiparticle}
\newacronym{qsgw}{QSGW}{quasiparticle self-consistent $GW$}
\newacronym{rixs}{RIXS}{resonant inelastic x-ray scattering}
\newacronym{rmsle}{RMSLE}{root mean square logarithmic error}
\newacronym{rpa}{RPA}{random phase approximation}
\newacronym{rt}{RT}{real-time}
\newacronym{rttddft}{RT-TDDFT}{real-time TDDFT}
\newacronym{scf}{SCF}{self-consistent field}
\newacronym{smape}{sMAPE}{symmetric mean absolute percentage error}
\newacronym{soc}{SOC}{spin-orbit coupling}
\newacronym{sr}{SR}{scalar-relativistic}
\newacronym{ste}{STE}{self-trapped exciton}
\newacronym{sv}{SV}{second-variational}
\newacronym{svlo}{SVLO}{second variation with local orbitals}

\newacronym{ta}{TA}{transient absorption}
\newacronym{tda}{TDA}{Tamm-Dancoff approximation}
\newacronym{tddft}{TDDFT}{time-dependent density-functional theory}
\newacronym{tdks}{TDKS}{time-dependent Kohn-Sham}
\newacronym{vdw}{vdW}{van-der-Waals}

\newacronym{xc}{xc}{exchange-correlation}
\newacronym{xas}{XAS}{X-ray absorption spectroscopy}
\newacronym{xps}{XPS}{X-ray photoelectron spectroscopy}
\newacronym{zpr}{ZPR}{zero-point renormalization}

%% file: sections/introduction.tex
Electronic and lattice excitations play a central role in a wide range of scientific and technological fields, from fundamental physics to applications in optoelectronics, energy harvesting, medical devices, and many more. Understanding, predicting, and modeling these excitations is crucial not only for the discovery of new materials but also for enhancing the performance and functionality of existing ones. In this context, \textit{ab initio} calculations are invaluable, offering a parameter-free and predictive framework to explore materials at the atomic scale. Among such approaches, \gls{dft} \cite{hohenberg1964,kohn1965} is the most widely used method, offering a good compromise between accuracy and computational cost. However, conventional \gls{dft} is intrinsically limited to ground-state properties. There are two different routes to describe electronic excitations. The first, \gls{tddft}, stays within the realm of \gls{dft} as its time-dependent variant~\cite{runge1984density,cui2025spin,jacquemin2009extensive,adamo2013calculations,casida2012progress}. The second is \gls{mbpt} with its Green-function-based methods~\cite{HedinPR1965,salpeter1951relativistic,hanke1980manyparticleeffects,hybertsen1986electron_correlation,strinati1988application,OnidaRMP2002}, for which \gls{dft} typically serves as a starting point. For lattice excitations, \gls{dfpt} is an efficient approach to provide phonon states~\cite{Zein1984,Baroni1987} and electron-phonon coupling parameters~\cite{Savrasov1994}. Together, all these concepts represent a rich portfolio that allows one to explore the vast space of materials and their properties and functions.  

In this work, we review the current status of the full-potential package \exciting, which covers a rich feature set through implementations on these methods. \exciting is based on the \gls{lapwlo} method~\cite{andersen1975lapw,singh2006lapw}, which is widely recognized as the gold standard in electronic-structure theory. Due to its all-electron nature, the code can be applied to all kinds of materials, irrespective of the atomic species involved, and it allows for exploring the physics of core electrons. Notably, we emphasize \exciting's high numerical precision, reaching the microhartree level \cite{Gulans2018}. As the name suggests, \exciting has a particular focus on excitations. It implements all of the aforementioned formalisms, \ie \gls{dft} and \gls{tddft} as well as \gls{mbpt} in terms of variants of the $GW$ approximation \cite{DmitriiPRB2016} and the \gls{bse} \cite{vorwerk2019bse}. For lattice excitations, it offers calculations by means of \gls{dfpt} \cite{TillackDFPT2026} as well as the frozen-phonon approach employing supercells. These methods can be used to tackle all kinds of spectroscopies, from optical to X-ray absorption, from infrared to Raman spectroscopy, from phonon to exciton dispersion, from electron-energy-loss to pump-probe spectroscopy, and more. Overall, \exciting has established itself as a benchmark tool for high-fidelity calculations and a source of reference data, offering the widest portfolio of theoretical spectroscopy features. 

The paper is organized into sections dedicated to the different methodologies. In each section, we first provide an overview of the current state of the art to contextualize the \exciting developments within the international landscape. Subsequent subsections describe the latest developments and implementations and provide examples. Additional sections cover workflows and tools, data handling and \gls{ml}, and more.

%% file: sections/code.tex
\input{sections/basis}

\subsection{\excitingb's portfolio}
\label{sec:portfolio_code}

\paragraph{Features.} As mentioned in the Introduction, \exciting offers a rich variety of features, including \gls{ks}-\gls{dft}, gKS-\gls{dft}, \gls{dfpt}, \gls{tddft}, and Green-function based approaches to obtain the \gls{qp} bands as well as all kinds of spectra. The most important features are summarized in Table \ref{tab:features}, including the developments and implementations of the last years, which will be described in the following sections. These new developments span all levels of methodology. At the \gls{dft} level, these are \gls{mgga} (\cref{sec:New_metaGGA_GS}), \gls{dft}-1/2 (\cref{sec:New_DFThalf_GS}), and hybrid functionals (\cref{sec:New_hybrids_GS}). \Gls{soc} can be treated very efficiently, by a method called \gls{svlo}~\cite{Vona2023} (\cref{sec:New_SVLO_GS}). This is not restricted to the \gls{gs}, but has been extended to optical spectra (\cref{sec:New_SVLO_BSE}). Moreover, \exciting covers implementations of \gls{cdft} (\cref{sec:New_cDFT_GS}), the newest version of Libxc~\cite{Lehtola2018,Chachiyo2020}, and an interface to the SIRIUS library~\cite{sirius_repo} (\cref{sec:New_SIRIUS_tools}). 
Lattice vibrations and \gls{epc} effects have been introduced to the code in many aspects. \exciting now features an implementation of \gls{dfpt} (\cref{sec:New_implementation_dftp}), providing phonon properties, \gls{epc} coupling constants, and related self-energies that renormalize the electronic properties and give rise to temperature effects. Obtaining Born effective charges also allows for treating vibrations of polar materials reliably.
\begin{table}[h]
\caption{Overview of the methods implemented in \exciting. The features marked with an asterisk are currently in development. The acronyms are explained in \cref{sec:Groundstate}.}
\label{tab:features}
\centering
\renewcommand{\arraystretch}{1.2}
\begin{tabular}{l l l c l}
\toprule
\textbf{Property} & \textbf{Method/module} & \textbf{Approximation} & \textbf{SOC} & \textbf{Comment} \\
\midrule
KS electrons & DFT & LDA, GGA, DFT-1/2,  & SV, SVLO, core: Dirac & direct or via Libxc\\
     & DFT & \gls{mgga}, OEP  & SV, SVLO, core: Dirac \\
gKS electrons & DFT & HF, PBE0, HSE & SV, SVLO* & parameters adjustable\\
\midrule
Quasiparticles & $GW$ & $G_0W_0$ & SV & various starting points\\
    & $GW$ & QSGW &  & various starting points\\
    & EPC & MBPT & -- & \makecell[tl]{temperature effects,\\ various starting points}\\
\midrule
\multirow{2}{*}{Phonon spectra} & DFPT & LDA, GGA & -- & \\
            & supercells & various & SV, SVLO* & \\
\midrule
Optical spectra & LR-TDDFT & IPA, RPA, various kernels & --  & \\
& BSE & IPA, RPA; TDA, full BSE & SV, SVLO &\\
Core spectra & BSE & IPA, RPA; TDA, full BSE & core: Dirac, cond. SV \\
Pump-probe spectra & BSE \& TDDFT & same as its components &  --&  \\
RIXS & BRIXS   & IPA, RPA, BSE &  as optical spectra & \\
Pumped RIXS & BRIXS & IPA, RPA, BSE & -- & like pump-probe\\
Raman spectra &  & as optical spectra & as optical spectra & \\
IR spectra &  & as optical spectra  & as  optical spectra & \\
MOKE &  & as optical spectra &  as  optical spectra & \\
Exciton-polarons & \gls{dfpt} \& BSE & as \gls{dfpt} / optical spectra &  & \\
STM & DFT & Tersoff-Hamann & -- &  \\
Boltzmann transport & DFT & constant relaxation time & -- & -- \\
\bottomrule
\end{tabular}
\end{table}
The \gw module (\cref{sec:GW}) contains a rich variety of new features, which are the direct computation of the polarizability, improvements in the calculation of the correlation self-energy, a precise treatment of long-range interactions in anisotropic and/or low-dimensional systems, a task-based \gw workflow with GPU porting, and an efficient evaluation of the screening of weakly-bound interfaces (\cref{sec:New_EAS_GW}). All above mentioned \gls{xc} functionals are available as a starting point. In addition, a \gls{qp} self-consistent version has been implemented \cite{Salas-Illanes2022} (\cref{sec:New_scf_GW}). 
On the spectroscopy side, \exciting now covers besides \gls{lrtddft} (\cref{sec:New_LR_TDDFT}) also \gls{rttddft} (\cref{sec:New_RT_TDDFT}), including Ehrenfest dynamics \cite{pela2022ehrenfest}. With a low-scaling implementation, the \gls{bse} code has experienced a dramatic speed-up~\cite{maurer2026fastBSE} (\cref{sec:New_Scaling_BSE}) and can be used also for non-equilibrium cases. Based on this, \gls{rixs} \cite{Vorwerk2020Brixs,Vorwerk2021Brixs} (\cref{sec:New_RIXS_BSE}) and pump-probe spectra (\cref{sec:New_PPS1_PP,sec:New_PPS2_PP}) can be computed, including even pumped \gls{rixs}. Linear-response functionality further includes the calculation of the \gls{moke}, enabling access to magnetic circular dichroism and complex magneto-optical constants. Furthermore, the scanning tunneling microscope (STM) module enables simulations of real-space surface properties using the Tersoff–Hamann approximation, with tunneling currents derived from the ground-state localized \gls{dos}.
The latest version of the code is accompanied by a number of tools, workflows, and interfaces, which all make \exciting calculations convenient to handle. \excitingworkflow, \excitingtools, and \excitingscripts, are \exciting-specific workflow tools (\cref{sec:New_WFs_tools}). Our new version of \elastic (\cref{sec:New_Elastic_tools}) to compute second- and higher-order elastic constants, supports also input from other codes. In addition to the  already mentioned interface to SIRIUS \cite{sirius_repo} (\cref{sec:New_SIRIUS_tools}), there is an interface to the \gls{cc} code Cc4s~\cite{Cc4s-code} (\cref{sec:New_CC_tools}). \exciting’s capability to compute transport properties by solving the electronic linearized \gls{bte} within the constant relaxation time approximation is complemented by an interface to \elphbolt~\cite{Protik2022a} (\cref{sec:New_elphbolt_tools}), enabling the calculation of transport properties beyond the RTA, including drag effects. The stand-alone code \cell~\cite{Rigamonti2024}, a {\python}-based \gls{ce} package, developed in the group, naturally works with \textit{ab initio} input from \exciting.
\cell also relates to \gls{ml}, since the \gls{ce} technique itself can be viewed as an \gls{ml} problem, \ie linear regression. One of the latest developments goes beyond by incorporating non-linearities \cite{Stroth2025} in the \gls{ce} models. A novel workflow combines \cell and \exciting by using the NOMAD infrastructure for building \gls{ce} models for disordered systems. This has been implemented in the NOMAD Oasis, which is organized by the \exciting developers from Berlin in collaboration with NOMAD (\url{https://nomad-lab.eu}). We report on this Oasis (\cref{sec:New_Oasis_data}), on \gls{ml} tools for error quantification (\cref{sec:New_error_data}), and on \fsvisual, a package for viewing Fermi surfaces (\cref{sec:New_FSvisual_data}). 

\paragraph{Tutorials.} Almost all of \exciting's features are covered by comprehensive, user-friendly tutorials, which not only teach new users the necessary know-how for running calculations with \exciting, but have also proven to be a valuable guidebook for experienced users. In their latest version, every tutorial is implemented in a Jupyter notebook \cite{jupyter}. The users can follow the instructions and execute accompanying commands in their web browsers, avoiding the need to switch between the command line and the tutorial. The tutorials can also be executed in a \gls{ci} pipeline (see the test suites below), complementing the test suite with further applications of the code. This rigorous regression testing is able to unveil bugs before users could be affected. Finally, the Jupyter framework allows direct conversion to an \gls{html} file, which is embedded into the \exciting website (\url{https://exciting-code.org/home/tutorials-jupyter}). Thus, users can still view the tutorials and follow along without running a Jupyter notebook on their machine. The relevant functionality is implemented in Python and distributed as part of the \exciting source code.

\paragraph{Test suite and continuous integration.} To ensure the high quality of \exciting results, the code undergoes extensive testing. We distinguish three tiers of tests. The first tier consists of \textit{unit tests} to verify that isolated functions and subroutines---the fundamental building blocks of \exciting---produce the expected results. The second tier involves \textit{regression tests}, which execute \exciting calculations and check their outputs against reference results to ensure numerical consistency. The third tier comprises \textit{workflow tests}, which automatically run the \exciting tutorials to verify that complex workflows involving multiple sequential runs function reliably. All tests are executed regularly within the development workflow through \gls{ci}. The \gls{ci} system builds the code and runs the entire test suite across different environments and build configurations, ensuring robustness and reproducibility.

\subsection{Code summary and availability}
\label{sec:New_availability_code}

\exciting is developed as an open-source package and is publicly available for download at: \url{https://exciting-code.org/} and \url{https://github.com/exciting/}. The code is primarily written in Fortran 2018, with performance-critical components optimized for use on high-performance computers. The code requires a Fast Fourier Transform library that supports the FFTW3 interface and BLAS/LAPACK implementation, \eg OpenBLAS, Intel MKL, Cray LibSci, etc. Also a Fortran compiler capable of MPI and OpenMP is required. The software is bundled with Libxc, FoX, and BSPLINE-FORTRAN, while also allowing the use of an externally provided Libxc installation. The build can optionally interface with HDF5 to enable scalable, platform-independent I/O. To accelerate performance, the code can interface with ELPA, ScaLAPACK, and SIRIUS. GPU-aware builds rely on vendor-specific accelerator libraries, such as Intel oneAPI, CUDA/cuBLAS, or ROCm, depending on the target platform. Moreover, an optimized binary is available for physically unified-memory systems, representing cutting-edge hardware where the CPU and GPU share the memory space.

Based on CMake, the build system supports a wide range of contemporary platforms and toolchains. Currently supported Fortran compilers include GNU Fortran (GCC~$\geq$~11), Intel oneAPI Fortran (IFX) [excluding the \gls{bse} module], Cray, and the AMD compiler Flang. Support for legacy compilers, including Intel IFORT, is currently being deprecated in favor of standards-compliant, actively maintained compiler toolchains. GPU support requires a performant compiler capable of OpenMP offloading, for which currently, Cray, AMD Flang, and IFX are supported.

The code is distributed under the GNU General Public License, ensuring free use, modification, and redistribution in accordance with open-source principles. The release series follows a naming convention based on elements of the periodic table, with the current version being \texttt{exciting-sodium}. All development within the core developer group follows a thorough peer-review prior to integration using an in-house git repository hosted on the gitlab servers of the Humboldt-Universit\"at zu Berlin. External developments and co-developments are welcome. If you are interested in contributing, please contact us.

%% file: sections/basis.tex
\subsection{The (linearized) augmented planewave + local-orbital method}
\label{sec:SOTA_LAPW}

The \exciting code is designed for the computation of excited-state properties of solids~\cite{gulans2014exciting}. At its core, there is a highly precise all-electron \gls{gs} framework based on the \gls{lapwlo} method~\cite{andersen1975lapw,singh2006lapw}. It provides systematically controllable numerical solutions of the \gls{ks} or \gls{gks} equations for periodic systems. The quality of the \gls{ks} eigenvalues and wavefunctions obtained at this stage is critical, as they form the starting point for higher-level methods considered in this review. In the \gls{lapwlo} method, the \gls{ks} equations are solved without shape approximations to the \gls{ks} potential, \ie without the use of pseudopotentials. The method combines a systematic path to basis-set completeness with an efficient all-electron representation of the \gls{ks} wavefunctions, allowing for high precision at manageable computational cost. This balance is essential for large-scale high-throughput and excited-state calculations. 

\begin{figure}
    \centering
    \includegraphics[width=0.35\textwidth]{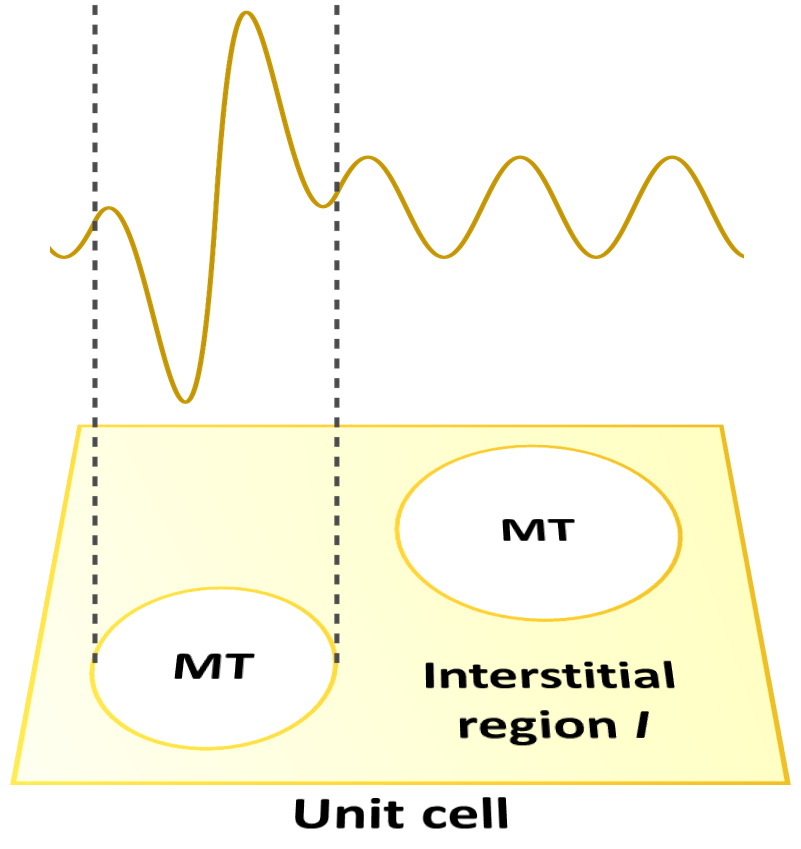}
    \includegraphics[width=0.35\textwidth]{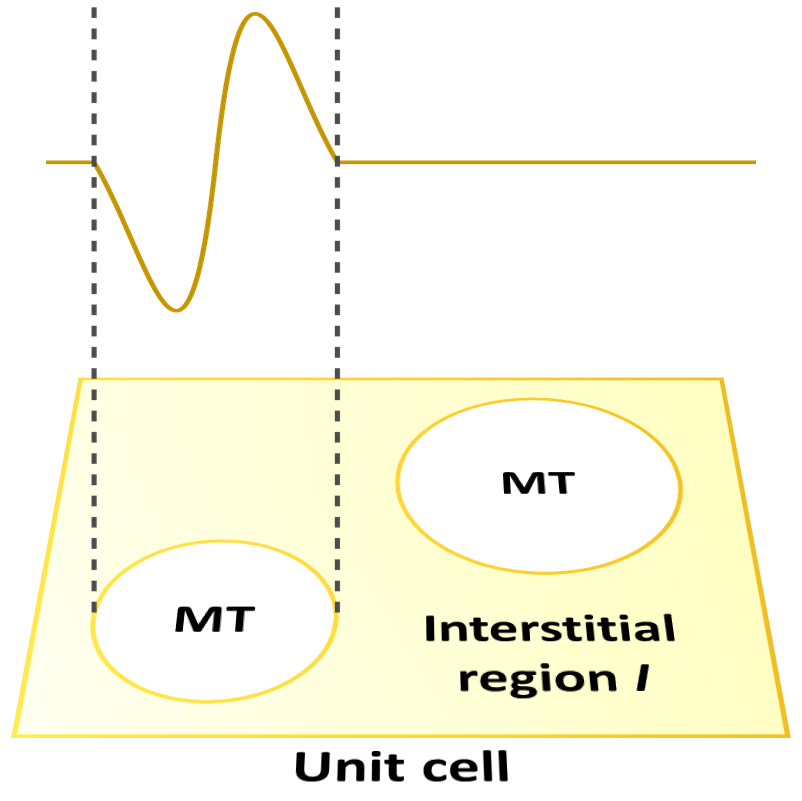}
    \caption{Schematic representation of a \acrfull{lapw} basis function (left) and a \acrfull{lo} (right).}
    \label{fig:LAPWLO}
\end{figure}

The central concept of the \gls{apw} framework is the spatial partitioning of the unit cell into non-overlapping \gls{mt} spheres with radii $R^{\atomidx}_{\ts\mathrm{MT}}$, centered at the atomic positions $\atompos_{\atomidx}$, and an interstitial region $I$~\cite{slater1937wave}. This partitioning reflects the distinct physical character of the electronic states, which are rapidly varying close to atomic nuclei and smooth in the interstitial region, and it enables the construction of basis functions tailored to each region as visualized in \cref{fig:LAPWLO}. The basis for the \gls{ks} wavefunctions $\kswf_{n\kk}(\pos)$, labeled by band index $n$ and Bloch wave vector $\kk$, consists of \glspl{apw}, which extend throughout the unit cell, and \glspl{lo}, which are strictly confined to individual \gls{mt} spheres,
\begin{equation}
\kswf_{n\kk}(\pos) 
=
\sum_{\GG}
\cwf_{n\GG}^{\kk} \,
\basis_{\GG}^{\kk}(\pos)
+
\sum_{\mu}
\cwf_{n\mu}^{\kk} \,
\basis_{\mu}(\pos).
\label{eq:KS-expansion}
\end{equation}
Here, $\GG$ runs over reciprocal lattice vectors, $\mu$ labels local orbitals, and $\cwf_{n\GG}^{\kk}$ and $\cwf_{n\mu}^{\kk}$ denote the expansion coefficients of the \gls{apw} and \gls{lo} functions, respectively.

\subsubsection*{Augmented planewaves and linearization}

The basis functions used in the \gls{lapw} framework are defined piecewise according to the spatial partitioning of the unit cell,
\begin{equation}
\basis_{\GG}^{\kk}(\pos) =
\begin{cases}
\displaystyle
\frac{1}{\sqrt{\unitCellVol}}\, {\rm e}^{\im(\GG+\kk)\cdot\pos},
& \pos \in I,\\[1.2em]
\displaystyle
\sum_{\ell m \xi}
\matching^{\GG+\kk}_{\ell m \atomidx \xi}\:
\radial_{\ell\atomidx\xi}(r_{\atomidx};\eigval_{\ell\atomidx}) \,
Y_{\ell m}(\hat{\pos}_{\atomidx}),
& \pos \in \mathrm{MT}_{\atomidx}\:,
\end{cases}
\label{eq:apw-def}
\end{equation}
where $\unitCellVol$ denotes the unit-cell volume and $\pos_{\atomidx}=\pos-\atompos_{\atomidx}$ defines local coordinates with respect to the position of atom $\atomidx$ inside $\mathrm{MT}_{\atomidx}$. For valence electrons, the radial functions $\radial_{\ell\atomidx\xi}(r_{\atomidx};\eigval_{\ell\atomidx})$ are obtained by solving an effective one-electron radial problem in the spherically averaged \gls{ks} potential $v_{0}^{\atomidx}(r)$. Its solution depends on the \gls{ks} energy, which implies that the radial functions entering the basis are explicitly energy dependent. This explicit energy dependence of the radial functions renders the \gls{ks} eigenvalue problem in the original \gls{apw} formulation \cite{slater1937wave} nonlinear. Within the \gls{lapw} framework~\cite{andersen1975lapw}, this difficulty is avoided by introducing fixed reference energies $\eigval_{\ell\atomidx}$, resulting in a generalized linear eigenvalue problem. This approximation introduces a linearization error that increases with the deviation of the true eigenvalue from the chosen reference, making a fixed-energy \gls{apw} basis insufficient for an accurate description of the energy range of one or more bands. This problem is fixed by enriching the radial part of the basis inside the \gls{mt} spheres with energy derivatives of the radial functions (captured in \cref{eq:apw-def} by the index $\xi$) evaluated at the reference energies. Each basis function has to satisfy specific matching conditions at the sphere boundary, which fix the coefficients $\matching^{\GG+\kk}_{\ell m \atomidx \xi}$. In the standard \gls{lapw} method, only the first energy derivative is used, giving rise to two matching conditions that enforce continuity of the basis function and its radial derivative at the sphere boundary. In \exciting, the reference energies $\eigval_{\ell\atomidx}$ are determined automatically from the initial potential following the Wigner-Seitz prescription~\cite{andersen1973}, \ie placing them in the center of the energy window spanned by the corresponding band, thus avoiding material-specific tuning. This linearization strategy defines the variational basis used for the description of valence states within the \gls{lapw} framework.

For completeness, we note that core states are treated separately within the all-electron framework. Owing to their strong localization and energetic separation from the valence manifold, core states are obtained by solving the radial Dirac equation inside the \gls{mt} spheres, using the respective spherically averaged \gls{ks} potential. These states do not have contributions in the interstitial region and thus do not enter the variational basis expansion in Eq.~\eqref{eq:KS-expansion}, although they are obtained from the same all-electron \gls{ks} Hamiltonian as the valence states. Being updated self-consistently in every iteration, they fully contribute to the self-consistent density and potential. They are also essential for core-level spectroscopies implemented in \exciting.

\subsubsection*{Local orbitals}

\glspl{lo} are basis functions that are strictly confined to a single \gls{mt} sphere and vanish identically in the interstitial region~\cite{singh1991lo},
\begin{equation}
\basis_{\mu}(\pos)=
\begin{cases}
\displaystyle
\sum_{\xi}
\locoeff_{\mu \xi}\,
\radial_{\mu\xi}(r_{\atomidx};\eigval_{\mu\xi})\,
Y_{\ell_\mu m_\mu}(\hat{\pos}_{\atomidx}),
& \pos\in\mathrm{MT}_{\atomidx},\\[0.8em]
0, & \pos\in I.
\end{cases}
\label{eq:lo-def}
\end{equation}
They can be used for an alternative way of linearizing the eigenvalue problem, as realized in the \gls{apwlo} method \cite{sjostedt2000apwlo}: By adding \glspl{lo} as additional basis functions, the energy derivatives required in the \gls{lapw} approach can be avoided, yielding a more flexible basis. Moreover, this explicit construction of additional radial solutions at selected energies enables an accurate treatment of semi-core states as well as a systematic improvement of the description of high-energy and high-angular-momentum components, which are essential for excited-state calculations. A sufficiently rich \gls{lo} basis also permits the use of larger \gls{mt} radii, which reduces the number of \glspl{apw} required and thus improves computational efficiency. Since all of these variants can be realized within a single calculation, we use the generic expression \gls{lapwlo} for the basis throughout this review unless explicitly addressing a particular aspect.

\subsubsection*{Basis-set completeness}

The \gls{lapwlo} basis offers two technically distinct routes to systematic convergence. Increasing the number of \glspl{apw} controls completeness in the interstitial region, while enriching the set of \glspl{lo} improves the quality of the basis inside the \gls{mt} spheres. This separation enables an internal consistency check of basis-set completeness. While the completeness of the \gls{apw} basis can be assessed straightforwardly by increasing the planewave cutoff, the convergence of the \gls{lo} basis is less transparent. \glspl{lo} do not form a naturally ordered hierarchy, and their completeness cannot be inferred from a single scalar cutoff parameter. To assess the quality of the \gls{mt} basis in a controlled manner, we employ a \gls{dbsv} procedure. The central idea behind is to use a well-converged interstitial planewave basis as an internal reference. This procedure involves two otherwise identical calculations that differ only in the choice of \gls{mt} radii. The resulting difference in the total energy per atom isolates the ability of the \gls{mt} basis to reproduce the physics of a spatial region that is otherwise accurately described by planewaves. A small energy difference indicates that the \gls{mt} basis is sufficiently complete and consistent with the converged interstitial reference. More details on the \gls{dbsv} methodology and its use for basis assessment and construction will be published elsewhere~\cite{excitingbasis}.

\subsubsection*{Automated and transferable basis-set selection}

While the \gls{lapwlo} basis provides a systematic route to numerical convergence, the practical construction of a basis set is often guided by user experience. In \exciting, recent developments aim to replace these heuristic choices by a transferable and largely automated procedure. Users specify a desired numerical precision, and \exciting determines consistent planewave cutoffs, \gls{mt} radii, and compact \gls{lo} sets automatically.

\paragraph*{Hierarchies of local orbitals from dual basis self-validation.}
In Ref.~\cite{excitingbasis}, \gls{dbsv} is used to construct hierarchies of local orbitals by quantifying their contribution to the completeness of the \gls{mt} basis. This makes it possible to identify near-redundant functions and to build the smallest \gls{lo} set required to reach a target precision without extensive manual convergence testing.

\paragraph*{Control of the \gls{apw} completeness.}
To control the number of \glspl{apw} for a desired precision in a material-independent way, \exciting introduces a numerical quality parameter derived from the conventional cutoff $R_{\ts\mathrm{MT}}|\mathbf{G}+\mathbf{k}|_{\max}$. Reference values of this parameter were tabulated for elemental solids, corresponding to a total-energy error of approximately $0.1\,\mathrm{meV/atom}$ with respect to an extrapolated complete-basis limit~\cite{Carbogno2022}. These values are used as species-dependent scaling factors, yielding a single parameter that controls the \gls{apw} basis precision in a transferable manner across different materials.

\paragraph*{Automatic and balanced choice of muffin-tin radii.}
The same reference data are further exploited to automate the choice of \gls{mt} sphere radii. For a given global planewave cutoff, individual radii are scaled such that different atomic species contribute to the overall basis-set error comparably. Combined with the \gls{dbsv}-based \gls{lo} hierarchies, this strategy provides a consistent, precision-driven basis specification suited for both \gls{gs} calculations and the excited-state methods built on top of them.

%% file: sections/groundstate.tex
\subsection{State of the art}
\label{sec:SOTA_GS}
The accurate description of the electronic \gls{gs} constitutes the conceptual and numerical basis for all excited-state methodologies discussed in this review. \gls{ks} theory~\cite{hohenberg1964,kohn1965} is the method of choice on which most \gls{gs} calculations for extended systems are based. Solving the \gls{ks} equations yields the \gls{ks} eigenvalues and wavefunctions, which are used not only to obtain the electronic \gls{gs} density and energy, but also to establish a well-defined starting reference for \gls{mbpt}, linear-response approaches, and spectroscopic simulations. 

The accuracy and efficiency of \gls{ks}-\gls{dft} depend critically on the level of approximation adopted for the \gls{xc} functional. Considerable efforts are being made to implement and use better functionals beyond the most widely used semilocal ones. Higher levels of sophistication can be reached by climbing up John Perdew's Jacob's ladder~\cite{Perdew2001jacob}. Among the most widely used beyond-semilocal approximations in solid-state \gls{dft} are \gls{mgga} and hybrid functionals, which improve upon local and semilocal functionals at a moderate additional computational cost. \gls{mgga} functionals~\cite{SunPRL2015,perdew2017,AschebrockPRR2019} often yield improved accuracy with only a moderate increase in numerical effort, sometimes even matching the accuracy of significantly more demanding hybrid \gls{xc} functionals~\cite{Liu2023}. A rather complete list of \gls{mgga} functionals is available in the Libxc library~\cite{Lehtola2018,Chachiyo2020}. 

The introduction of nonlocal contributions to the effective potential renders the resulting single-particle potential explicitly orbital-dependent. While the \gls{oep} method constructs a local potential~\cite{Goerling1996} and thus remains within \gls{ks}-\gls{dft}, the inclusion of \gls{hf} exchange requires solving the \gls{gks} equations~\cite{Seidl1996,Perdew1996a}. By partly curing the spurious self-interaction effects introduced in semilocal functionals, hybrid functionals aim at improving single-particle states. While \gls{ghs} like B3LYP~\cite{becke1993} or PBE0~\cite{Perdew1996a} contain a constant \gls{hf} contribution, more recent developments aim at considering spatial variations of \gls{xc} effects. In \glspl{lh} functionals~\cite{LhReview_2019}, a real-space position-dependent \gls{hf} admixture is used, controlled by a local mixing function. \Gls{lh} functionals explicitly incorporate terms to deal with static correlation as well as with delocalization errors~\cite{Kaupp2024}. In range-separated (screened) hybrids, the exchange interaction is partitioned into short- and long-range parts; the mixing fraction \(\alpha\) is typically chosen as a constant, while a range-separation (screening) parameter controls which length scales are treated with \gls{hf} exchange. In the case of screened hybrids such as HSE~\cite{heyd2003,Heyd2006}, the Coulomb operator is split and \gls{hf} exchange is only used for short-range interactions to screen the long-range Coulomb singularity. This is highly effective for reducing computational costs in periodic solids. Reducing the self-interaction error compared to semilocal \gls{dft}, hybrid functionals yield significantly improved orbital energies, including band gaps and ionization potentials, and provide a more reliable single-particle electronic structure as a starting point for excited-state theories~\cite{becke1993,heyd2003,Cui2018,ghosh2024accurate,Sagredo2025}. For periodic systems, HSE06~\cite{heyd2003,Heyd2006} is among the most widely used cost-effective hybrid functionals, employing a fixed mixing \(\alpha=1/4\) and a standard screening parameter \(\omega\approx 0.11\,\mathrm{bohr}^{-1}\). Some developments for solids aim to connect the amount and/or range of \gls{hf} exchange to the screening, for example in dielectric-dependent and optimally tuned range-separated hybrid functionals~\cite{Skone2014,refaelyabramson2015solidstate,Wing2021,Zheng2019}.

By explicitly including information from unoccupied \gls{ks} orbitals, the \gls{xc} functionals on the fifth (top) rung of Jacob's ladder enable the description of nonlocal electron-correlation effects. The two representative classes of fifth-rung functionals that have been actively developed in recent years are the \gls{rpa} and \glspl{dha}. Both approximation types can be derived within the framework of the \gls{ac} approach to \gls{dft}~\cite{Gunnarsson1976,Langreth1975}. In particular, \gls{rpa} can resolve subtle energy differences in various chemical environments~\cite{Casadei2016,Yang_2022} and systems with relevant \gls{vdw} interactions~\cite{Gao2020rpa,Bjorkman2012} with very high accuracy compared to lower-rung functionals. Both \gls{rpa} and \glspl{dha} are reviewed in detail in Ref.~\cite{Ying_Zhang_2025}.

All functionals up to the fourth rung, \ie  local, semilocal, \gls{mgga}, and hybrid functionals, do not capture long-range dispersion without an explicit nonlocal correlation term or an additional dispersion correction. While semiempirical dispersion schemes (including atom-pairwise and many-body variants)~\cite{Grimme2011,Tkatchenko2009,Tkatchenko2012,Hermann2020} provide an efficient description, these corrections are typically added on top of the underlying \gls{xc} functional. A more fundamental approach is offered by nonlocal vdW-DF functionals~\cite{Dion2004,Tran2019vdw,Shukla2022}, in which dispersion interactions are determined directly from the electron density. These functionals have recently been implemented in the code-agnostic library libvdwxc~\cite{Larsen_2017}, enabling a consistent density-based treatment of long-range correlations.

In parallel to classical approaches to developing \gls{xc} functionals, machine-learned functionals have emerged as a data-driven route to improved accuracy, ranging from neural-network density functionals to physics-informed constructions that target known exact constraints and challenging regimes such as fractional charges~\cite{Pederson2022,Dick2020,Kirkpatrick2021}.

Beyond the choice of \gls{xc} functional, the numerical realization of the \gls{ks} or \gls{gks} equations plays a decisive role in the reliability of \gls{gs} results, including the representation of wavefunctions and potentials, as well as the treatment of core and valence electrons. All-electron methods~\cite{singh2006lapw,gulans2014exciting} avoid pseudoization and the frozen-core approach and treat core and valence electrons on the same footing, thereby ensuring transferability across the periodic table and enabling direct access to core-level properties. In-depth comparisons of different implementations and codes are still extremely rare~\cite{Kurt2016Science} and are mostly restricted to simple materials, selected properties, and semilocal functionals. In view of this, all-electron implementations like \exciting can play a major role in providing benchmark data for advanced functionals and higher-level methods using \gls{dft} as a starting point.

\subsection{Methodology}
\label{sec:Methodology_GS}
In the following, we sketch the \gls{dft} formalism only to the extent needed to follow the remainder of this work. For more general considerations and details, we refer to the ample literature on this topic.

The practical realization of the \gls{ks}-\gls{dft} formalism relies on the numerical solution of the \gls{ks} equations, a set of Schr\"odinger-like single-electron equations,
\begin{equation}\label{eq:kseqs}
\hKS\,\kswf_{i\kk}(\pos)=
\left[
-\frac{\nabla^2}{2} + \vKS(\pos)
\right]
\!\kswf_{i\kk}(\pos) = \eigval_{i\kk}\,\kswf_{i\kk}(\pos)\:,
\end{equation}
which provide the electron density
\begin{equation}\label{eq:rhogs}
\nKS(\pos)=\sum\limits_{i,\kk} \kptweight {\kk }\,\occ{i}{\kk}\,\vert\kswf_{i\kk}(\pos)\vert^2\:,
\end{equation}
where $\occ{i}{\kk}$ denotes the occupation factors, and $\kptweight {\kk }$ represents the \kk-point weights. \Gls{mb} effects are included in a mean-field manner in the \gls{ks} potential, 
\begin{equation}\label{eq:kspot}
\vKS(\pos)=v_{\rm ext}(\pos)+v_{\ts\rm H}(\pos)+v_{\rm xc}(\pos)\:\qquad\textrm{with}\qquad v_{\rm xc}(\pos)=
\frac{\delta E_{\textrm{xc}}[n]}
{\delta n(\pos)}\:.
\end{equation}
It consists of the interaction with the nuclei, $v_{\rm ext}(\pos)$, the classical Hartree contribution, $v_{\ts\rm H}(\pos)$, and the \gls{xc} potential, $v_{\rm xc}(\pos)$. Both $v_{\rm xc}(\pos)$ and the \gls{xc} energy functional $E_{\textrm{xc}}[n]$ must be approximated. The lowest-level approximations are (semi)local functionals. The \gls{lda} depends on the \gls{gs} electron density only. \glspl{gga} incorporate also the density gradient. By climbing up a further step of Jacob's ladder~\cite{Perdew2001jacob}, we arrive at the \gls{mgga} functionals that include the second derivative of the electron density in the form of the Laplacian $\nabla^2n(\pos)$ and/or the \gls{ked}
\begin{equation}
    \tau(\pos) = \frac{1}{2}\sum\limits_{i,\kk} \kptweight {\kk }\,\occ{i}{\kk}\,|\nabla \kswf_{i\kk}(\pos)|^2 \;.
\end{equation}
Since the \gls{ked} is not an explicit functional of the density, functional derivatives with respect to the \gls{ks} orbitals are used. This gives rise to an additional non-multiplicative contribution to the \gls{xc} potential
\begin{equation}
    \hat{v}_{\rm xc}^{\rm mGGA}\, \kswf_{i\kk}(\pos) = v_{\rm xc}^{\rm mult}(\pos)\, \kswf_{i\kk}(\pos) 
    - \frac{1}{2} \nabla \cdot \left[ \frac{\partial \varepsilon_{\rm xc}^{\rm mGGA}}{\partial \tau(\pos)}\, \nabla \kswf_{i\kk}(\pos) \right] \;.
\end{equation}

The next rung of Jacob's ladder incorporates also occupied orbitals into the \gls{xc} functional. In hybrid functionals, the Fock-exchange operator is included with a fraction $\alpha$, while the remaining exchange effects are treated on the semilocal level,
\begin{equation}
    \label{eq:hybrids}
    E^\mathrm{hyb}_{\mathrm{xc}} =\alpha E^\mathrm{HF}_{\mathrm{x}}+ (1-\alpha)E^\mathrm{GGA}_{\mathrm{x}}+ E^\mathrm{GGA}_{\mathrm{c}}.
\end{equation}
 $E^\mathrm{HF}_{\mathrm{x}}$ and $E^\mathrm{GGA}_{\mathrm{x}}$ are the Fock and \gls{gga} exchange energies, respectively. A prominent example of such a hybrid functional is PBE0~\cite{Perdew1996a}, with the mixing parameter $\alpha$=0.25. The nonlocal Fock exchange requires to solve the \gls{gks} equations, which in the most general case are written as
\begin{equation}\label{eq:gkseqs}
\left[
-\frac{\nabla^2}{2} + v_{\rm ext}(\pos) + v_{\ts\rm H}(\pos) + 
\hat{v}^{\mathrm{NL}}_{\rm x}[\{\kswf_{i\kk}\}]+{v}^{\mathrm{SL}}_{\mathrm{xc}}[n](\pos)
\right]
\!\kswf_{i\kk}(\pos) = \eigval_{i\kk}\,\kswf_{i\kk}(\pos)\:,
\end{equation}
where the nonlocal operator $\hat{v}^{\mathrm{NL}}[\{\kswf_{i\kk}\}]$ depends on all occupied orbitals, and ${v}^{\mathrm{SL}}_{\mathrm{xc}}[n](\pos)$ is a semilocal potential, which formally includes the residual \gls{xc} interactions. Extensions to the simple form of \cref{eq:hybrids} aim to improve the long-range properties of the approximate exchange hole by screening the exchange interaction at longer distances. The most prominent example of range-separated hybrid functionals for solids is HSE~\cite{heyd2003,Heyd2006}, reading
\begin{equation}
    \label{eq:HSE}
    E^\mathrm{HSE}_{\mathrm{xc}} = 
    E^\mathrm{PBE}_{\mathrm{xc}} + \alpha 
    \bigg[
    E^\mathrm{HF,sr}_{\mathrm{x}}(\omega)-
    E^\mathrm{PBE,sr}_{\mathrm{x}}(\omega)
    \bigg],
\end{equation}
where $E^\mathrm{PBE}_{\mathrm{xc}}$ is the PBE \gls{xc} energy~\cite{PerdewPRL1996} and $\omega$ is a range-separation parameter that determines the spatial division between long-range and short-range exchange contributions.

\subsection{\Acrlongpl*{mgga} in \excitingb}
\label{sec:New_metaGGA_GS}

\begin{figure}[h]
\centering
\includegraphics[width=0.45\textwidth]{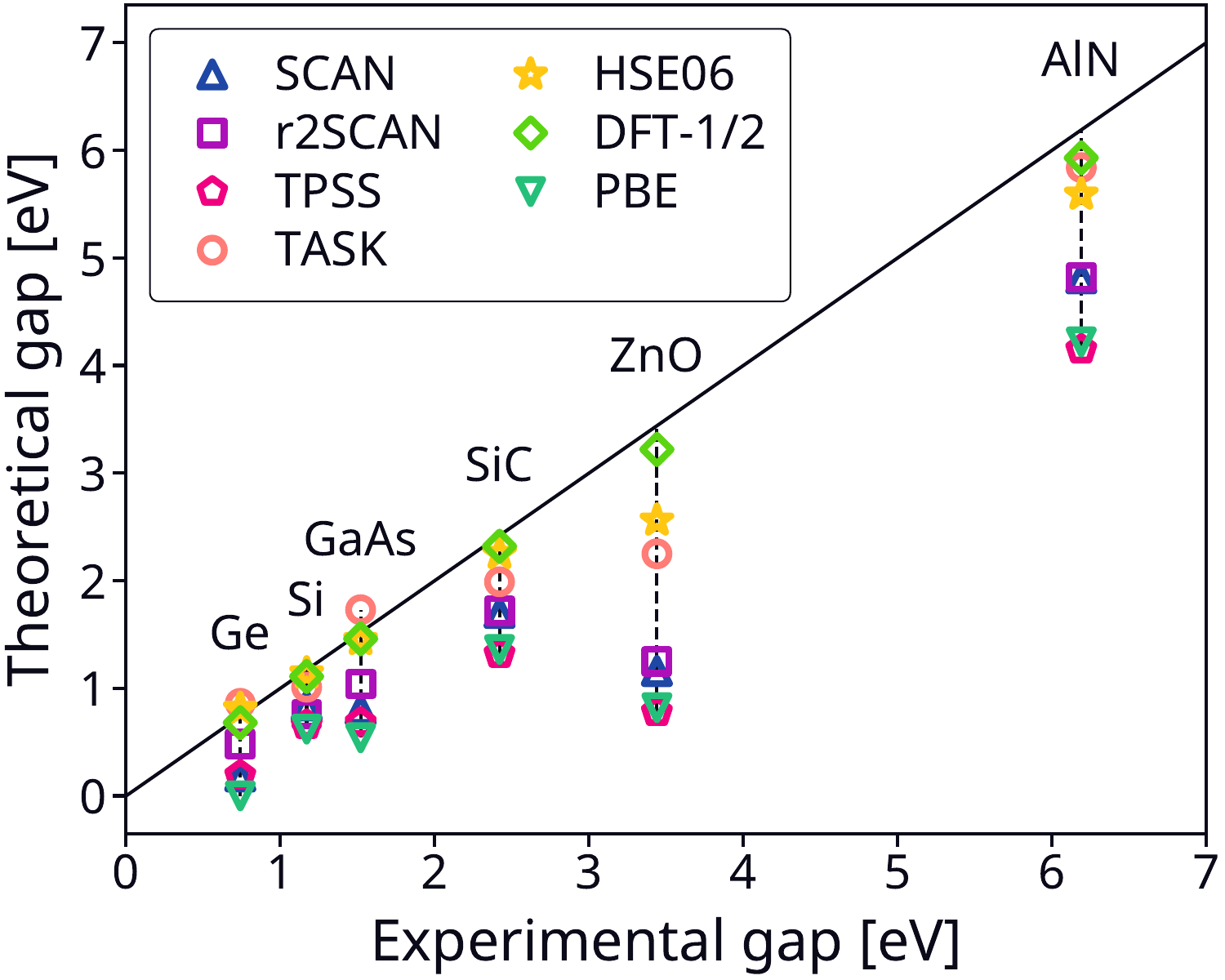}
\caption{Band gaps obtained with different metaGGA functionals. For comparison, PBE~\cite{Uzulis2025}, HSE06~\cite{Uzulis2025}, and DFT-1/2 results are shown. All calculations were carried out with \exciting.}
\label{fig:mgga_band_gaps}
\end{figure}

\exciting implements \gls{mgga} functionals using the Libxc library~\cite{Lehtola2018,Chachiyo2020}. The functionals SCAN~\cite{SunPRL2015}, r2SCAN~\cite{Furness2020}, TASK~\cite{AschebrockPRR2019}, TPSS~\cite{Tao2003}, and HLE17~\cite{Verma2017} come preinstalled. All other functionals provided by Libxc can be easily added without additional programming effort. \gls{mgga} can improve over the \gls{lda} or \glspl{gga} in the prediction of band gaps at only a slight increase in computational cost, yielding results of similar quality as hybrid functionals. For the prediction of band gaps, TASK proves favorable over other \gls{mgga} functionals as shown in~\cref{fig:mgga_band_gaps}, while SCAN or r2SCAN yield lattice constants closest to experiment. 

Concerning the implementation, complications arise when solving the radial Schr\"odinger and Dirac equations for obtaining the radial parts of the \gls{apw} basis functions and the core states, respectively. The standard outward integration approach does not work because the action of the (spherical) potential on the radial functions is unknown prior to integration but depends on the yet unknown radial functions due to the presence of a non-multiplicative contribution to the potential. 
Therefore, \exciting follows Refs.~\cite{Doumont2022,Tran2019} and circumvents this issue by computing a \gls{gga} potential in each iteration and using its spherical part for the update of core states and radial basis functions. As a consequence, a \gls{mgga} calculation in \exciting starts with a \gls{gga} iteration to produce an initial set of wavefunctions, from which the \gls{ked} is computed. From then onward, the \gls{mgga} potential is used to compute the Hamiltonian, the wavefunctions, and the electron density, which is then used to update both the \gls{mgga} and \gls{gga} potentials. In principle, there is an optimal type of \gls{gga} functional best suited for each type of \gls{mgga} functional~\cite{Tran2019}. In practice, however, it turns out that PBE performs well for all flavors of \gls{mgga}.

\subsection{\Acrshort*{dft}-1/2}
\label{sec:New_DFThalf_GS}

While methods such as \gw or hybrid functionals produce significantly improved band gaps compared to semilocal \gls{dft}, they are computationally much more demanding and may become impractical for large or complex systems. The \gls{dft}-1/2 approach~\cite{ferreira2008approximation} offers a compelling compromise: band gaps, ionization potentials, and defect levels can be obtained with an accuracy close to that of higher-level methods while retaining the low computational cost of semilocal functionals~\cite{ferreira2008approximation, pela2015, pela2011accurate, pela2012, Pela2017, ferreira2013, Pela2018, RodriguesPelaPRB2016, pela2024electronic}. This makes \gls{dft}-1/2 a highly efficient alternative when more demanding methods are not feasible.

The \gls{dft}-1/2 method resembles Slater's transition-state technique, in which the ionization energy of a given single-particle state is approximated by enforcing its half-occupation. In \gls{dft}-1/2, however, half-occupations are not treated explicitly. Instead, a modified set of \gls{ks} equations is solved:
\begin{equation}\label{eq:dfthalf}
\left[
-\frac{\nabla^2}{2} + \vKS(\pos) - {v}_{\ts\mathrm{S}}(\pos)
\right]\!\kswf_{i\kk}(\pos) = \eigval_{i\kk}\,\kswf_{i\kk}(\pos),
\end{equation}
where ${v}_{\ts \mathrm{S}}$ is the so-called ``self-energy potential'' named for its analogy with the classical self-energy. The potential ${v}_{\ts \mathrm{S}}$ is constructed to incorporate the effect of half-occupation. For a compound, it is typically derived from \gls{ks} calculations of the isolated constituent atoms and subsequently trimmed to remove the $1/r$ divergence~\cite{ferreira2008approximation}. As evident from~\cref{eq:dfthalf}, \gls{dft}-1/2 retains the simplicity of semilocal functionals. At the same time, it can yield electronic properties with an accuracy comparable to that of \glspl{mgga}, hybrid functionals, or even \gw. This is illustrated in \cref{fig:mgga_band_gaps} for the band gaps of a variety of materials. Therefore, \gls{dft}-1/2 also provides an appealing starting point for $G_0W_0$. In this case, the resulting \gls{qp} shifts are expected to be small, so the perturbative treatment is well justified~\cite{RodriguesPelaPRB2016}. Accordingly, the \gls{dft}-1/2 method is available as a starting-point for $G_0W_0$~\cite{RodriguesPelaPRB2016} in \exciting. A detailed description of the implementation can be found in Ref.~\cite{Pela2017}.

\subsection{Hybrid functionals in \excitingb}
\label{sec:New_hybrids_GS}

\exciting implements the hybrid functionals PBE0~\cite{Perdew1996a} and HSE06~\cite{heyd2003,Heyd2006}. Both require the nonlocal exchange defined as
\begin{equation}\label{eq:exop}
\hat{v}^\mathrm{HF}_{\mathrm{x}} \psi_{n \mathbf{k}}(\mathbf{r})=-\frac{1}{N_{\mathbf{k}}}\sum_{n^\prime\mathbf{k}^\prime} \psi_{n^\prime\mathbf{k}^\prime}(\mathbf{r})\!
\int\! \psi^*_{n^\prime\mathbf{k}^\prime}(\mathbf{r}^\prime)\, \psi_{n\mathbf{k}}(\mathbf{r}^\prime)\,
v(|\mathbf{r}-\mathbf{r}^\prime|)\, \mathrm{d}\mathbf{r}^\prime,
\end{equation}
where $v(r)=1/r$ is the Coulomb kernel that becomes $v(r)=\mathrm{erfc}(\omega r)/r$ in the screened case required for HSE06. Including this term in a calculation requires evaluating the matrix elements $\langle \phi^{\mathbf{k}}_{\mathbf{G}}|\hat{v}^\mathrm{HF}_{\mathrm{x}} |\phi^{\mathbf{k}}_{\mathbf{G}^\prime} \rangle$, which is implemented in \exciting in two ways~\cite{Vona2022,Zavickis2022}. The first one follows the approach introduced in Ref.~\cite{Betzinger2010} and approximates the exchange operator as
\begin{equation}
\label{eq:hybrids_mb}
    \hat{v}^\mathrm{MB}_{\mathrm{x}} = \sum\limits_{nn^\prime\mathbf{k}}  
    | \psi_{n\mathbf{k}} \rangle\langle \psi_{n\mathbf{k}} |
    \hat{v}^\mathrm{HF}_{\mathrm{x}} | \psi_{n^\prime\mathbf{k}} \rangle\langle \psi_{n^\prime\mathbf{k}} | .
\end{equation}
To obtain the matrix elements $M_{nn^\prime}=\langle \psi_{n\mathbf{k}} | \hat{v}^\mathrm{HF}_{\mathrm{x}} | \psi_{n^\prime\mathbf{k}} \rangle$, products of wavefunctions are expressed in terms of the mixed product basis that is described in more detail in~\cref{sec:New_Implementation_GW} (hence the label MB in~$\hat{v}^\mathrm{MB}_{\mathrm{x}}$). 

The projection of the Fock exchange on the wavefunctions in~\cref{eq:hybrids_mb} introduces a dependence of the total energy and the band energies on the number of empty bands used in the calculation. We overcome this issue and gain control over the precision of hybrid calculations with an alternative implementation \cite{Zavickis2022} that adopts the \gls{ace}~\cite{Lin2016}. It constructs a low-rank representation of the Fock exchange operator in the subspace spanned by the occupied states, 
\begin{equation}
\label{eq:ace2}
    \hat{v}^\mathrm{ACE}_{\mathrm{x}}= \sum\limits_{nn^{\prime}\mathbf{k}} | W_{n\mathbf{k}} \rangle 
    \left( M^{-1}_\mathbf{k} \right)_{nn^\prime} \langle W_{n^\prime\mathbf{k}}| ,
\end{equation}
with $W_{n\mathbf{k}}(\mathbf{r})=\hat{v}^\mathrm{HF}_{\mathrm{x}} \psi_{n \mathbf{k}}(\mathbf{r})$ and converges to the exact result for the given basis once self-consistency is reached. Note that the radial functions are generated using a local multiplicative potential even though the potential in hybrid functionals has a nonlocal contribution. The usual approach in \gls{lapw} is to precompute the radial basis as well as core orbitals with \gls{gga} and use them unmodified in hybrids~\cite{Betzinger2010,Tran2011}. We addressed this deficiency by implementing a radial solver that generates radial functions and core orbitals consistent with hybrid functionals~\cite{Uzulis2022,Uzulis2026}. This approach employed together with \gls{ace} recovers the micro-Ha precision for absolute total energies in calculations with hybrid functionals.

In the \gls{ace} code, we follow~\cref{eq:exop} directly and evaluate the convolution integral for every pair of wavefunctions using the pseudocharge method~\cite{Weinert1981} and its modifications~\cite{Uzulis2025} for the bare and screened Coulomb kernels, respectively. With $N_\mathrm{at}$ being the number of atoms, the computational effort scales as $\bigO(N_\mathrm{at}^4)$ as in the other implementation. In the \gls{ace} code, this scaling derives from calculating the convolution integral that has to be evaluated for all $\bigO(N_\mathrm{at}^2)$ pairs of wavefunctions, requiring $\bigO(N_\mathrm{at}^2)$ floating-point operations for periodic systems. Reference~\cite{Uzulis2025} discusses modifications of the pseudocharge method that allowed us to reduce the computational effort of its most time-consuming steps and thus make first steps towards the overall $\bigO(N_\mathrm{at}^3 \log N_\mathrm{at})$ scaling in \gls{ace} calculations.

\subsection{\Acrshort*{svlo}: An efficient basis for spin-orbit coupling}
\label{sec:New_SVLO_GS}
In materials with sizeable \gls{soc}, the conventional \gls{sv} treatment~\cite{KoellingHarmon1977} may require a prohibitively large number of unoccupied \gls{sr} states to converge \gls{soc}-induced splittings and derived properties. To overcome this limitation, \exciting implements the \gls{svlo} approach, which accelerates \gls{soc} calculations by explicitly enriching the \gls{sv} basis in the vicinity of the atomic nuclei, where relativistic effects are strongest. A detailed description and benchmarks can be found in Ref.~\cite{Vona2023}.

Starting from the \gls{sr} \gls{ks} eigenstates $\kswf^{\mathrm{SR}}_{j\kk}(\pos)$, the conventional \gls{sv} spinors are written as
\begin{equation}
\bm\psi^{\mathrm{SV}}_{n\kk}(\pos)
=
\sum_{\sigma}
\sum_{j=1}^{N_b^{\mathrm{SV}}}
C^{\mathrm{SV}}_{n j\sigma}(\kk)\,
\kswf^{\mathrm{SR}}_{j\kk}(\pos)\Ket{\sigma},
\label{eq:sv_spinor_expand_gs}
\end{equation}
where the \gls{sv} subspace size $N_b^{\mathrm{SV}} = N_{\mathrm{occ}} + N_{\mathrm{unocc}}$ denotes the number of \gls{sr} states entering the \gls{sv} step. Here, $\sigma \in \{\uparrow,\downarrow\}$ is the spin index and $\Ket{\sigma}$ is the corresponding spin-basis state. For systems with strong \gls{soc}, the required number of unoccupied states $N_{\mathrm{unocc}}$ can become prohibitively large. The \gls{svlo} scheme addresses this bottleneck by augmenting the \gls{sv} basis with explicit \glspl{lo}, including also Dirac-type \glspl{lo} that accurately capture the near-nuclear behavior of relativistic states. In particular, when heavy-element $p$ states dominate the band edges, introducing $p_{1/2}$-type \glspl{lo} becomes essential~\cite{Kunes2001}. In the \gls{svlo} approach, the spinors are constructed as 
\begin{equation}
\bm{\psi}^{\mathrm{SVLO}}_{n\kk}(\pos)
=
\sum_{\sigma}
\sum_{m=1}^{N_b^{\mathrm{SVLO}}}\!
C^{\mathrm{SVLO}}_{n m\sigma}(\kk)\,
\chi_{m\kk}(\pos)\Ket{\sigma},
\label{eq:svlo_spinor_expand_gs}
\end{equation}
using $N_b^{\mathrm{SVLO}} = N_{\mathrm{occ}} + N_{\mathrm{unocc}} + N_{\mathrm{LO}}$ basis functions $\chi_{m\kk}(\pos)$ that comprise (i) the \gls{sr} eigenstates with the \gls{lo} contributions omitted and (ii) the \glspl{lo}. 

\begin{figure}[h]
  \centering
  \includegraphics[width=0.8\textwidth]{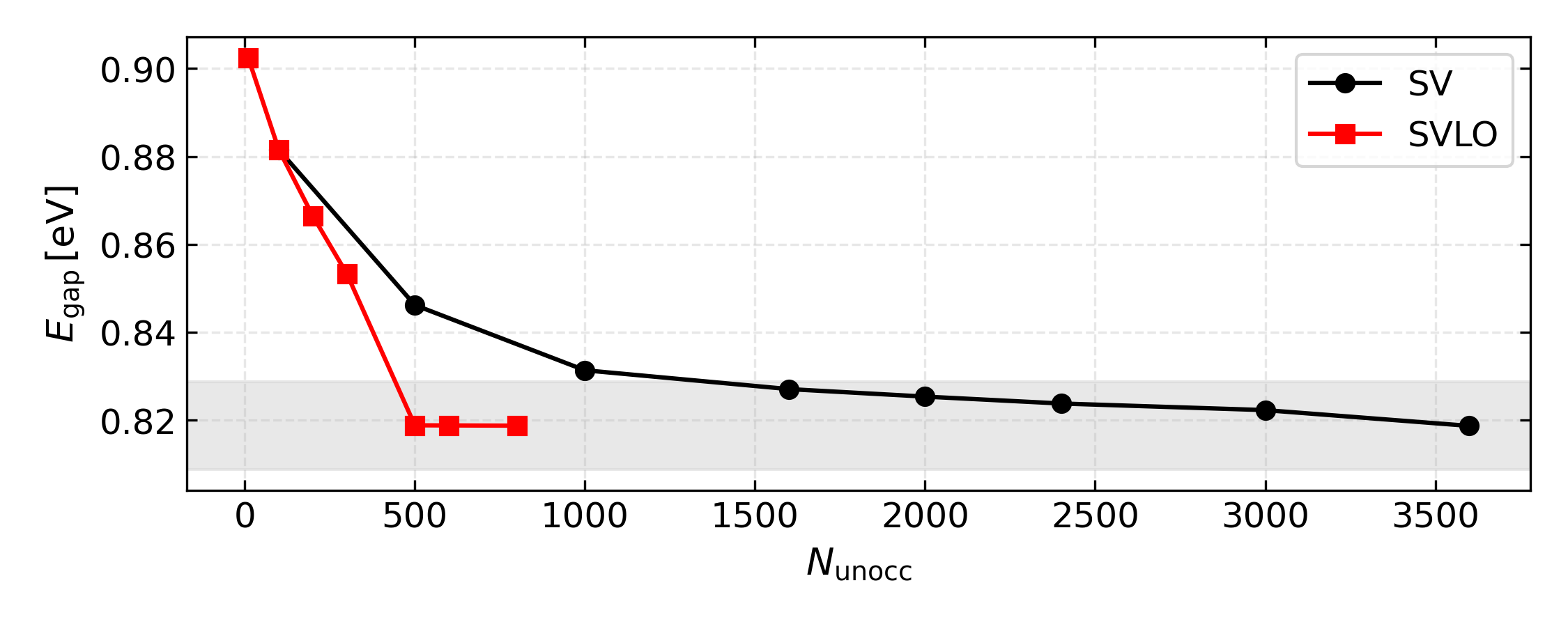}
  \vspace*{-5mm}
  \caption{Convergence of the \gls{ks} band gap of \(\gamma\)-CsPbI\(_3\) with the number of unoccupied states \(N_{\mathrm{unocc}}\), comparing \gls{sv} and \gls{svlo}. The gray shaded area indicates a tolerance window of $\pm 0.01\,\mathrm{eV}$ around the converged value.}
  \label{fig:svlo_gs_nempty_cspbi3}
\end{figure}

In practice, \gls{svlo} enables a substantial reduction of the \gls{sv} subspace, allowing for highly precise calculations. For example, \cref{fig:svlo_gs_nempty_cspbi3} shows for $\gamma$-CsPbI$_3$ that the \gls{svlo} basis with $N_{\mathrm{occ}} = 228$ and $N_{\mathrm{LO}} = 496$ reaches convergence of the band gap already with about 500 unoccupied states, while the conventional \gls{sv} basis requires several thousand to achieve the same value. This reduction of the \gls{sv} subspace shows a sizable speedup by a factor of $3.6$ at comparable precision~\cite{Vona2023}.

The analogous formalism has been implemented to accelerate \gls{bse} calculations in the presence of \gls{soc} (see \cref{sec:New_SVLO_BSE}).

\subsection{Constrained \acrshort*{dft}}
\label{sec:New_cDFT_GS}

The \gls{bse} is considered the state-of-the-art method for modeling neutral excitations, as it explicitly addresses electron–hole interactions. However, its high computational cost and unfavorable scaling with system size hampers its use for complex materials. In such cases, \gls{cdft} offers an attractive alternative. By imposing occupation constraints to mimic excited-state configurations, it can capture essential excitation characteristics with an accuracy comparable to \gls{bse} at a fraction of the computational expense, in particular if the electron-hole pairs are confined, \ie don't require large supercells. This approach has already proven rather successful in the so-called supercell core-hole approach, but is not limited to core excitations. It has been applied to a variety of systems, including molecules, organic dyes, and perovskites~\cite{gavnholt2008, barca2018simple, carterfenk2020statetargeted, kowalczyk2011assessment, gilbert2008selfconsistent, luo2018efficient}, making it a practical and scalable option when full \gls{bse} calculations are not feasible. 

In \exciting, \gls{cdft} serves as an efficient approach for modeling excitations following picosecond time delays in pump-probe experiments (see \cref{sec:New_PPS2_PP}). The core idea in \gls{cdft} is to solve the \gls{ks} equations while keeping the electronic occupations fixed in a non-equilibrium configuration. Initially, a standard \gls{gs} calculation is performed to determine the equilibrium occupation numbers $\occ{n}{\kk}$ and the corresponding electronic density $\nKS(\pos)$ following \cref{eq:rhogs}. The occupation factors are then constrained based on physical considerations, such as mimicking a specific excitonic state identified with higher-order theories (\eg \gls{bse}) or simulating the electron-hole pairs generated in a pump-probe experiment. For a given number of excited carriers per unit cell, $N_\mathrm{exc}$, the change in the carrier distribution, $\Delta\occ{n}{\kk}$, must satisfy
\begin{equation}\label{eq:cdft_nexc}
\sum_{n\kk}^{\mathrm{val}}\kptweight {\kk }\,\Delta\occ{n}{\kk} = \sum_{n\kk}^{\mathrm{cond}}\kptweight {\kk }\,\Delta\occ{n}{\kk} = N_\mathrm{exc},
\end{equation}
where the summations are performed over valence and conduction states, respectively, to model excited holes and electrons. The electron density $\nKS'$ under the presence of excitations is constructed as:
\begin{equation}
\nKS'(\pos) = 
\sum_{\kk} \kptweight {\kk }\!
\left[
\sum_n \occ{n}{\kk}\, |\kswf_{n\kk}'(\pos)|^2 + 
\sum_n^{\mathrm{cond}} \Delta\occ{n}{\kk}\, |\kswf_{n\kk}'(\pos)|^2 -
\sum_n^{\mathrm{val}} \Delta\occ{n}{\kk}\, |\kswf_{n\kk}'(\pos)|^2 
\right],
\end{equation}
where $\kswf_{n\kk}'$ denotes the non-equilibrium \gls{ks} states. This excited-state density is subsequently used to build the \gls{ks} Hamiltonian. Following the usual procedure, the \gls{ks} equations are solved iteratively until self-consistency is achieved. Further details regarding implementation and applications can be found in Refs.~\cite{Rossi2025,Qiao2025}.

\subsection{Wannier interpolation}
\label{sec:New_Wannier_GS}
Wannier functions $w^{\!\textrm{WA}}_{n\latvec}(\pos)$ provide an alternative to the representation of a subspace of Bloch bands $\kswf_{n\kk}(\pos)$ and are labeled by band-like index $n$ and a unit-cell $\latvec$ within the \gls{bvk} supercell. They can be defined by a Fourier-like transform of rotated Bloch states 
\begin{equation}
    \label{eq:wannier-def}
    w^{\!\textrm{WA}}_{n\latvec}(\pos) = \frac{1}{N_{\kk}} \sum\limits_{\kk} {\rm e}^{-\im\kk\cdot\latvec}\, \sum\limits_{m} U_{mn}(\kk)\, \kswf_{m\kk}(\pos) \;,
\end{equation}
and the unitary rotations $U(\kk)$ can be tuned in order to find \glspl{mlwf}. The inversion of \cref{eq:wannier-def} for any arbitrary point $\tilde{\kk}$ allows for the interpolation of the \gls{ks} wave functions in reciprocal space. The corresponding unitary matrices $U(\tilde{\kk})$ can be found as the eigenvectors of the interpolated Hamiltonian
\begin{equation}
    H_{mn}(\tilde{\kk}) = \sum\limits_{\latvec} {\rm e}^{\im\tilde{\kk}\cdot\latvec} \braket{w^{\!\textrm{WA}}_{m\boldsymbol{0}} | \hKS | w^{\!\textrm{WA}}_{n\latvec}} \;,
\end{equation}
whose eigenvalues give the corresponding interpolated electron energies $\eigval_{n\tilde{\kk}}$. The reasons for the efficiency of this interpolation approach are the following: (i) The \glspl{mlwf} $w^{\!\textrm{WA}}_{n\latvec}(\pos)$ and hence the real space Hamiltonian in the Wannier representation ${H(\latvec) = \braket{w^{\!\textrm{WA}}_{\boldsymbol{0}} | \hKS | w^{\!\textrm{WA}}_{\latvec}}}$ are $\kk$-independent and thus only need to be calculated once prior to the interpolation; (ii) The interpolation to arbitrary points $\tilde{\kk}$ is obtained by a simple Fourier transform; (iii) Diagonalizing the interpolated Hamiltonian $H(\tilde{\kk})$ is quick, since the \glspl{mlwf} provide a minimal basis, \ie there is only one basis function for each band $n$ inside the subspace of interest (\eg a few bands around the Fermi level), and thus the Hamiltonian in Wannier representation is typically much smaller than the Hamiltonian in the original basis; (iv) Due to the strong localization of the \glspl{mlwf}, the real space Hamiltonian $H(\latvec)$ typically decays rapidly (exponentially for valence bands in insulators), and thus the sum over lattice vectors $\latvec$ converges quickly within the \gls{bvk} super cell.

\begin{figure}[h]
\centering
\includegraphics[width=0.6\textwidth]{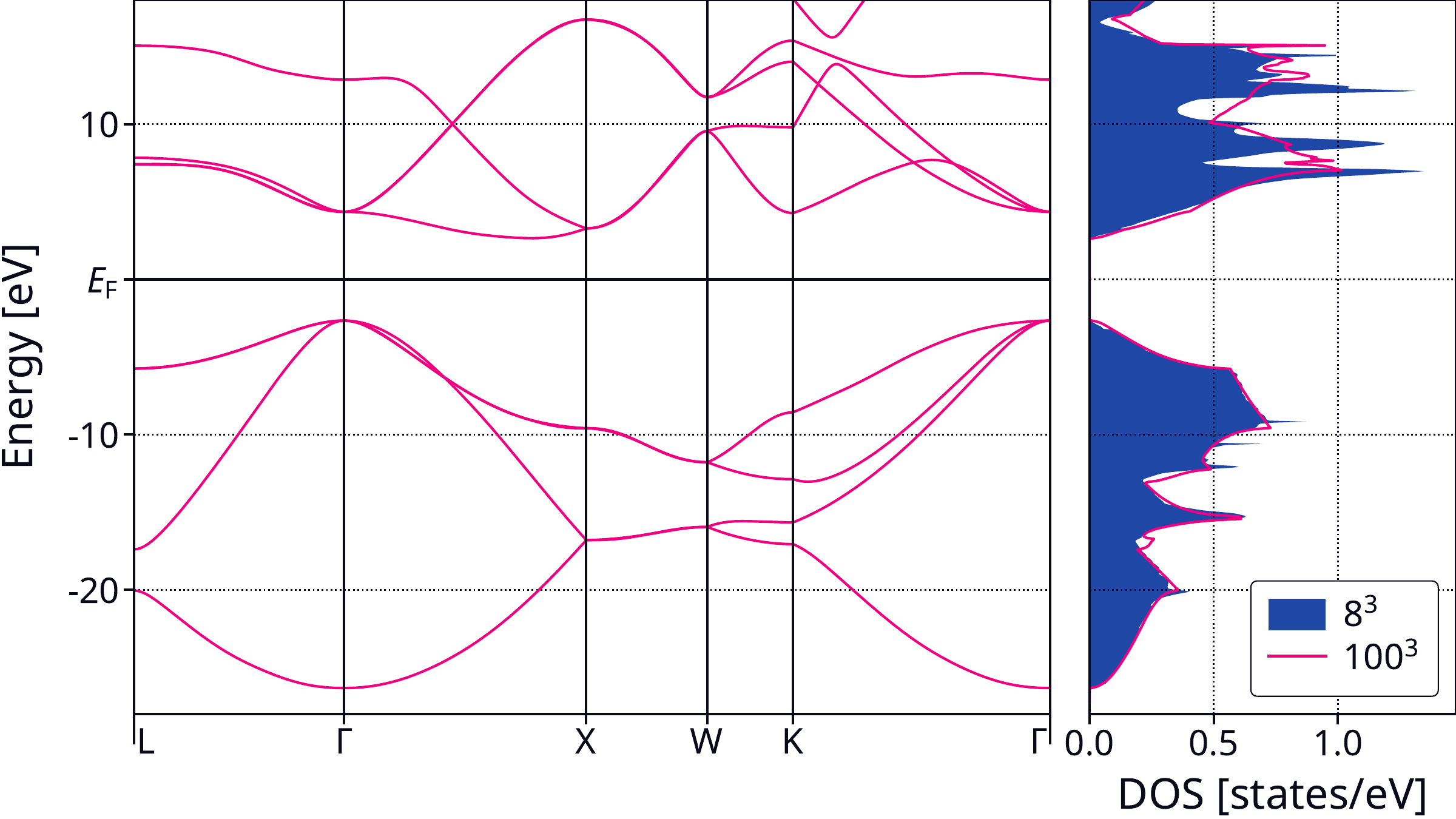}
\caption{Left: Band structure of diamond obtained using Wannier interpolation of eigenvalues from an $8^3$ $\kk$-grid using \gls{gdft} with the HSE06 functional. Right: \Gls{dos} obtained from a coarse grid (blue) and converged result obtained using Wannier interpolation on a grid of $100^3$ points (red).}
\label{fig:wannier_interpolation}
\end{figure}

\exciting allows for the calculation of \glspl{mlwf} representing both isolated~\cite{Marzari1997} (\eg valence bands in insulators) and entangled~\cite{Souza2001} (\eg conduction bands or bands in metals) subspaces of bands. In addition to the two-step procedure for entangled subspaces described in~\cite{Souza2001}, \exciting also implements the variational formalism described by \citet{Damle2019}. A special feature of \exciting is that there is no need to manually provide a set of projection functions in order to obtain an initial guess for the gradient-based optimization of $U(\kk)$. We automatically generate such an initial guess by finding optimized projection functions as a linear combination of automatically generated \glspl{lo}. The only input required from the user is the subspace of bands for which \glspl{mlwf} are to be computed. This can either be a range of band indices or a given energy window. For a detailed description of the implementation within \exciting we refer to Refs.~\cite{Tillack2020,Tillack2025}. \glspl{mlwf} can be calculated from either conventional or generalized \gls{ks}-\gls{dft}, or \gw input and hence allow for the interpolation of wavefunctions and eigenenergies, enabling the calculation of accurate band structures and \gls{dos} on different levels of sophistication (see \cref{fig:wannier_interpolation}). Further, \glspl{mlwf} can be used for computing energy-band derivatives, \ie group velocities and effective masses, and for the automatic search of band extrema away from high-symmetry points. Beyond the interpolation of wavefunctions and energies, Wannier interpolation can also be used to interpolate matrix elements as we do for the electron-phonon matrix $\elphmatel(\kk,\qq)$ for the calculation of electron renormalization due to electron-phonon interaction (see \cref{sec:DFPT}).

\subsection{Exploiting symmetry}
\label{sec:gs_symmetry_GS}

The exploitation of crystal symmetry is essential for achieving high efficiency at fixed precision. Its systematic implementation in \exciting significantly reduces computational cost and memory consumption, particularly for large unit cells and dense \gls{bz} samplings. Inside the \gls{mt} spheres, the \gls{ks} potential is expanded in symmetry-adapted angular functions (lattice harmonics). This enforces the crystal symmetry by construction and reduces the number of independent expansion coefficients. As a result, both the generation of the effective potential and the subsequent setup of the \gls{ks} eigenvalue problem are accelerated. For inversion-symmetric systems, the \gls{ks} eigenvalue problem can be formulated in terms of real symmetric matrices, requiring a real solver only. This substantially reduces the computational effort required for diagonalization. In practice, \exciting achieves a speedup of roughly a factor of four compared to the complex formulation consistently with results reported for other \gls{lapwlo} codes~\cite{blaha2020wien2k}.

%% file: sections/dfpt.tex

\subsection{State of the art}
\label{sec:SOTA_dfpt}

\Gls{dfpt} \cite{Zein1984,Baroni1987} is nowadays the method of choice for most phonon calculations, being implemented in many electronic-structure codes. Since it does not involve supercells, it is not restricted to certain high-symmetry $\qq$-points and is typically computationally efficient. It naturally provides also access to the response to perturbations other than atomic displacements such as electric fields or mechanical strain. However, the finite-difference approach to phonons is still appealing due to its conceptual simplicity. It also proves advantageous for systems with large unit cells, reduced symmetry or defects, or in highly anharmonic materials. The self-consistent phonon method~\cite{Zacharias2023} combines \gls{dfpt} with a series of special finite displacements in order to capture anharmonic effects such as temperature-dependent phonon softening and phase stabilization. Abandoning the Born-Oppenheimer approximation, recent work~\cite{Ponce2025_TotalEnergy} has studied the phonon contribution to the total energy, which is essential for the correct description of phase diagrams of materials with polymorphs that differ little in total energy. 

The first calculations of \acrfull{epc} based on \gls{dfpt} emerged more than three decades ago~\cite{Savrasov1994}. Nowadays, electron and phonon lifetimes and band structure renormalization, combining \gls{dfpt} and \gls{mbpt} are implemented in different codes such as \abinit~\cite{VerstraeteTJCP2025} or \quantumespresso~\cite{giannozzi2009quantum} + EPW~\cite{EPW2010}. However, all-electron full-potential results are still scarce, and \exciting is filling this gap. The reproducibility of the \gls{zpr} within different methods and implementations has been verified recently \cite{Ponce2025_Verification}. Current developments in the calculation of electron self-energies target the partially self-consistent calculation of spectral functions including vertex corrections for a more accurate description of \gls{epc}-induced features like kinks and satellites~\cite{Lihm2025}. Attempts to treat electron-phonon and electron-electron interaction simultaneously in a self-consistent fashion have been made by combining the perturbative description of \gls{epc} with the $GW$ method (see also~\cref{sec:GW}). 

Increasing attention has also been drawn on the interaction of phonons with other quasi-particles like magnon-phonon interaction~\cite{Luo2025} or exciton-phonon coupling (see also~\cref{sec:BSE}). One phenomenon where the established perturbative approach breaks down, is the formation of polarons, which are localized electron or hole states caused by local lattice distortions. Theories and practical implementations for calculating polaron properties like formation energies, self-trapping, their spatial distribution, and their effect on \gls{zpr} have been developed~\cite{Lafuente-Bartolome2022, Lafuente-Bartolome2022a, Vasilchenko2025, Luo2025}. Finally, phonons and \gls{epc} also play a crucial role in the description of Raman \cite{Draxl2002} and infrared spectroscopy, charge and heat transport, as well as superconductivity within the Migdal-Eliashberg formalism. Several software packages utilize phonons, \gls{epc} constants, and other results from various first-principle codes as input for the calculation of transport-properties, \eg in the framework of the \glspl{bte}. Examples are EPW~\cite{EPW2016}, Perturbo~\cite{Perturbo2021}, and elphbolt~\cite{Protik2022a}. Converged transport calculations typically require ultra-fine \gls{bz} samplings to accurately capture small energy differences around the chemical potential. Significant performance gains in this area have been achieved by the use of compressed representations of electron-phonon and phonon-phonon coupling constants \cite{Luo2024,Luo2025a}.

\subsection{Methodology}
\label{sec:Methodology_dftp}
The calculation of the linear response of the electronic system to an external perturbation lies at the heart of \gls{dfpt}. The \gls{ks} equations are replaced by their first-order counterpart, the so-called Sternheimer equation,
\begin{equation}
    \label{eq:sternheimer}
    \left[\order{\hKS}{0} - \order{\eigval}{0}_{n\kk}\right]\!{\order{\kswf}{1}_{n\kk}}(\pos) = - \left[\order{\hKS}{1} - \order{\eigval}{1}_{n\kk} \right]\! {\order{\kswf}{0}_{n\kk}}(\pos) \;,
\end{equation}
which, for a given first-order response of the \gls{ks} Hamiltonian $\order{\hKS}{1}$, can be solved for the first-order response of the wavefunctions $\order{\kswf}{1}_{n\kk}(\pos)$ and eigenenergies $\order{\eigval}{1}_{n\kk}$. Similar to the \gls{ks} equations, \cref{eq:sternheimer} has to be solved self-consistently for the first-order response of the density $\order{\nKS}{1}(\pos)$ and the potential $\order{\vKS}{1}(\pos)$. In the case of lattice dynamics, the external perturbation is a collective coherent displacement of the nuclei from their equilibrium positions,
\begin{equation}
    \atompos_{\cellidx \atomidx} = \order{\atompos_{\cellidx \atomidx}}{0} + {\rm e}^{\im \qq \cdot \latvec^{\phantom{I}}_{\!\cellidx}}\, \atomdisp_{\!\atomidx} \;,
\end{equation}
and~\cref{eq:sternheimer} is solved for the corresponding first-order responses $\order{\kswf}{\qq\atomidx i}$, $\order{\eigval}{\qq\atomidx i}$, $\order{\nKS}{\qq\atomidx i}$, and $\order{\vKS}{\qq\atomidx i}$, where $\qq$ is the phonon wavevector. Eventually, the dynamical matrix $\phdmat$ is obtained from the first-order response of the atomic forces
\begin{equation}
    \sqrt{M_\atomidx\, M_{\atomidx'}}\: \phdmat_{\atomidx i, \atomidx' j}(\qq) = - \sum\limits_\cellidx {\rm e}^{\im\qq\cdot\latvec^{\phantom{I}}_{\!\cellidx}}\, \frac{\partial F_{\atomidx' j}}{\partial \atompos_{\atomidx i \cellidx}} = - \order{F_{\atomidx' j}}{\qq\atomidx i} \;,
\end{equation}
and the vibrational eigenmodes are given by its eigenvalues and eigenvectors:
\begin{equation}
    \phdmat(\qq)\cdot\pheig_{\phmode\qq} = \freq_{\phmode\qq}^2\, \pheig_{\phmode\qq} \;.
\end{equation}

A complete description of lattice dynamics must also take into account the long-range dipole interactions that may be induced in polar materials by the atomic displacements. They lead to an additional non-analytic contribution to the dynamical matrix, which allows for the correct description of the splitting of longitudinal and transverse optical phonon modes. The key ingredients needed to fully capture this potentially anisotropic effects are the Born effective charge tensors $\born_\atomidx$, which describe the macroscopic polarization induced by the displacement of atom $\atomidx$, and the dielectric constant $\boldsymbol{\diel}$, which is the connection between the macroscopic displacement field and an external electric field $\efield$~\cite{King-Smith1993,Resta2007}. The Born effective charges are computed as the derivatives of the macroscopic polarization with respect to nuclear displacements,
\begin{equation}
    \born_\atomidx = \unitCellVol \,\frac{\partial\polarization}{\partial\atomincell_\atomidx} \;,
\end{equation}
 the dielectric constant is computed from the derivatives of the electronic contribution to the polarization with respect to a static electric field:
\begin{equation}
    \boldsymbol{\diel}^\infty = \boldsymbol{1} + 4\pi \,\frac{\partial\polarization^{\rm el}}{\partial\efield} \;.
\end{equation}
All necessary derivatives are obtained within \gls{dfpt}.

The interaction of electrons with the vibrational modes may renormalize the electronic structure. To capture this effect, Green-function based methods from \gls{mbpt} are employed. The key ingredient are the \gls{epc} constants,
\begin{equation}
    \elphmatel_{mn,\phmode}(\kk,\qq) = \sum\limits_{\atomidx,i} \frac{1}{\sqrt{2M_\atomidx\, \freq_{\phmode\qq}}}\, \textrm{w}_{\atomidx i,\phmode\qq} \braket{\kswf_{m\kk+\qq} | \order{\vKS}{\qq \atomidx i} | \kswf_{n\kk}} \;,
\end{equation}
describing the transition probability for the scattering of the electron from the initial state $\kswf_{n\kk}$ into the final state $\kswf_{m\kk+\qq}$ by the interaction with a phonon in mode $\nu\qq$. For the electron self-energy due to \gls{epc}, two contributions are typically considered. The first is the Fan-Migdal self-energy,
\begin{equation}
    \label{eq:fm-self-energy}
    \selfenergy^{\rm FM}_{nn'}(\freq,\kk,T) = \sum\limits_{\phmode,m} \int\limits_{\rm BZ} \frac{{\rm d}\qq}{\unitCellVol_{\rm BZ}}\, \elphmatel_{mn,\phmode}^\ast(\kk,\qq)\: \elphmatel_{mn',\phmode}(\kk,\qq)\! \left[ \frac{f_{m\kk+\qq}(T) + n_{\phmode\qq}(T)}{\freq - \eigval_{m\kk+\qq} + \freq_{\phmode\qq} + \im\eta} + \frac{1 - f_{m\kk+\qq}(T) + n_{\phmode\qq}(T)}{\freq - \eigval_{m\kk+\qq} - \freq_{\phmode\qq} + \im\eta} \right] ,
\end{equation}
where $f(T)$ ($n(T)$) is the temperature dependent fermionic (bosonic) occupation of the electrons (phonons) and $\eta$ is a smearing parameter. The second term is the static Debye-Waller self-energy,
\begin{equation}
    \label{eq:dw-self-energy}
    \selfenergy^{\rm DW}_{nn'}(\kk,T) = \sum\limits_{\phmode} \int\limits_{\rm BZ} \frac{{\rm d}\qq}{\unitCellVol_{\rm BZ}}\, \elphmatel^{\rm DW}_{nn',\phmode}(\kk,\qq)\, \left[n_{\phmode\qq}(T) + \frac{1}{2} \right] ,
\end{equation}
with the Debye-Waller matrix elements
\begin{equation}
    \label{eq:dw-matrix-elements}
    \elphmatel^{\rm DW}_{nn',\phmode}(\kk,\qq) = \sum\limits_{\atomidx,i,\atomidx',j} \frac{1}{\sqrt{2M_\atomidx\, \freq_{\phmode\qq}}}\, \textrm{w}_{\atomidx i,\phmode\qq}^\ast\, \braket{\kswf_{n\kk} | \order{\vKS}{-\qq\atomidx i,\qq\atomidx' j} | \kswf_{n'\kk}}\, \textrm{w}_{\atomidx' j,\phmode\qq}\, \frac{1}{\sqrt{2M_{\atomidx'}\, \freq_{\phmode\qq}}} \;,
\end{equation}
including the second-order potential response $\order{\vKS}{-\qq\atomidx i,\qq\atomidx' j}$.
The knowledge of the self-energy allows for the calculation of temperature dependent renormalized quasi-particle energies via the solution of the Dyson equation,
\begin{equation}
    \eigval^{\rm QP}_{n\kk}(T) = \eigval_{n\kk} + \selfenergy_{nn}(\eigval^{\rm QP}_{n\kk},\kk,T) \;.
\end{equation}

\gls{epc} is also the driving mechanism behind conventional phonon-mediated superconductivity. A commonly used parameter is the dimensionless electron-phonon coupling strength,
\begin{equation}
    \label{eq:eph-coupling-strength}
    \lambda(\eigval) = 2 \int\limits_0^\infty \frac{{\rm d}\freq}{\freq}\, \alpha^2F(\eigval, \freq) \;,
\end{equation}
which is employed to estimate the critical temperature. The Eliashberg function $\alpha^2F(\eigval,\freq)$ can be viewed as a weighted phonon \gls{dos},
\begin{equation}
    \label{eq:a2F}
    \alpha^2F(\eigval,\freq) = \sum\limits_\phmode \int\limits_{\rm BZ} \frac{{\rm d}\qq}{\unitCellVol_{\rm BZ}}\, \freq_{\phmode\qq}\, \lambda_{\phmode\qq}(\eigval)\, \delta(\freq - \freq_{\phmode\qq}) \;,
\end{equation}
where the weights are given by the phonon-mode resolved coupling strength,
\begin{equation}
    \label{eq:ph-coupling-strength}
    \lambda_{\phmode\qq}(\eigval) = \frac{1}{N(\eigval)\, \freq_{\phmode\qq}} \sum\limits_{mn} \int\limits_{\rm BZ} \frac{{\rm d}\kk}{\unitCellVol_{\rm BZ}}\, |\elphmatel_{mn,\phmode}(\kk,\qq)|^2\, \delta(\eigval - \eigval_{n\kk})\, \delta(\eigval_{n\kk} - \eigval_{m\kk+\qq}) \;,
\end{equation}
and $N(\eigval)$ is the electron \gls{dos} at energy $\eigval$. While $\lambda_{\phmode\qq}(\eigval)$ provides a measure for how strong a specific phonon mode couples to electrons with a given energy $\eigval$, it is also possible to define an equivalent electron-band resolved coupling parameter
\begin{equation}
    \label{eq:el-coupling-strength}
    \lambda_{n\kk} = 2\int\limits_0^\infty \frac{{\rm d}\freq}{\freq} \sum\limits_{m,\phmode} \int\limits_{\rm BZ} \frac{{\rm d}\qq}{\unitCellVol_{\rm BZ}}\, |\elphmatel_{mn,\nu}(\kk,\qq)|^2\, \delta(\freq - \freq_{\phmode\qq})\, \delta(\eigval_{n\kk} - \eigval_{m\kk+\qq}) \;,
\end{equation}
which describes the coupling of a specific electronic state to any phonon mode. It is also called the mass-enhancement parameter, because the effective mass of a quasi-particle in that state is enhanced by a factor $(1+\lambda_{n\kk})$. Even though~\cref{eq:eph-coupling-strength,eq:a2F,eq:ph-coupling-strength} are commonly used in the context of superconductivity, where the electron energy of interest is typically the Fermi level, \ie, ${\eigval=\eigval_{\rm F}}$, they can also be used to study the coupling to electronic states around any energy, even in insulating materials.

\subsection{Phonons and electron-phonon coupling effects in \excitingb}
\label{sec:New_implementation_dftp}

\begin{figure}[h]
\centering
\includegraphics[width=0.6\textwidth]{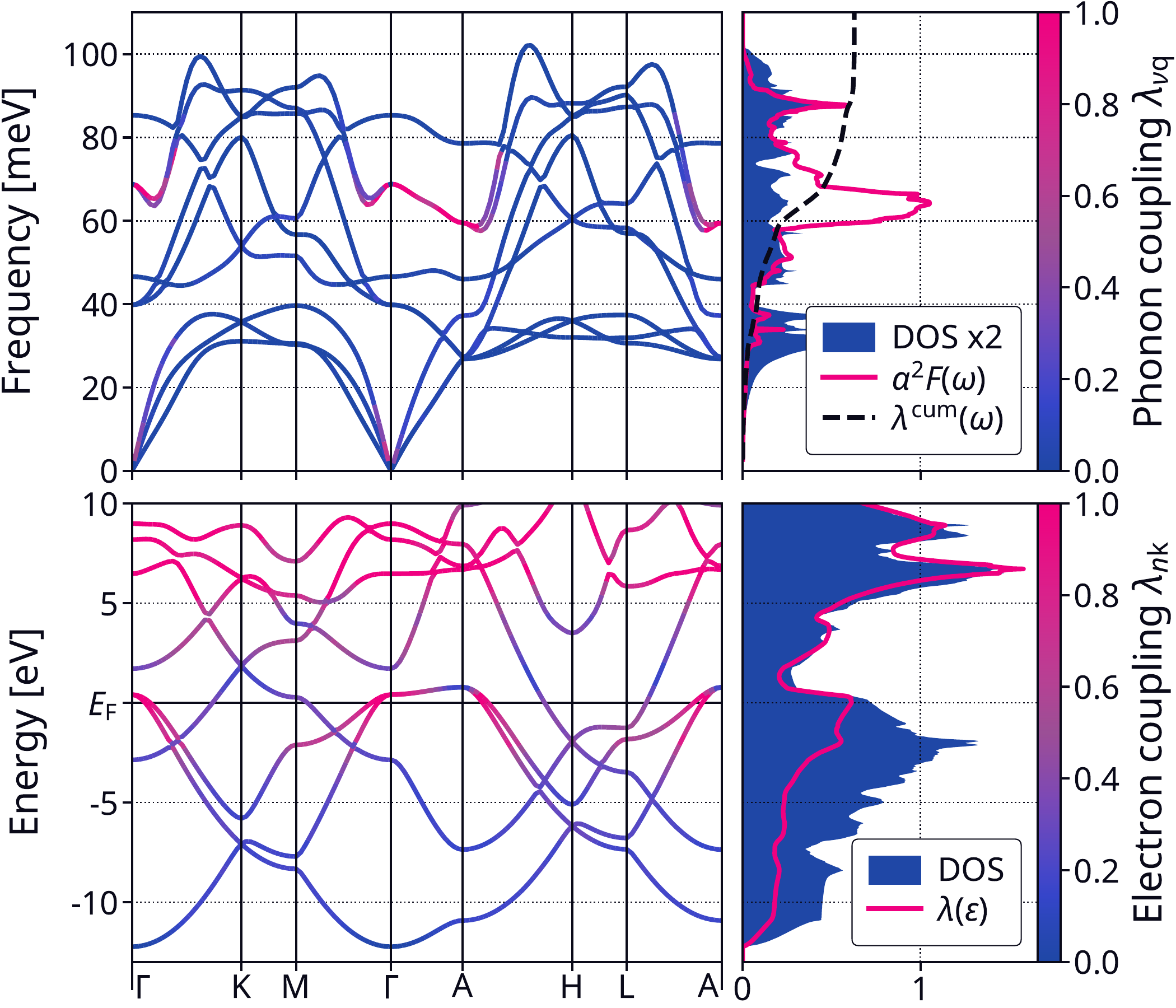}
\caption{Electron-phonon coupling in MgB$_2$. 
Top: Phonon dispersion and mode resolved coupling strength $\lambda_{\phmode\qq}$ together with the phonon DOS (blue), the Eliashberg function $\alpha^2F$ (magenta), and the cumulative coupling strength $\lambda^{\rm cum}$ (dashed line). Bottom: Electron dispersion and band-resolved coupling strength $\lambda_{n\kk}$ together with the electron DOS (blue) and coupling strength $\lambda$ (magenta).}
\label{fig:eph-coupling-strength}
\end{figure}

\exciting allows for the calculation of phonons using both \gls{dfpt} and finite differences. The \gls{dfpt} implementation fully exploits crystal symmetries by the use of irreducible representations. \Gls{dfpt} calculations are highly parallelized over both symmetry reduced $\kk$- and $\qq$-points as well as displacement patterns. \Gls{dfpt} currently supports all \gls{lda} and \gls{gga} functionals provided by Libxc. Spin-polarization is not yet implemented. For further details on the finite-differences and \gls{dfpt} implementations, we refer to Refs. \cite{gulans2014exciting} and \cite{TillackDFPT2026}, respectively.

The electron-phonon part of \exciting can capture the impact of \gls{epc} on the electronic structure, \ie the phonon-renormalized quasi-particle energies. In order to converge \gls{bz} integrals such as in~\cref{eq:fm-self-energy,eq:dw-self-energy}, \exciting employs Wannier-Fourier interpolation of electron energies, the dynamical matrices, and the electron-phonon matrix elements~\cite{Giustino2007}, including the long-range coupling in polar materials~\cite{Verdi2015} onto a dense integration grid. Furthermore, tetrahedron integration is used, which does not require a smearing parameter and allows for faster convergence with respect to the integration grid. The Debye-Waller matrix elements are computed within the rigid-ion approximation, which avoids the evaluation of the second-order potential response. Self-energy calculations are MPI parallelized over both electron ($m$) and phonon ($\phmode$) bands and $\kk$-points. Thanks to the Wannier-Fourier interpolation, the electron wavevector $\kk$ is not restricted to the grid used in the underlying calculation of the electronic structure. In particular, $\kk$ can be set to be a high-symmetry path or the locations of band extrema. The single-particle energies $\epsilon$ can come from either \gls{ks}, \gls{gks}, or $GW$ calculations. In~\cref{eq:ph-coupling-strength,eq:el-coupling-strength}, the quasi-elastic approximation is considered for the coupling parameters $\lambda_{\phmode\qq}$ and $\lambda_{n\kk}$, \ie the phonon energies are assumed to be much smaller than the typical electron energies, and the difference between phonon absorption and emission is neglected. However, in \exciting, also the coupling parameters for either absorption or emission processes are implemented.

We demonstrate our implementation by calculating phonons and \gls{epc} for the high-$T_{\rm c}$ superconductor MgB$_2$. In the top panels of~\cref{fig:eph-coupling-strength}, we show the phonon dispersion as obtained with \gls{dfpt} and the mode-resolved coupling strength $\lambda_{\phmode\qq}$. Mainly the $E_{2g}$ modes along the $\BZpoint{\Gamma}$-$\BZpoint{A}$ line show a strong coupling strength, which results in a clear peak between 60 and 70 meV in the Eliashberg function, which does not appear in the phonon \gls{dos}. The cumulative coupling strength
\begin{equation}
    \lambda^{\rm cum}(\freq) = 2 \int\limits_0^\freq \frac{{\rm d}\freq'}{\freq'}\, \alpha^2F(\freq')
\end{equation}
reaches a maximum value of ${\lambda^{\rm cum} = 0.63}$. The bottom panels of~\cref{fig:eph-coupling-strength} shows the electron dispersion with the band-resolved coupling strength $\lambda_{n\kk}$ as well as the electron \gls{dos}, and the integrated energy-dependent \gls{epc} strength $\lambda(\eigval)$. At the Fermi level, its value is $0.63$, the same as obtained from the Eliashberg function. The band-resolved coupling strength near the Fermi level shows that the mass-enhancement parameter for the $\sigma$-bonding bands reaches values between $0.8$ and $1.1$, \ie more than twice as big as the values between $0.3$ and $0.4$ for the $\pi$-bonding bands. All these findings are consistent with previously published results~\cite{Choi2003}.

\begin{figure}[h]
\centering
\includegraphics[width=0.48\textwidth]{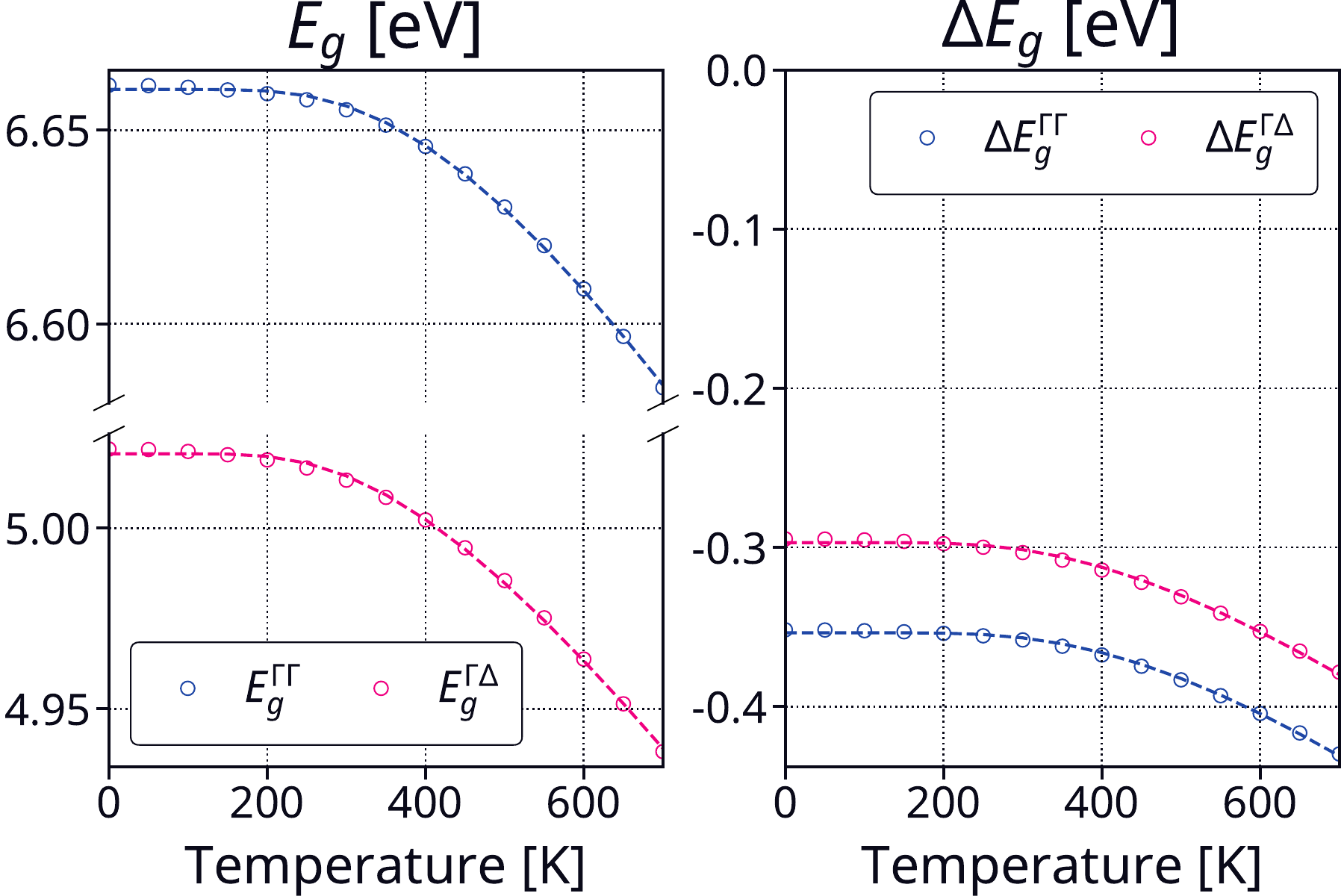}\hspace{0.04\textwidth}\includegraphics[width=0.48\textwidth]{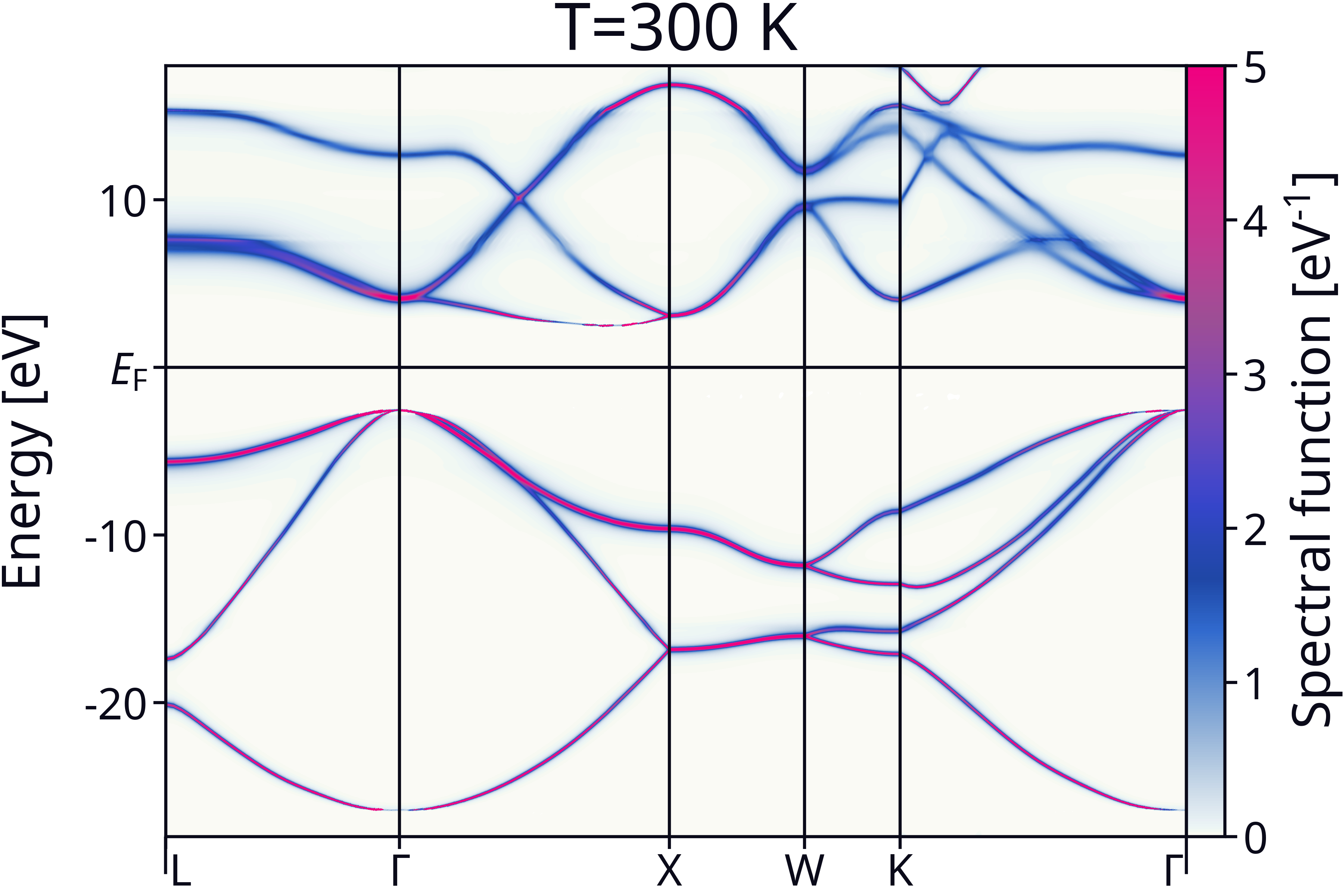}
\caption{Electron-phonon coupling in diamond. 
Left: Direct (blue) and indirect (magenta) band gap $E_g(T)$ and its renormalization ${\Delta E_g(T) = E_g(T) - E_g^{\rm DFT}}$ as a function of temperature. Right: Band structure with the color code indicating the electron spectral function at \SI{300}{\K}.}
\label{fig:eph-renormalization}
\end{figure}
As a second example, the temperature-dependent \gls{qp} energies from the electron self-energy are computed for diamond. Phonons and the first-order change in potential are computed with PBE, electron eigenvalues and wavefunctions with HSE06, demonstrating the possibility to use different starting points for the energy renormalization due to \gls{epc}. In \cref{fig:eph-renormalization}, we present the resulting direct and indirect band gap and its \gls{epc}-based renormalization as a function of temperature as well as the electron spectral function at room temperature. The base HSE06 results for the direct (indirect) band gap are \SI{7.01}{\eV} (\SI{5.32}{\eV}) and a \gls{zpr} of \SI{-352}{\milli\eV} (\SI{-295}{\milli\eV}). 

%% file: sections/gw.tex

\subsection{State of the art}
\label{sec:SOTA_GW}
While \gls{ks}-\gls{dft} is, in principle, exact for \gls{gs} densities and energies, it is not designed to predict excited-state properties. In particular, the \gls{ks} eigenvalues, which were introduced as Lagrange multipliers, cannot be interpreted as \gls{qp} energies. Although they often provide a reasonable first approximation~\cite{LudersJPCM2001,OnidaRMP2002}, they can differ significantly from the true many-body spectrum~\cite{PerdewPRL1996,MoriSanchezPRL2008,SunPRL2015,ReiningWCMS2018,AschebrockPRR2019}. Hybrid functionals partially remedy these discrepancies, however, charged excitations cannot be captured by a single \gls{dft} calculation with a fixed particle number. An accurate description requires \gls{mbpt}, and in particular, the self-consistent solution of Hedin’s equations~\cite{HedinPR1965}. They account for the nonlocal and dynamical screening effects induced by electron addition and removal, thereby providing a direct connection to experiments such as \gls{xps} or \gls{arpes}.

An approach to reduce this set of equations to a more tractable problem is the so-called $GW$ approximation. The name comes from the form of the lowest-order approximation to the self-energy operator, $\Sigma = iGW$, where $G$ is the single-particle Green's function and $W$ is the screened Coulomb interaction. It is obtained by approximating the vertex function as $\Gamma \approx 1$. The most common and computationally affordable implementation of the $GW$ approximation is the \textit{single-shot GW} or $G_0W_0$. In this scheme, the self-energy is computed perturbatively, starting from a preceding \gls{dft}, \gls{hf}, or hybrid-functional calculation. $G_0W_0$ has been widely implemented across a variety of codes, such as \abinit~\cite{VerstraeteTJCP2025}, \exciting~\cite{DmitriiPRB2016}, \vasp~\cite{ShishkinPRB2006}, \gpaw~\cite{HuserPRB2013}, FLAPWMBPT~\cite{KutepovPRB2009}, YAMBO~\cite{SangalliJPCM2019}, BerkeleyGW~\cite{DeslippeCPC2012}, CP2K~\cite{GramlJCTC2024}, and FHI-aims \cite{BlumCPC2009}. The wide variety of existing implementations differ in their usage of basis functions, different frequency-integration schemes, etc. This makes cross-validation challenging. A recent study \cite{AziziCMS2025} has shown that for a selection of seven materials, the $G_0W_0$ band gaps obtained from four different codes, \ie \abinit, \exciting, FHI-aims, and \gpaw agree to within 0.1~eV, with the two all-electron codes---\ie \exciting and FHI-aims---showing the best agreement and consistency.

The perturbative nature of the $G_0W_0$ approximation introduces an undesirable dependence on the mean-field starting point \cite{BrunevalPRL2005,GattiPRL2007,VidalPRL2010,vanSettenJCTC2015,RodriguesPelaPRB2016,RodriguesPelaNPJCM2024,TalPNAS2024,WenJCTC2024,jana2025nonempirical}. Addressing this limitation has driven intense research efforts towards self-consistent $GW$ schemes~\cite{BrunevalPRB2006,SchilfgaardePRL2006,KotaniPRB2007,KutepovPRB2012,GrumetPRB2018,Salas-Illanes2022,HauleCPC2024}. They often result in an overestimation of band gaps, due to the neglect of vertex corrections and thus under-screening. Vertex effects can be included to some extent via the cumulant approach, improving the description of plasmonic satellites~\cite{LischnerPRL2013,CarusoPRL2015,GumhalterPRB2016,LoosFD2024}, even if the full vertex is required for an accurate spectral representation~\cite{MejutoZaeraJPC2021}. While diagrammatic approaches can incorporate the vertex explicitly~\cite{KutepovPRB2016,KutepovPRB2022}, a more practical strategy is to include electron-hole interaction effects via an \gls{xc} kernel into the polarizability and self-energy. The quality of this approach has been shown to depend critically on the kernel~\cite{SchmidtPRB2017,OlsenNPJCM2019}, which must capture both the correct long-range behavior to describe exciton binding and energy gaps~\cite{KimPRL2002,ShishkinPRL2007,ChenPRB2015,BergerPRL2015,rigamonti2015estimating,TalPRB2021,SchmidtPRB2017,OlsenNPJCM2019} and the proper short-range behavior to ensure accurate \gls{qp} energies and ionization potentials~\cite{GruneisPRL2014,TalPRB2021,SchmidtPRB2017,OlsenNPJCM2019}. Notably, renormalization of the electronic structure by \gls{epc}, \ie zero-point vibrations and temperature effects \cite{AllenPRB1981,GiustinoRMP2017}, is crucial for quantitative comparison with experiment. Also lattice screening plays a role in polar materials \cite{BechstedtPRB2005,BottiPRL2013}.

Achieving reliable precision requires addressing critical numerical bottlenecks. A major problem here is the sensitivity of excited-state properties to the quality and completeness of the basis set, in particular in \gls{lapwlo} methods \cite{FriedrichPRB2011,DmitriiPRB2016}. To mitigate this issue, an \gls{ibc} has been proposed~\cite{BetzingerPRB2015}. Furthermore, the treatment of long-range interactions in the ${\mathbf{q} \to 0}$ limit is critical, particularly for anisotropic or low-dimensional materials where standard expressions fail \cite{FreysoldtCPC2007,RasmussenPRB2016,RodriguesPelaNPJCM2024}. To ensure numerical accuracy across diverse systems, code-agnostic libraries such as \texttt{GreenX} \cite{greenXrepo,IDieLrepo} have been developed to rigorously treat the ${\mathbf{q} \to 0}$ limit \cite{FreysoldtCPC2007,RasmussenPRB2016}. The practical application of advanced $GW$ methods is often limited by the inherent quartic scaling with the number of atoms. This bottleneck has been addressed through algorithmic improvements like the space-time method, leading to cubic scaling ~\cite{RojasPRL1995,LiuPRB2016,KutepovCPC2017,AziziPRB2025} or linear-scaling stochastic $GW$~\cite{NeuhauserPRL2014,VlcelsePRB2018}. For weakly bound van der Waals heterostructures, the \gls{eas} method \cite{XuanJCTC2019,LiuJCTC2019,SchebekPRB2025} has been proven to be efficient: By approximating the total polarizability of the heterostructure as a superposition of its individual components, the computing time for interfaces between organic molecular layers and 2D substrates has been reduced by over 50\%  \cite{SchebekPRB2025}.

\subsection{Methodology}
\label{sec:Methodology_GW}
The essence of the $GW$ method and the many variants has already been outlined in \cref{sec:SOTA_GW}. Some aspects will be summarized when describing the implementation of \gw in \exciting, in \cref{sec:New_Implementation_GW}. Providing more details on the entire formalism would go beyond the scope of this article. Instead, we refer to review articles and books \cite{hybertsen1986electron_correlation,aryasetiawan1998gw,OnidaRMP2002,inkson2012many,martin2016interacting,ReiningWCMS2018}.

\subsection{Implementation in \excitingb}
\label{sec:New_Implementation_GW}

The current implementation of $G_0W_0$ in the \exciting code, to a large extent based on the approach of Ref.~\cite{JiangCPC2013}, is described in detail in Ref.~\cite{RayaMorenoGW2026}. Here, we briefly summarize the main aspects. Starting from a \gls{dft} or \gls{gdft} reference, the QP energies are obtained from the linearized \gls{qp} equation:
\begin{equation}
    \re{[\epsilon_{n\kk}^{\text{QP}}]} \approx \eigval_{n\kk} + Z_{n\mathbf{k}}\braket{\kswf_{n\mathbf{k}} | \re{[\Sigma(\mathbf{r},\mathbf{r}',\eigval_{n\kk})}]-\vXCnohat(\mathbf{r})\,\delta(\mathbf{r}-\mathbf{r}')|\kswf_{n\mathbf{k}}},
\label{eq:QPGW}
\end{equation}
where $\eigval_{n\mathbf{k}}$ and $\kswf_{n\mathbf{k}}$ are the \gls{ks} eigenvalues and wavefunctions, $v_\mathrm{XC}(\pos)$ is the exchange-correlation potential of the reference calculation, and $Z_{n\mathbf{k}}$ is the \gls{qp} renormalization factor.

Single-particle states are represented using the all-electron \gls{lapwlo} basis, while two-particle quantities such as the polarizability and dielectric function are expanded in an auxiliary mixed-product basis~\cite{AryasetiawanPRB1994,JiangCPC2013}. Products of \gls{ks} states are expressed as
\begin{equation}
\kswf_{n\kk}(\pos) \,
\kswf^{*}_{m\kk-\qq}(\pos)
=
\sum_i
M^{\shup{i}}_{nm}(\kk,\qq)\:
\mbasis_i^{\qq}(\pos)\:,
\end{equation}
where $\mathcal{B}_i^\qq(\pos)$ are the mixed-product basis functions and $M^{\shup{i}}_{nm}(\kk,\qq)$ are the corresponding expansion coefficients. Computing these coefficients represents one of the main computational bottlenecks of the method. In the mixed-product basis, the dielectric matrix in the \gls{rpa} reads
\begin{align}
\diel_{ij}(\qq,\freq)
&= 1
- \frac{2}{N_\kk}
  \sum_{n m \kk}
   f_{n\kk}
  \bigl(1 - f_{m\kk-\qq}\bigr)\;\tilde{M}^{\shup{i}}_{nm}(\kk,\qq)\,
\bigl[
  \tilde{M}^{\shup{j}}_{nm}(\kk,\qq)
\bigr]^{\!*}
\nonumber \\[4pt]
&\quad \times
\left[
  \frac{1}{
    \freq - (\epsilon_{m\kk-\qq}-\epsilon_{n\kk}) + \im\delta
  }
  -
  \frac{1}{
    \freq + (\epsilon_{m\kk-\qq}-\epsilon_{n\kk}) - \im\delta
  }
\right],
\end{align}
where $\delta$ is a positive infinitesimal, $N_{\kk}$ denotes the number of $\kk$ points, and $ \tilde{M}^i_{nm}(\kk,\qq)$ is:
\begin{equation}
    \tilde{M}^{\shup{i}}_{nm}(\kk,\qq) = \sum_{l} v_{il}^{\frac{1}{2}}(\qq) \,M^{\shup{l}}_{nm}(\kk,\qq)\:,
\end{equation}
with $v_{ij}(\qq)$ being the bare Coulomb interaction expanded in the mixed-product basis. All quantities are transformed to the $v_{ij}(\qq)$ eigenbasis to isolate the long-range $\mathbf{q} \to 0$ behavior. This properly addresses the long-range limit while enabling an efficient truncation of the dielectric matrix to reduce computational cost.

The screened Coulomb interaction is then obtained as
\begin{equation}
\scrcoul(\mathbf{q},\omega) =
\varepsilon^{-1}(\mathbf{q},\omega)\,v(\mathbf{q})\:.
\end{equation}
Using $\scrcoul(\mathbf{q},\omega)$, the correlation part of the self-energy is given by
\begin{equation}
\Sigma^{\textrm{c}}_{nl\mathbf{k}}(\omega) = 
\frac{1}{N_\mathbf{q}}
\sum_{\mathbf q}
\sum_m
\frac{\im}{2\pi}
\int_{-\infty}^{\infty} \!\textrm{d}\freq'\,
\frac{\sum_{ij}
\left[M^{\shup{i}}_{nm}(\kk,\qq)\right]^{*}
\,\scrcoul^{\textrm{c}}_{ij}(\qq,\freq’)\,
M^{\shup{j}}_{lm}(\kk,\qq)}
{\freq + \freq' - \varepsilon_{m\kk-\qq}}\:,
\end{equation}
where $\scrcoul^{\textrm{c}}_{ij}(\qq,\freq) = \scrcoul_{ij}(\qq,\freq) - v_{ij}(\qq)$. The exchange contribution is evaluated as
\begin{equation}
\Sigma^{\textrm{x}}_{nl\kk}
=
-\frac{1}{N_\qq}
\sum_{\mathbf q}
\sum_{i,j}
\sum_{m}^{\mathrm{occ}}
\left[\tilde{M}^{\shup{i}}_{nm}(\kk,\qq)\right]^{\!*}
\tilde{M}^{\shup{j}}_{lm}(\kk,\qq)\:.
\end{equation}
The total self-energy,
$\Sigma = \Sigma^{\mathrm{x}} + \Sigma^{\mathrm{c}}$,
is then used in \cref{eq:QPGW} to compute the \gls{qp} energies.

The implementation follows a task-based workflow considering symmetry, multi-level MPI parallelism, and GPU offloading (see~\cref{sec:New_HPC_GW} for further information). This approach allows for scalable all-electron $G_0W_0$ calculations for both bulk and low-dimensional systems. Moreover, the task-based formalism naturally extends to several flavors of self-consistent $GW$ schemes, and enables seamless integration with the \gls{eas} for treating  \gls{vdw} stackings. Further details of the current implementation are provided in \cref{sec:New_EAS_GW}. 

\subsection{Computational advancements: HPC-friendly \gwb}
\label{sec:New_HPC_GW}

\begin{figure}
    \centering
    \input{figures/tools/workflow_GW}
    \caption{Workflow diagram illustrating the tasks required for a $GW$ calculation. Color coding denotes MPI parallelization over $\qq$ points, $\kk$ points, and unoccupied states ($n$). Tasks with thick outlines support GPU acceleration. Solid arrows represent the preferred workflow, whereas dashed arrows indicate non-recommended alternatives. The \texttt{Coulomb} task evaluates the bare Coulomb interaction, and \texttt{vxc} provides the diagonal matrix elements of the exchange-correlation potential. The \texttt{polarizability} task constructs the irreducible polarizability, which is then used by \texttt{epsilon} to form the dielectric matrix within the \gls{rpa}. Its inversion is carried out by \texttt{invertEpsilon}, followed by \texttt{irreducibleMapping}, which exploits crystal symmetries to reconstruct the inverse dielectric matrix over the full \gls{bz} from the irreducible one in case crystal symmetry was used in \texttt{epsilon}. The self-energy is separated into exchange and correlation contributions, evaluated by \texttt{sigmax} and \texttt{sigmac}, respectively. Finally, \texttt{QPEigenvalues} combines these components to obtain the \gls{qp} energies.} 
    \label{fig:gw_workflow}
\end{figure}
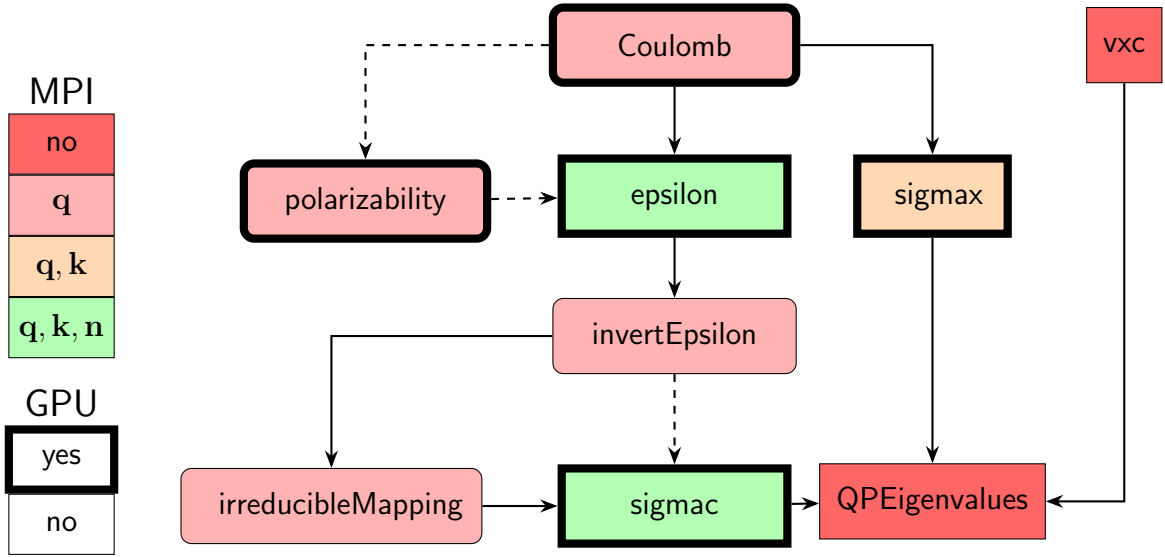

While the implementation based on Ref.~\cite{JiangCPC2013} allowed for some parallelization, it offered limited scalability and lacked GPU support, hindering efficient use of modern supercomputers. The recent version introduced a new HPC-optimized $GW$ implementation based on a task-based workflow in which each task allows for parallelism not only over \textbf{k/q}-points but also over bands and frequencies. This leads to substantial improvements in both strong and weak scaling. To further mitigate computational bottlenecks, GPU acceleration has been incorporated via a hybrid strategy: linear algebra operations are offloaded to vendor-optimized libraries, while compute-intensive loops are parallelized using directives, \ie OpenMP offload. This approach achieves 4-10$\times$ speedups across a variety of systems and workloads, enabling large-scale, high-precision $GW$ calculations that fully leverage modern HPC resources. In Fig.~\ref{fig:gw_workflow}, we present a schematic overview of the general workflow of the new implementation. Details on this implementation are provided in Ref.~\cite{RayaMorenoGW2026}.

\subsection{Accurate treatment of long-range interactions}
\label{sec:New_LR_GW}

As already noted, the computation of the self-energy operator $\Sigma$ requires an integration over the \gls{bz}, which converges too slowly due to the long-range nature of the Coulomb interaction. Several techniques have been developed to mitigate this issue, including the use of auxiliary functions~\cite{CarrierPRB2007}, Monte Carlo approaches~\cite{MariniCPC2009,daJornadaPRB2017}, and Coulomb-truncation schemes~\cite{RenPRMat2021,SpencerPRB2008}, among others. Most of these techniques---except for truncation-based methods---face the nontrivial challenge of assigning a well-defined value to the inverse dielectric matrix and/or the screened Coulomb potential at $\mathbf{q}=0$. Methods that avoid this issue typically do so at the expense of introducing an additional external parameter, such as a truncation cutoff, which must be carefully chosen and converged~\cite{RenPRMat2021}. This difficulty arises from the non-analytic behavior of these quantities, which are in general---except for cubic Bravais lattices---not properly defined at $\mathbf{q} = 0$ but rather in the limit $\mathbf{q} \to 0$~\cite{BaroniPRB1986,DraxlCPC2006}.

\begin{figure}[h]
    \centering
    \includegraphics[width=0.65\textwidth]{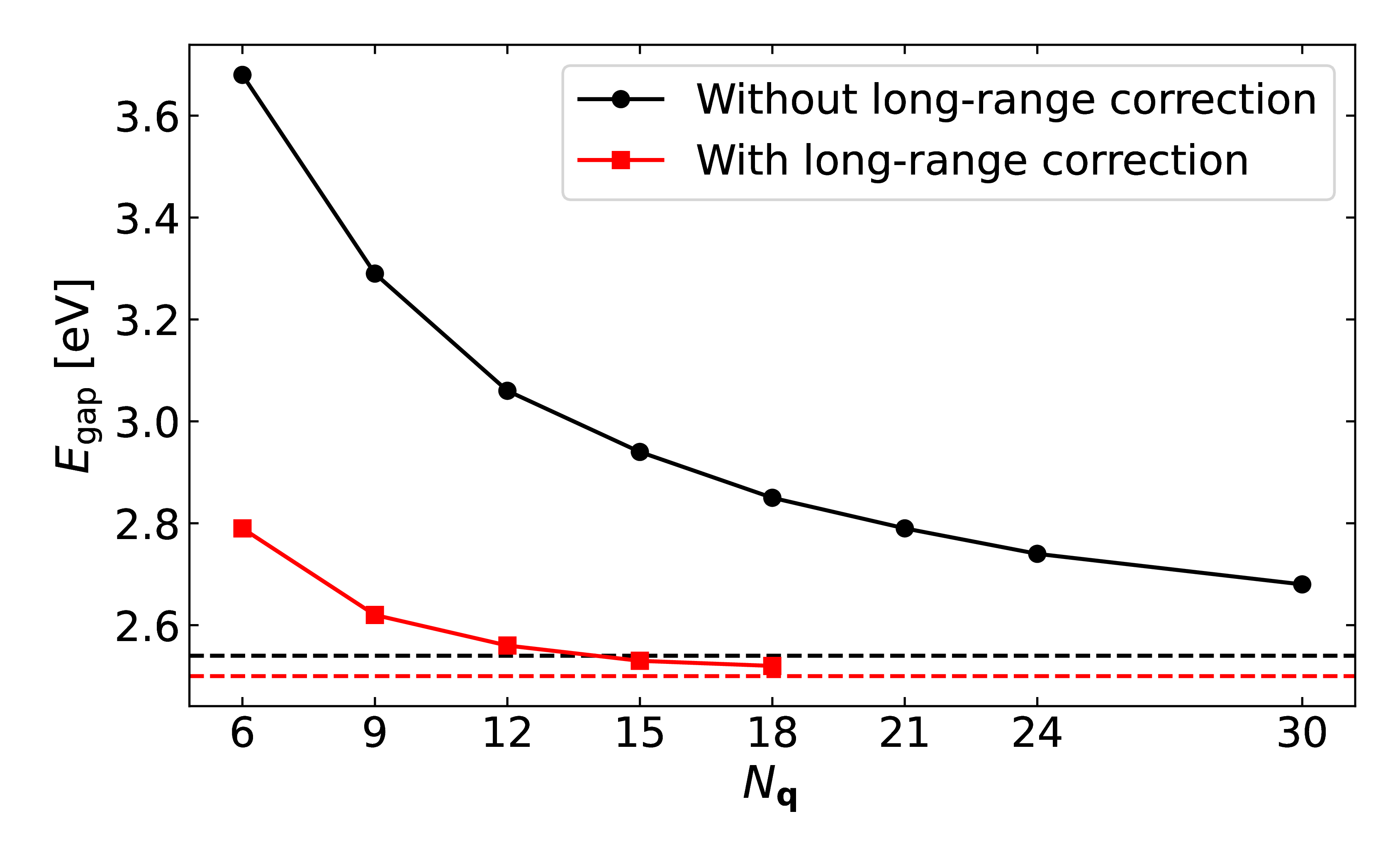}
    \caption{Direct band gap at the K point of MoS\textsubscript{2} for different \qq-point grids, $N_\qq$$\times$$N_\qq$$\times$1. The values are obtained with (red) and without (black) considering corrections for the long-range contribution. Dashed lines indicate extrapolation to the infinitely dense meshes.}
    \label{fig:2Dgw}
\end{figure}

This ambiguity complicates the application of convergence-acceleration techniques, particularly for 2D materials. To address this, Rasmussen \textit{et al.}~\cite{RasmussenPRB2016} employed a Coulomb-cutoff method to derive expressions with the proper 2D $\qq \to 0$ limit for the screened Coulomb potential and proposed a \gls{bz} integration scheme that combines numerical and isotropic analytical contributions. This approach ensures a physically consistent determination of the $\qq = 0$ limit and accelerates the convergence with the \qq-mesh by roughly a factor of three~\cite{RasmussenPRB2016,BenAlexThesis} (see Fig.~\ref{fig:2Dgw}). This methodology has been implemented in \exciting~\cite{BenAlexThesis}, enabling efficient $G_0W_0$ calculations for low-dimensional materials such as MoS\textsubscript{2}~\cite{RodriguesPelaNPJCM2024}.

For anisotropic systems, \exciting is interfaced with the code-agnostic library \texttt{IDieL}~\cite{greenXrepo,IDieLrepo}, which enables an accurate and parameter-free evaluation of the $\mathbf{q} \to 0$ limit. This library implements a formalism suggested in Ref.~\cite{FreysoldtCPC2007}, in which the dielectric matrix is averaged over a scaled-down $\Gamma$-centered \gls{bz} using spherical expansions of head and body, with Lebedev grids for angular integration. For example, the inverse of the head element can be written as
\begin{equation}
\varepsilon^{-1}_{00}(\mathbf{0}) = \frac{1}{\unitCellVol} \int_{\!\bm{\mathcal{S}}} \!\drm\hat{\mathbf{q}} \: \frac{1}{\hat{\mathbf{q}} \cdot \mathbf{L} \cdot \hat{\mathbf{q}}} \int_0^{q_\mathrm{max}(\hat{\mathbf{q}})}\!\! \drm q \, q^2\:,
\end{equation}
where $\mathbf{L}$ is the macroscopic dielectric matrix including local-field effects, $\bm{\mathcal{S}}$ is the solid angle, and $q_\mathrm{max}$ is the
direction-dependent upper cutoff in the magnitude of $\mathbf q$. The angular integral over $\hat{\mathbf{q}}$ can be efficiently evaluated using a spherical harmonic expansion, which separates the expression into two contributions, to accelerate convergence:
\begin{align}
\varepsilon^{-1}_{00}(0) &\sim \sum_{l=0}^{40} \frac{1+(-1)^l}{2}  \sum_{m=-l}^{l} c_{lm} \:
\underbrace{\frac{1}{\Omega_{\ts\Gamma{\mathrm{BZ}}}} \int \mathrm{d}\hat{\mathbf{q}} \: Y_{lm}(\hat{\mathbf{q}}) \int_0^{q_{\mathrm{max}}(\hat{\mathbf{q}})} \!\!\drm q \, q^2}_{\text{geometry contribution}}\:, \\
c_{lm} &= \underbrace{\int \mathrm{d}\hat{\mathbf{q}} \: [Y_{lm}(\hat{\mathbf{q}})]^* \,\frac{1}{\hat{\mathbf{q}} \cdot \mathbf{L} \cdot \hat{\mathbf{q}}}}_{\text{fast-converging contribution}}\:,
\end{align}
where $\Omega_{\ts\Gamma{\mathrm{BZ}}}$ is the volume of the scaled-down $\Gamma$-centered \gls{bz}.
The \textit{geometry contribution} converges slowly because the integration domain is a parallelepiped, requiring a high~$l$ (at least 131) for accurate results. However, it can be precomputed for all angles and contributions (head, wings, and body). The \textit{fast-converging contribution}, which depends on the dielectric matrix, requires only a few terms. A similar approach, using circular expansions and cubic splines, generalizes the method by Rasmussen \etal~\cite{RasmussenPRB2016} to fully anisotropic systems, ensuring an accurate and efficient evaluation of the ${\qq \to 0}$ limit.

\begin{figure}[h]
    \centering
    \includegraphics[width=0.65\textwidth]{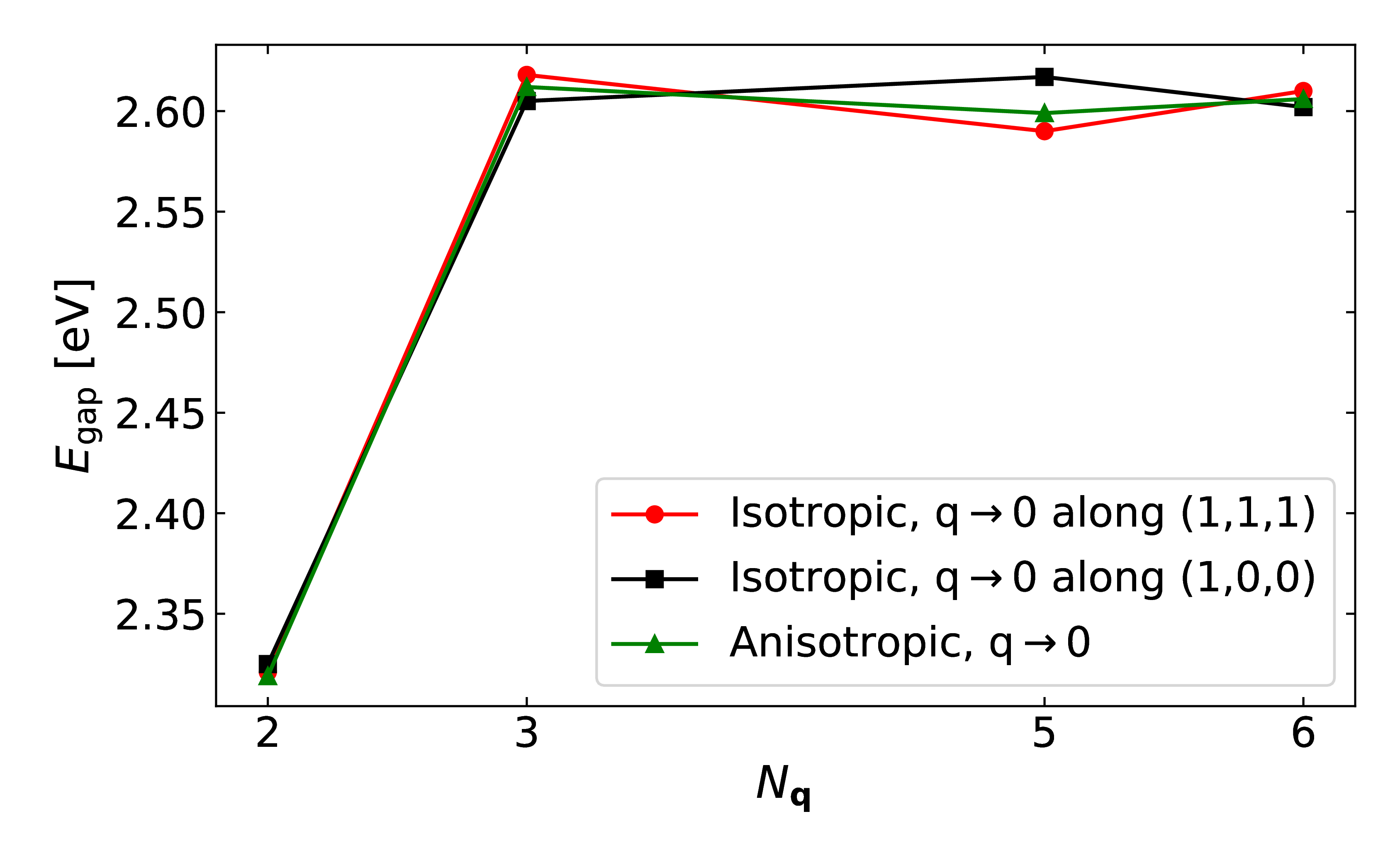}
    \caption{\gls{qp} gap at $\Gamma$ of black phosphorus for different $\mathbf{q}$-point grids ($2\times2\times2$), ($3\times3\times4$), ($5\times5\times6$), and ($6\times6\times8$), comparing the isotropic approach for different directions for the $\qq \rightarrow 0$ limit to the anisotropic formalism. All calculations performed with \exciting and the \texttt{IDieL} library.}
    \label{fig:IDieL}
\end{figure}

The effect of this anisotropic treatment on \gls{qp} energies is illustrated in Fig.~\ref{fig:IDieL}, which compares the band gap of black phosphorus computed with and without the anisotropic $\qq \to 0$ approach. Including the full anisotropy removes the need to select specific directions, eliminating errors associated with free parameters, whilst remaining valid for isotropic materials.

\subsection{Beyond the One-Shot Approximation}
\label{sec:New_scf_GW}

\begin{figure}
    \centering
    \includegraphics[width=0.95\textwidth]{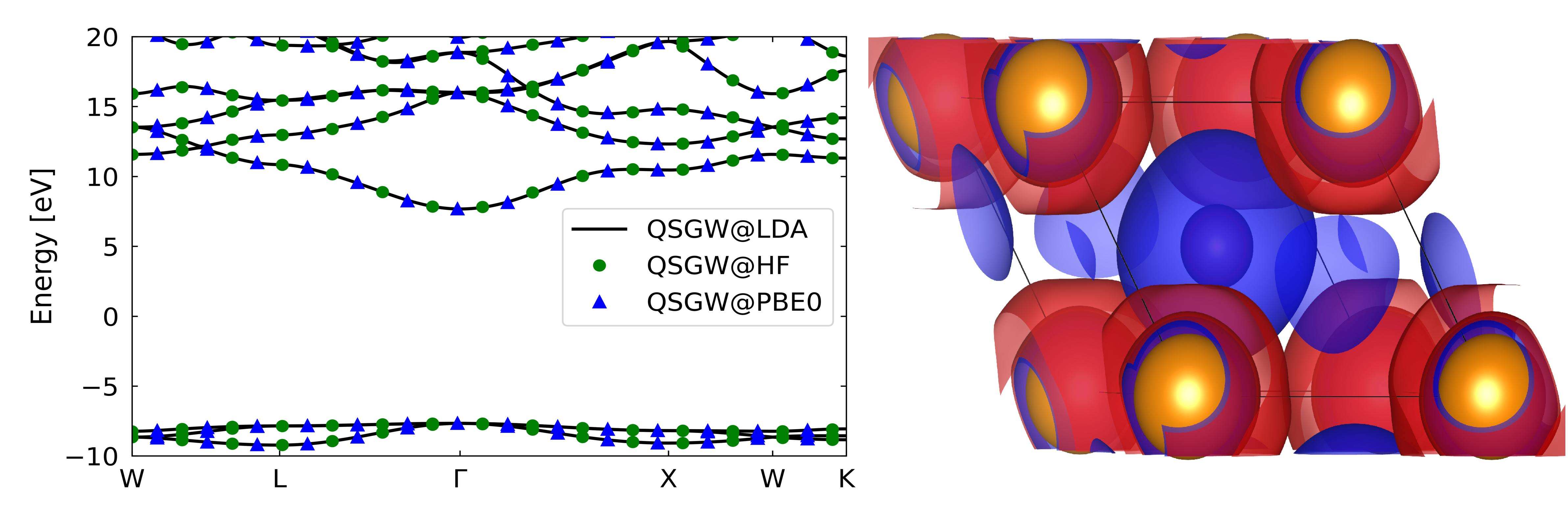}
    \caption{Left: QSGW bandstructure of Ar obtained with different starting points: LDA (solid-black line), HF (green circles), and PBE0 (blue triangles). Right: Charge-density difference in CaO  between \gls{qsgw} and LDA results at an isovalue of $\pm \num{1.5e-3}$ \unit{e.a_0^{-3}}. Positive values are indicated in red, negative ones in blue. Orange and white spheres depict calcium and oxygen atoms, respectively. Plots adapted from Ref.~\cite{Salas-Illanes2022} (copyright American Physical Society).}
    \label{fig:QSGW_image}
\end{figure}

As noted, one of the main drawbacks of the one-shot $GW$ approach is its starting-point dependence. This limitation can be overcome by self-consistency. Among the various self-consistent $GW$ schemes, one of the most widely used is the \gls{qsgw} method~\cite{SchilfgaardePRL2006,KotaniPRB2007,Salas-Illanes2022}. The central idea of \gls{qsgw} is to replace the dynamical, energy-dependent self-energy $\Sigma(\mathbf{r},\mathbf{r}',\omega)$ with an optimal static, nonlocal \gls{xc} potential $\vXCnohat^{\mathrm{opt}}$ that best represents the quasiparticle excitations. This potential is constructed such that the independent-particle Green function $G_0$, generated from $v_{\rm xc}^{\mathrm{opt}}$, approximates the fully interacting Green's function $G$ as closely as possible at the \gls{qp} level. Within this framework, the optimized \gls{xc} potential is defined as~\cite{KotaniPRB2007}:
\begin{equation}
\vXCnohat^{\mathrm{opt}}(\kk)
=
\frac{1}{2}
\sum_{n l}
\left| \kswf_{n\kk} \right\rangle
\left[
\re\!\left(\Sigma_{n l \mathbf{k}}(\epsilon_{n\kk})\right)
+
\re\!\left(\Sigma_{n l \mathbf{k}}(\epsilon_{l\kk})\right)
\right]
\left\langle \kswf_{l\kk} \right|.
\end{equation}
This formalism is implemented in \exciting~\cite{Salas-Illanes2022}, showing that the starting-point dependence can be overcome and that improved electronic densities can be obtained. Both aspects are demonstrated in Fig.~\ref{fig:QSGW_image}. A careful analysis of the charge-density redistribution reveals that it systematically reduces the delocalization error inherent to semilocal \gls{xc} functionals~\cite{MoriSanchezPRL2008}. 

Finally, we emphasize the importance of the all-electron nature of \exciting in this context. By treating valence, semicore, and core states on an equal footing, the method accurately captures core-valence exchange and correlation contributions to the self-energy. In contrast, neglecting these contributions, as is common in pseudopotential-based approaches, is already a source of error in $G_0W_0$ calculations \cite{Gomez-Abal2008,Li2012}, and self-consistent iterations can further amplify this error, potentially leading to larger biases in \gls{qp} energies and charge distributions, particularly in materials where semicore or core states play an active role. 

\subsection{Computational advancements: expansion and addition screening}
\label{sec:New_EAS_GW}

The computational cost of calculating the non-interacting polarizability represents a major bottleneck in both $G_0W_0$ and \gls{bse} calculations, scaling as $\bigO(N_\mathrm{at}^4)$ with the number of atoms $N_\mathrm{at}$ in the unit cell. This challenge becomes particularly severe for heterostructures and interface systems, which typically require large supercells compared to their bulk constituents. To address this, we have implemented the \gls{eas} approach, extending the formalism originally developed for plane-wave basis sets~\cite{XuanJCTC2019,LiuJCTC2019} to the mixed-product basis~\cite{SchebekPRB2025}. The key idea of the \gls{eas} is an additive ansatz, \ie the polarizability of the heterostructure is obtained by summing the contributions of the individual components. This comes with the advantage that the polarizability of a constituent does not need to be computed for that supercell; instead can be obtained from the respective unit cell by a folding (expansion) procedure. 

Our implementation of this formalism in the \exciting code~\cite{SchebekPRB2025} extends the original formulation \cite{XuanJCTC2019} in two key aspects. First, we have developed the necessary transformations for the mixed-product basis used in the \gls{lapwlo} method, accounting for the dual representation in \gls{mt} spheres and interstitial regions. Second, we have extended the \gls{eas} to optical excitations within \gls{bse} (\cref{sec:BSE}), where the polarizability plays an analogous role in screening the electron-hole interaction. We refer the reader to Ref.~\cite{SchebekPRB2025} for a full account of the implementation details. 

\begin{figure}[h]
    \centering
    \includegraphics[width=0.65\textwidth]{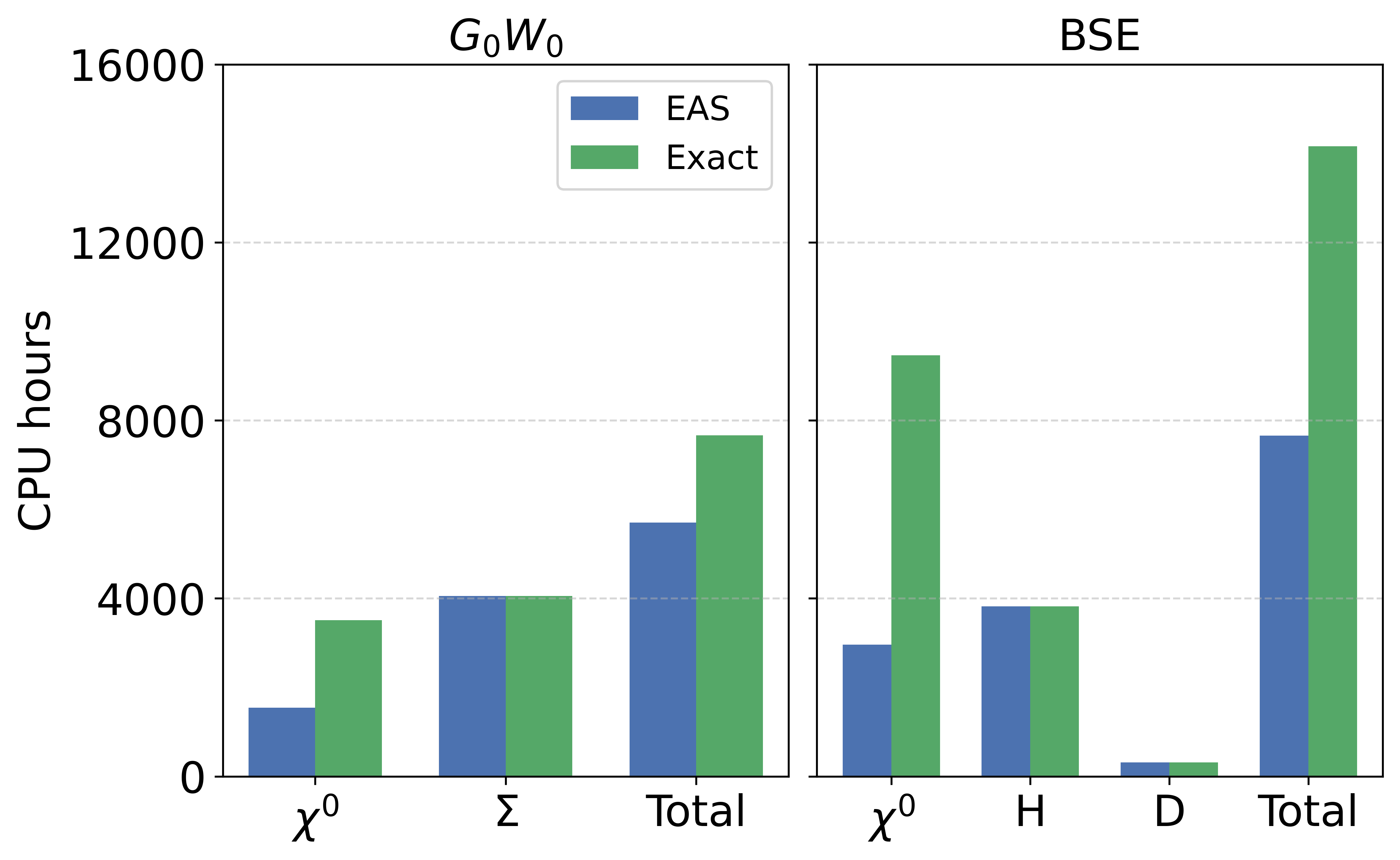}
    \caption{Computational effort for $G_0W_0$ (left) and BSE (right) calculations of pyridine@MoS$_2$ using the EAS method (blue) versus the exact calculations (green). The breakdown shows CPU hours for different computational steps: polarizability ($\pola$), self-energy ($\selfenergy$), Hamiltonian set up (H), and diagonalization (D).}
    \label{fig:eas_timings}
\end{figure}

The \gls{eas} is ideally suited for weakly bound \gls{vdw} heterostructures, in which covalent interactions are negligible~\cite{GonzalezOlivaPRMat2022,GonzalezOlivaPSSA2024,BenneckeNATUREPHYS2025}, allowing the system to be partitioned into two distinct components. \cref{fig:eas_timings} summarizes the computational savings reported in Ref.~\cite{SchebekPRB2025} for an organic/inorganic hybrid material, formed by MoS$_2$ and a monolayer of pyridine. The \gls{eas} reduces the computational time for the polarizability by 56\% for $G_0W_0$ and 69\% for \gls{bse}, resulting in total speedups of 25\% and 46\%, respectively. In the context of \gls{bse}, the combination of \gls{eas} with the \gls{isdf} method~\cite{henneke2020fast} (see~\cref{sec:New_Scaling_BSE}) provides a highly efficient framework.

%% file: figures/tools/workflow_GW.tex
	\begin{tikzpicture}[
	node distance=.8cm,
	every node/.append style={font=\sffamily\large},
	arrow/.style={thick,->,>=Stealth}
	]
	
	\tikzstyle{redbox}   = [rectangle, minimum width=3cm, minimum height=1cm,
	text centered, draw=black, fill=red!60]
    \tikzstyle{redbox2}   = [rectangle, minimum width=1cm, minimum height=1cm,
	text centered, draw=black, fill=red!60]
	\tikzstyle{greenbox} = [rectangle, minimum width=3cm, minimum height=1cm,
	text centered, draw=black, fill=green!30]
    \tikzstyle{greenbox2} = [rectangle, minimum width=1cm, minimum height=1cm,
	text centered, text width=1cm, draw=black, fill=green!30]
	
	\tikzstyle{orangebox} = [rectangle, rounded corners, minimum width=2cm,
	minimum height=1cm, text centered, text width=3cm, draw=black, fill=red!30]
    \tikzstyle{orangebox2} = [rectangle, rounded corners, minimum width=4cm,
	minimum height=1cm, text centered, text width=3cm, draw=black, fill=red!30]
	\tikzstyle{yellowbox} = [rectangle, minimum width=1cm, minimum height=1cm,
	text width=1cm, text centered, draw=black, fill=orange!30]
    \tikzstyle{yellowbox2} = [rectangle, minimum width=2cm, minimum height=1cm,
	text width=1cm, text centered, draw=black, fill=orange!30]
	
	\tikzstyle{legend}   = [rectangle, minimum width=1.4cm, minimum height=.8cm,
	text centered, draw=black]
	
	\node (coulomb)   [orangebox,line width=3.1pt,]                 {Coulomb};
	\node (polar)     [orangebox, line width=3.1pt,below left=of coulomb, xshift=-.2cm, yshift=-.4cm] {polarizability};
	\node (epsilon)   [greenbox,  line width=3.1pt, below=of coulomb, yshift=-.1cm]{epsilon};
	\node (invert)    [orangebox, below=of epsilon]               {invertEpsilon};
	\node (sigmac)    [greenbox,  line width=3.1pt, below=of invert, yshift=-.4cm]     {sigmac};
	\node (irred)     [orangebox2, left=of sigmac, xshift=-0.2cm]
	{irreducibleMapping};
	\node (sigmax)    [yellowbox2, line width=3.1pt,right=of epsilon, xshift=.02cm]  {sigmax};
	\node (qp)        [redbox, below=of sigmax, yshift=-2.2cm]
	{QPEigenvalues};
	\node (vxc)       [redbox2, right=of coulomb, xshift=3.0cm]         {vxc};
	
	\draw[arrow] (coulomb) -- (epsilon);
	\draw[arrow] (coulomb) -| (sigmax);
	\draw[arrow,dashed] (coulomb) -| (polar);
	\draw[arrow,dashed] (polar) -- (epsilon);
	
	\draw[arrow] (epsilon) -- (invert);
	\draw[arrow] (invert) -| (irred);
	\draw[arrow,dashed] (invert) -- (sigmac);
	\draw[arrow] (irred) -- (sigmac);
	
	\draw[arrow] (sigmac) -- (qp);
	\draw[arrow] (sigmax) -- (qp);
	\draw[arrow] (vxc) |- (qp);
	
	\node (gpu) [above left=of irred, xshift=-.4cm] {\Large GPU};
	
	\node (gpu1) [legend, line width=3.1pt, fill=white, below=0pt of gpu] {yes};
	\node (gpu2) [legend, fill=white, below=0pt of gpu1] {no};

	\node (mpi) [above=of gpu, yshift=2.75cm] {\Large MPI};
	
	\node (mpi1) [legend, fill=red!60,  below=0pt of mpi]  {no};
	\node (mpi2) [legend, fill=red!30, below=0pt of mpi1] {$\mathbf{q}$};
	\node (mpi3) [legend, fill=orange!30, below=0pt of mpi2] {$\mathbf{q,k}$};
	\node (mpi4) [legend, fill=green!30, below=0pt of mpi3] {$\mathbf{q,k,n}$};
	
\end{tikzpicture}

%% file: sections/bse.tex
\subsection{State of the art}
\label{sec:SOTA_BSE}
 Excitonic effects often leave clear signatures in absorption spectra. To account for these two-body interactions, the \gls{bse} is the method of choice and the workhorse of theoretical spectroscopy~\cite{salpeter1951relativistic,hanke1980manyparticleeffects,strinati1988application,albrecht1998abinitio_excitonic,rohlfing2000electron-hole,OnidaRMP2002} Solving the \gls{bse} may, however, be computationally demanding. The central task involves the construction and diagonalization of the \gls{bsh} in the basis of electronic transitions between occupied and unoccupied states. Owing to the two-body nature of the electron–hole interaction, the Hamiltonian contains dense off-diagonal elements, and achieving quantitative convergence typically requires a dense sampling of the \gls{bz}. Together, these factors lead to the generation of very large, dense Hamiltonian matrices and, consequently, to high computational cost \cite{OnidaRMP2002,rohlfing2000electron-hole,vorwerk2019bse,alvertis2023importance}. Significant progress has been achieved by expressing the interaction kernels using \acrfull{isdf} and combining this representation with an iterative Lanczos solver, substantially accelerating both matrix construction and the calculation of spectra \cite{shao2018structure,hu2018accelerating,henneke2020fast,maurer2026fastBSE}.

Beyond algorithmic advances, recent work has increasingly focused on extending the \gls{bse} formalism itself. In particular, going beyond the widely adopted approximation of static screening has emerged as an important direction, as dynamical screening effects are required to capture frequency-dependent correlations and can significantly improve the description of excitation energies and spectral features in a range of materials \cite{loos2020dynamical,zhang2023effect}. At the same time, extensions of the \gls{bse} framework have enabled access to magnetic excitations, allowing for the description of spin-flip processes and collective spin modes \cite{olsen2021magnons}. These developments highlight the versatility of generalized \gls{bse} approaches in describing a broader class of collective excitations beyond conventional excitonic physics.

The \gls{bse} has also been successfully applied to non-equilibrium situations. By combining \gls{rt}-\gls{tddft} simulations with subsequent \gls{bse} calculations, it has become possible to model spectroscopic responses of systems driven out of equilibrium by external perturbations \cite{perfetto2015nonequilibrium,sangalli2016nonequilibrium}, including pump-probe spectroscopy \cite{Rossi2025,Qiao2025,Farahani2024PRB} (see \cref{sec:PP}). In addition, \gls{bse}-based methods have been extended to the theoretical description of \acrfull{rixs}. These applications require an accurate treatment of core-level excitations and their coupling to valence states, posing both conceptual and computational challenges within the many-body framework \cite{Vinson2017RIXS,Vorwerk2020Brixs,Vorwerk2022Brixs}. 

Another emerging area of research concerns the coupling between excitonic states and lattice vibrations \cite{marini2008finitTexcitons,filip2021phonon,chan2023exciton,alvertis2024phonon,schebek2025phonon}. For example, phonon-mediated screening impacts exciton binding energies and spectral features in polar materials \cite{alvertis2024phonon,schebek2025phonon}. A consistent treatment of \gls{exph} further gives rise to phenomena such as \gls{ste} and excitonic polarons, which have been observed in a variety of materials \cite{yang2022novalApproach,BaiPRLAbInitioSTE,Dai2024ExcitonicPolarons,Dai2024Theory_of_excitonic_polarons}.

\subsection{Methodology}
\label{sec:Methodology_BSE}
In practice, the \gls{bse} is typically reformulated as an effective two-particle eigenvalue equation. The corresponding two-particle wavefunction is expanded in a transition basis constructed from products of single-particle wavefunctions of the form
\begin{align}
&\kswf_{o\kk_+}(\rr)\,\kswf_{u\kk-}(\rr')\,, \label{eq:bse:resonant} \\
&\kswf_{u(-\kk_-)}(\rr)\,\kswf_{o(-\kk_+)}(\rr')\,, \label{eq:bse:aresonant}
\end{align}
where $o$ and $u$ label occupied and unoccupied electronic states, respectively, and $\kk$ denotes a \gls{bz} wavevector. The single-particle wavefunction $\kswf_{i\kk}(\rr)$ corresponds to electronic state $i$ at crystal momentum $\kk$ and position $\rr$. The shifted wavevectors are defined as $\kk_\pm = \kk \pm \qq/2$, where $\qq$ is the momentum transfer. \cref{eq:bse:resonant} reflects resonant (excitation) transitions, while \cref{eq:bse:aresonant} corresponds to anti-resonant (de-excitation) transitions. Using this transition basis, solving the \gls{bse} is equivalent to solving the eigenvalue problem of an effective Hamiltonian of the form
\begin{align}
H^{\mathrm{BSE}}(\qq) =
\begin{pmatrix}
\phantom{-}A(\qq) & \phantom{-}B(\qq) \\
-B^*\hspace{-0.7pt}(\qq) & -A^*\hspace{-0.7pt}(\qq)
\end{pmatrix}\,.
\end{align}
The diagonal and off-diagonal blocks are given by
\begin{align}
A_{ou\kk,\:o'u'\kk'}(\qq)
&= D_{ou\kk, \: o'u'\kk'}(\qq)
+ \gamma_\mathrm{x} \, V_{ou\kk, \: o'u'\kk'}(\qq)
- W_{ou\kk, \: o'u'\kk'}(\qq)\,, \\
B_{ou\kk, \: o'u'\kk'}(\qq)
&= \gamma_\mathrm{x} \, V_{ou\kk, \: o'u'\kk'}(\qq)
- W_{ou\kk, \: o'u'\kk'}(\qq)\,,
\end{align}
where $D$ denotes the diagonal (independent-particle) term, $V$ the exchange interaction, and $\scrcoul$ the screened direct interaction. The factor $\gamma_\mathrm{x}$ allows to distinguish between spin singlet ($\gamma_\mathrm{x} = 2$) and spin triplet ($\gamma_\mathrm{x} = 0$) excitations. If the screened Coulomb interaction $W$ is neglected, the resulting eigenvalue problem reduces to \gls{rpa}.

A widely used approximation that substantially reduces the computational complexity of the problem is the \acrfull{tda} \cite{OnidaRMP2002}. In this approximation, the coupling between resonant and anti-resonant transitions is neglected, such that the off-diagonal block satisfies $B(\qq) \approx 0$. As a result, the \gls{bsh} becomes block diagonal, and the problem reduces to the diagonalization of the matrix $A(\qq)$. In this case, the effective Hamiltonian is given by
\begin{align}
H^{\mathrm{TDA}}(\qq) = A(\qq).
\end{align}
For simplicity, the \gls{tda} is assumed throughout the following discussion. Nevertheless, the \exciting code is capable of going beyond the \gls{tda} and explicitly accounting for the coupling terms when required \cite{vorwerk2019bse}.

The diagonal term is given by the energy differences between the participating occupied and unoccupied electronic states,
\begin{align}\label{eq:bsh_diagonal} 
D_{ou\kk, \: o'u'\kk'}(\qq) = \left[\epsilon_{u\kk_-} - \epsilon_{o\kk_+} \right] \delta_{oo'}\delta_{uu'}\delta_{\kk\kk'}\:. 
\end{align}
Neglecting the interaction kernels altogether reduces the problem to the \acrfull{ipa}, which is computationally efficient and widely used, but inherently incapable of capturing excitonic effects. The matrix elements of the exchange interaction are given by
\begin{align}\label{eq:bse:exchange_kernel_marix_elements} 
\begin{split} 
V_{ou\kk,\: o'u'\kk'}(\qq) 
&=  
\frac{1}{N_{\kk}^2} \iint_{\unitCellVol\times\unitCellVol} \mathrm{d} \rr \, \mathrm{d} \rr' \: \kswf^{\phantom{*}}_{o\kk_+}\!(\rr)\,\kswf^*_{u\kk_-}\!(\rr) \:v^\pI_{\ts\rm C}(\rr, \rr')\, \kswf^*_{o'\kk'_+}\!(\rr')\,\kswf^{\phantom{*}}_{u'\kk'_-}\!(\rr') \:, 
\end{split} \end{align}
where $v^\pI_{\ts\rm C}(\rr,\rr')$ denotes the bare Coulomb interaction. The matrix elements of the direct interaction are given by
\begin{align}\label{eq:bse:screened_kernel_marix_elements} 
\begin{split} 
\scrcoul_{ou\kk, \: o'u'\kk'}(\qq) 
&= 
\frac{1}{N_{\kk}^2} \iint_{\unitCellVol\times\unitCellVol} \mathrm{d} \rr \, \mathrm{d} \rr' \,\psi^{\phantom{*}}_{o\kk_+}\!(\rr')\,\psi^*_{o'\kk_+'}\!(\rr')\: \scrcoul(\rr, \rr')\: \psi^*_{u\kk_-}\!(\rr)\,\psi^{\phantom{*}}_{u'\kk'_-}\!(\rr) \:, 
\end{split} 
\end{align}
where $\scrcoul(\rr,\rr')$ denotes the statically screened Coulomb interaction:
\begin{align} 
\scrcoul(\rr, \rr') 
= 
\int \mathrm{d} \rr''\: v^\pI_{\ts\rm C}(\rr, \rr'')\:\varepsilon^{-1}(\rr'', \rr') \:. 
\end{align}
with $\diel^{-1}$ being the inverse dielectric function.

The real-space integrals entering the exchange and direct interaction kernels can be evaluated efficiently in reciprocal space. In this representation, the matrix elements of the exchange and screened interactions take the form
\begin{align} 
V_{ou\kk,\: o'u'\kk'}(\qq) 
&= 
\sum_\GG \hat{v}_\GG^\pI(\qq)\: M^*_{ou\kk_-}\!(\GG, \qq)\:M_{o'u'\kk'_-}\!(\GG, \qq) \:, \\[0.5em] 
\scrcoul_{ou\kk,\: o'u'\kk'}(\qq) 
&= 
\sum_{\GG, \GG'} \hat{\scrcoul}_{\GG\GG'}^\pI(\kk\! -\! \kk') \:M^*_{o'o\kk'_+}\!(\GG, \kk\! -\! \kk') \:M_{u'u\kk'_-}\!(\GG', \kk\! -\! \kk')\:, 
\end{align}
where the planewave matrix elements are defined as
\begin{align}\label{eq:plane_wave_matrix_elements} 
M_{mn\kk}(\GG, \qq) 
= 
\langle n\kk | \,\ee^{-\im(\GG + \qq)\rr} | m(\kk\! +\! \qq)\rangle \:. 
\end{align}
These matrix elements are computed within the \gls{lapwlo} basis using single-particle states obtained from a \gls{gs} calculation. The Fourier transforms of the bare and screened Coulomb interactions are given by
\begin{align} 
v_{\GG}^\pI(\qq) 
&= 
\frac{1}{\Omega} \frac{4\pi}{|\GG + \qq|^2} \:, \\[0.5em]
\scrcoul_{\GG, \GG'}^\pI(\qq)
&= 
v_{\GG}^\pI(\qq)\:\diel^{-1}_{\GG\GG'}(\qq, \freq=0) \:. 
\end{align}
The dielectric matrix $\diel_{\GG\GG'}^\pI(\qq,\freq)$ is evaluated within the random-phase approximation (\gls{rpa}) \cite{ehrenreich1959self} and reads
\begin{equation}\label{eq:dielectric_matrix_rpa} 
\varepsilon^{\mathrm{RPA}}_{\GG\GG'}(\qq, \omega) 
= 
\delta_{\GG,\GG'}^\pI - \frac{\hat{v}_\GG^\pI(\qq)}{N_\kk}\sum_{ij\kk}\frac{\occ{j}{\kk+\qq}-\occ{i}{\kk}}{\eigval_{j\kk+\qq}-\eigval_{i\kk}-\freq}\:M^*_{ij\kk\!}(\GG, \qq) \: M_{ij\kk\!}(\GG', \qq)\:, 
\end{equation}
where $\occ{i}{\kk}$ denotes the occupation of an electronic state with energy $\eigval_{i\kk}$.

\begin{figure}[h]
\centering
\includegraphics[width=0.8\textwidth]{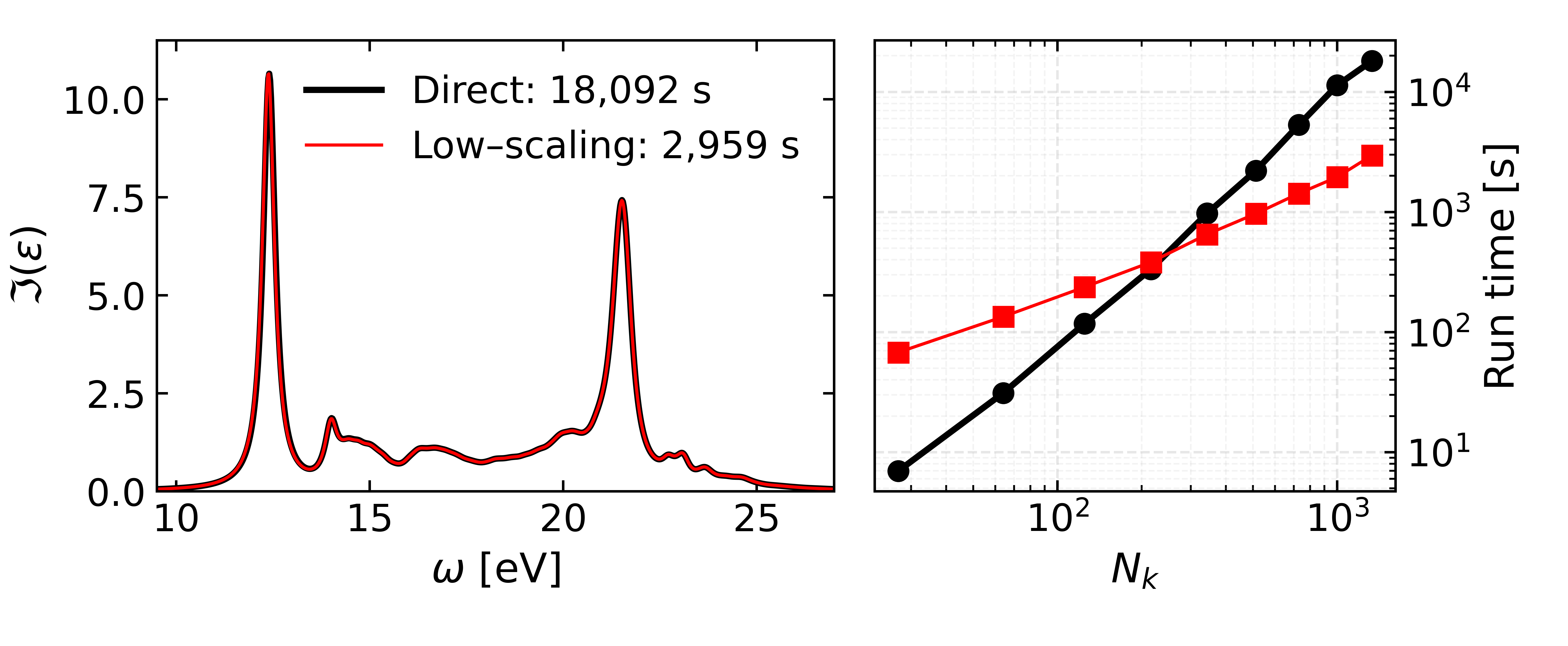}
\vspace*{-5mm}
\caption{Left: Optical absorption spectrum of LiF on an $11^3$ $\kk$-point grid obtained from the direct solution of the \gls{bse} (black) and from the \gls{isdf}–Lanczos \gls{bse} (red); the corresponding run times differ by a factor of 6. Right: Run time of the direct (black line, round markers) and \gls{isdf}–Lanczos \gls{bse} (red line, square markers) implementations as a function of the number of $\kk$-points.}
\label{fig:fbse_spectra_timing}
\end{figure}

\subsection{Low-scaling \acrshort*{bse} implementation}
\label{sec:New_Scaling_BSE}
For complex materials, solving the \gls{bse} is a computationally demanding task and often not feasible in practice. This is particularly true when obtaining precise exciton binding energies require dense $\kk$-grids, such as for surfaces and 2D materials or materials with rather weak electron-hole interaction. The construction of the \gls{bse} matrix scales as $\bigO(N_o^2N_u^2N_{\kk}^2)$, and solving the resulting eigenvalue problem scales as $\bigO(N_o^3N_u^3N_{\kk}^3)$, quickly becoming prohibitive with increasing system size. To reduce the computational complexity associated with computing the matrix elements, we approximate pair products of wavefunctions using the \gls{isdf}. Given a set of wavefunctions $\kswf_{i\kk}(\rr)$, where $\rr$ denotes points on a discrete real-space grid of size $N_{\rm g}$ for sampling the unit cell, any pair product can be approximated as an interpolation of the wavefunctions evaluated at a small number $N_\mu$ of special interpolation points \cite{shao2018structure,hu2018accelerating}:
\begin{equation}\label{eq:bse:fast:isdf_me} \kswf^*_{i\kk}(\rr)\,\kswf^{\phantom{*}}_{j\kk'}(\rr) \approx \sum_{\mu=1}^{N_\mu}\zeta_\mu(\rr)\,\kswf^*_{i\kk}(\rr_\mu)\,\kswf_{j\kk'}(\rr_\mu)\:, \end{equation}
where $\zeta_\mu(\rr)$ are interpolation functions. The interpolation points $\left\{\rr_\mu\right\}$ themselves are obtained efficiently using a \acrfull{cvt} \cite{dong2018cvt}. By replacing the wavefunction pair products in \cref{eq:bse:exchange_kernel_marix_elements} and \cref{eq:bse:screened_kernel_marix_elements} with the \gls{isdf} representation, the interaction kernels are reformulated such that they can be applied efficiently to a vector without explicitly constructing the full matrix. When combined with a Lanczos solver, this approach significantly reduces the overall computational scaling to $\bigO(N_o N_u N_{\kk}\log N_{\kk})$ \cite{henneke2020fast,maurer2026fastBSE}. In \cref{fig:fbse_spectra_timing}, we demonstrate for the example of LiF that the new implementation in \exciting reproduces the spectrum in perfect agreement with the direct approach, but with improved scaling  with respect to the number of $\kk$-points. While in this simple example, the speedup is a factor of six, it can be even two orders of magnitude for larger unit cells or systems requiring very dense \gls{bz} samplings.

\subsection{\Acrshort*{svlo} in the \acrshort*{bse}}
\label{sec:New_SVLO_BSE}

\begin{figure}[h]
  \centering
  \includegraphics[width=0.7\textwidth]{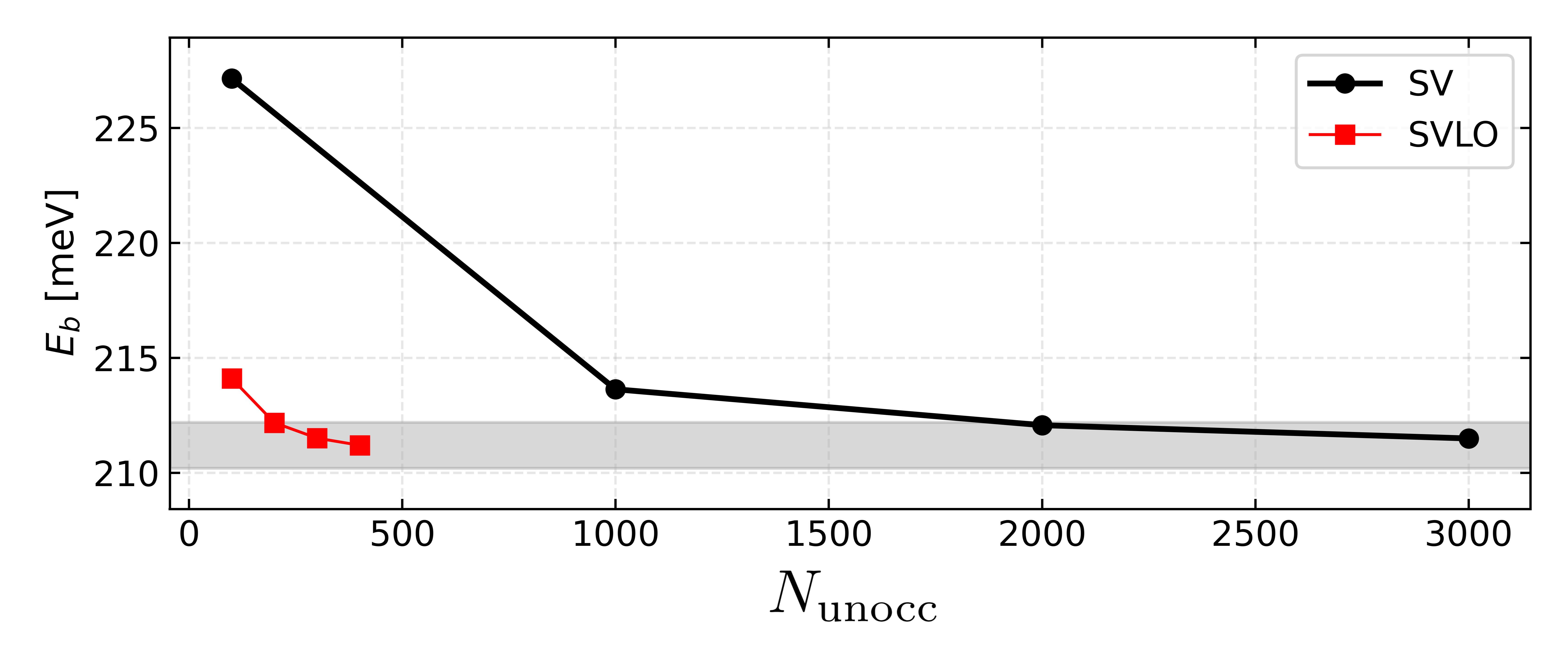}
  \vspace*{-3mm}
  \caption{Binding energy of the lowest-lying exciton in the optical spectrum of $\gamma$-CsPbI$_3$ as a function of the number of unoccupied states, $N{_\mathrm{unocc}}$. The gray shaded region indicates the convergence threshold of 2~meV.}
  \label{fig:svlo_bse_cspbi3}
\end{figure}

Precise \gls{bse} calculations require high-quality \gls{gs} calculations as a starting point. This is particularly challenging for systems with strong \gls{soc}, where convergence typically requires a large number of unoccupied states, $N_{\mathrm{unocc}}$. The \gls{svlo} basis, introduced in \cref{sec:New_SVLO_GS}, achieves convergence of \gls{gs} calculations with a number of orders of magnitude smaller compared to the conventional \gls{sv} basis. We extend this concept to \gls{bse} calculations by consistently implementing both planewave and momentum matrix elements in the \gls{svlo} basis. \cref{fig:svlo_bse_cspbi3} illustrates the impact of the two basis sets on a representative \gls{bse} calculation by showing the binding energy of the lowest exciton in $\gamma$-CsPbI$_3$ as a function of $N{_\mathrm{unocc}}$. The \gls{svlo} basis exhibits a substantially faster convergence, reaching the target accuracy with substantially fewer unoccupied states.

\subsection{Non-equilibrium \acrshort*{bse}}
\label{sec:New_NonEq_BSE}

The standard implementation of the \gls{bse} provides an accurate description of neutral excitations in materials. However, it is limited to steady-state conditions and fails to account for the non-equilibrium dynamics arising under ultrafast excitations. In pump-probe spectroscopy, for instance, a laser pulse drives the system far from equilibrium, resulting in transient modifications to the electronic structure and optical response. To capture these effects, we have implemented a non-equilibrium variant of \gls{bse} by including photoinduced carrier distributions in the evaluation of both the momentum matrix elements and the screened Coulomb interaction \cite{Rossi2025,Qiao2025}. This modification enables the calculation of transient excitonic states and the corresponding spectra in the presence of non-equilibrium carrier populations. In practice, the approach involves solving the \gls{bse} using a time-dependent photoexcited carrier occupation, typically obtained from \gls{cdft} (see \cref{sec:New_cDFT_GS}) or \gls{tddft} (see \cref{sec:New_fsDynamics_TDDFT}). For an application to pump-probe spectroscopy, we refer to \cref{sec:New_PPS2_PP}.

\subsection{Resonant inelastic x-ray scattering}
\label{sec:New_RIXS_BSE}
\begin{figure}[htb]
    \centering
    \vspace*{-2mm}
    \includegraphics[width=0.7\textwidth]{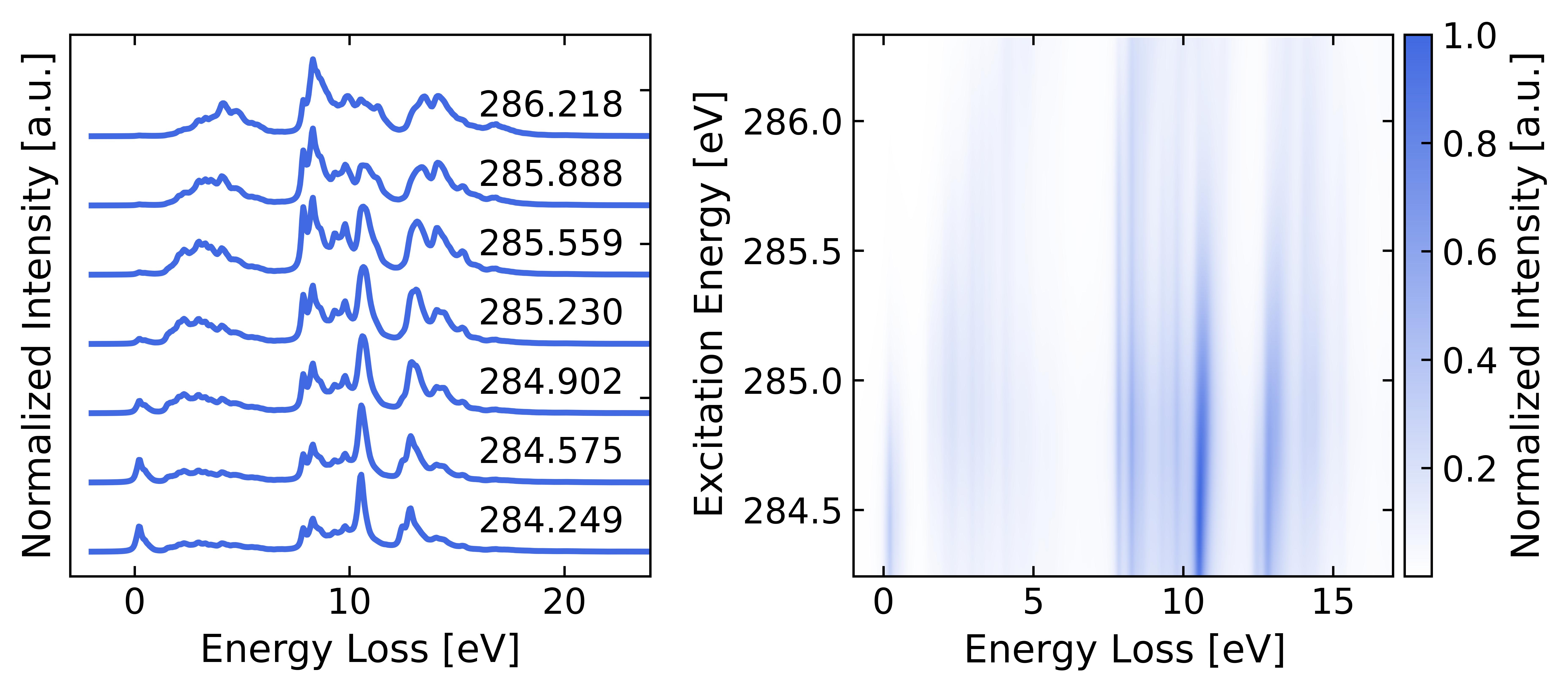}
    \vspace*{-2mm}
    \caption{Left: \gls{rixs} spectra for the carbon K-edge of graphite as a function of the energy loss for different excitation energies $\omega_1$. Right: Normalized double-differential \gls{rixs} cross section as function of excitation energy $\omega_1$ and energy loss $\omega$. The incident beam angle $\alpha$ is $60^\circ$, with perpendicular polarization relative to the beam direction, described by $(\cos \alpha, 0, \sin \alpha)$. The emitted photons are detected at an angle of $90^\circ$, relative to the incident beam, with corresponding polarization given by $(-\sin \alpha, 0, \cos \alpha)$.}
    \label{fig:RIXS}
\end{figure}

\gls{rixs} is a two step scattering process. An incident X-ray photon with energy $\omega_1$ is absorbed by the system, creating an intermediate excited state by exciting a core electron into the conduction band. In a next step, the core hole is filled by a valence electron, yielding the emission of a photon with energy $\omega_2$. The final excited state thus consists of a hole in the valence band and an excited electron in the conduction band. While energies of both incoming and outgoing photons lie in the X-ray region, the energy loss $\omega=\omega_1-\omega_2$ is in the range of only a few eV. The process is described by the Kramers-Heisenberg formula, following the theoretical description proposed in Refs.~\cite{Vorwerk2020Brixs,Vorwerk2022Brixs} and implemented in the \brixs package. Recent extensions concern the generalization in terms of arbitrary polarization vectors of the incoming and outgoing photon \cite{GraphiteRIXS}. Accurate \gls{rixs} spectra are computed based on the results of two \gls{bse} calculations, one for the core, and one for the optical excitations. \brixs uses the exciton eigenstates and eigenvalues from both calculations to determine the oscillator strengths and the excitation pathways. In a subsequent step, these quantities are combined to calculate the \gls{rixs} spectra for selected excitation energies. The present framework can be straightforwardly extended to non-equilibrium \gls{rixs} by employing non-equilibrium \gls{bse} calculations, as described in \cref{sec:New_NonEq_BSE}. For more details on the implementation of \gls{rixs} in \brixs, we refer to Ref.~\cite{Vorwerk2021Brixs}. 

We showcase the capabilities of \brixs for the carbon K-edge of graphite. Graphite is a layered material with strong covalent in-plane bonds and weak inter-plane \gls{vdw} interactions. While $\pi$-orbitals are oriented predominantly perpendicular to the graphene layers, $\sigma$-orbitals lie within the plane. Thus, graphite exhibits strong angular dependencies in its \gls{rixs} response. We consider an incident angle of $\alpha=\ang{60}$ with perpendicular polarization relative to the beam direction, while the emitted photons are detected at $\ang{90}$ relative to the incident beam. Lifetime broadenings of $\SI{0.5}{\eV}$ and $\SI{0.1}{\eV}$ are applied for the intermediate state and the final state, respectively. The resulting \gls{rixs} spectra, shown in \cref{fig:RIXS}, reveal distinct angular dependencies. The left panel of \cref{fig:RIXS} displays the RIXS intensity as a function of energy loss for several excitation energies $\omega_1$. For the chosen incident angle, the spectral feature region of $\SIrange{7}{11}{\eV}$ is dominating, indicating favored transitions into lower-lying $\sigma^*$ states. In contrast, transitions into $\pi^*$ states, lying closer to the Fermi energy, are suppressed and exhibit significantly weaker spectral features for low energy loss. However, the relative intensity of these $\pi$-related features becomes more pronounced at higher $\omega_1$. The right panel presents the normalized double-differential \gls{rixs} cross section as a function of both excitation energy $\omega_1$ and energy loss $\omega$. It highlights the same dominant spectral region between $\SIrange{7}{11}{\eV}$, while features at lower energy loss appear with reduced intensity. In addition, systematic shifts in the $\SIrange{3}{5}{\eV}$ energy-loss range towards higher energy loss with increasing excitation energy are observed. This behavior reflects the dispersion of the $\pi$ band involved in the corresponding excitation and de-excitation pathways.

Since \gls{rixs} spectra are obtained from preceding \gls{bse} calculations, the \brixs package was designed as a stand-alone tool, \ie it is not directly implemented in \exciting. It its seamlessly integrated within \excitingworkflow, but can also accept input from other \gls{bse} codes. The \gls{rixs} DDCS is computed with the Python package \pybrixs. Both packages are available on GitHub \cite{git-brixs,git-pybrixs}.

\subsection{Lattice screening in the optical spectra of polar materials}
\label{sec:New_screening_BSE}

\begin{figure}[ht]
    \centering
    \includegraphics[width=.8\textwidth]{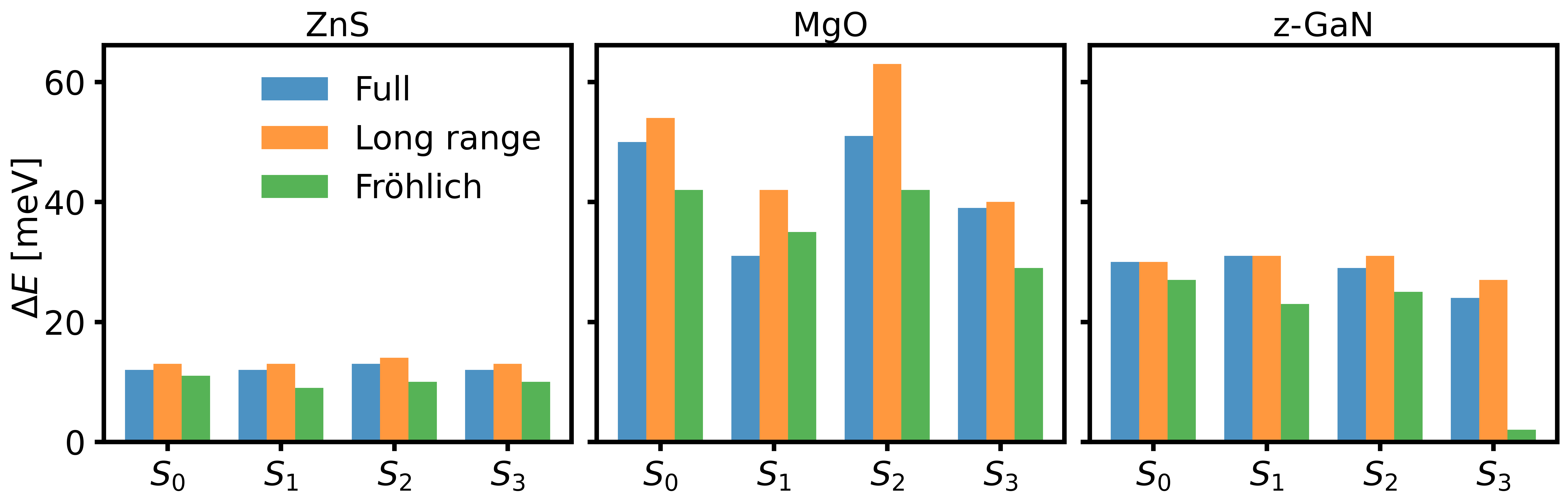}
\caption{Renormalization of excitation energies due to polar lattice screening for ZnS, MgO, and z-GaN. The results correspond to the most prominent peaks in the absorption spectra. The calculations are performed using the full \emph{ab initio} electron–phonon matrix elements (blue), their long-range contribution only, and the Fr\"ohlich approximation~\cite{Frhlich1954}.}
\label{fig:lattice_screening}
\end{figure}

Exciton–phonon coupling gives rise to a wide range of well-established physical phenomena (see also \cref{sec:New_ex_ph_BSE}). It may, for example, play a role in the screening of the Coulomb interaction between electrons and holes. In practice, however, most implementations of the \gls{bse} only account for the electronic contribution. One notable consequence of exciton–phonon coupling is the reduction of exciton binding energies, which originates from a weakened Coulomb attraction due to vibrational contributions to the screening. This effect is especially important in ionic materials, where long-range optical phonon modes can generate strong macroscopic electric fields \cite{Giannozzi1991Ab_initio_phonon_dispersion}. Within the \gls{bse} framework, such effects have previously been incorporated by modeling lattice screening through approximations to the dielectric function \cite{bechstedt2019Influence_of_screening,Schleife2018Optical_properties,umari2018Infrared,Fuchs2008Efficient_ONsqr,Adamska2021BSE_approach,bokdam2016Role_of_polar,BechstedtPRB2005}. More recent first-principles approaches instead explicitly include a phonon-induced contribution to the screened Coulomb interaction \cite{alvertis2024phonon,filip2021phonon,schebek2025phonon}.

In a recent work \cite{schebek2025phonon}, some of us implemented phonon-assisted screening effects into the screened Coulomb interaction within the \gls{bse} framework. Applications to polar semiconductors such as ZnS, MgO, and GaN  have shown that vibrational screening not only leads to a renormalization of exciton binding energies at the absorption onset, but also to comparable red shifts of higher-energy absorption features in the order of $\SI{50}{\milli\eV}$ (see Fig.~\ref{fig:lattice_screening}). These effects are primarily caused by long-range  coupling~\cite{Verdi2015,Frhlich1954} to polar longitudinal optical phonons, while other vibrational modes contribute minimally. We also find that the Fröhlich model~\cite{Frhlich1954}---assuming an isotropic system with a single longitudinal optical phonon mode---largely captures the observed renormalization effects in these systems.

\subsection{Exciton-phonon coupling and exciton-polarons} 
\label{sec:New_ex_ph_BSE}

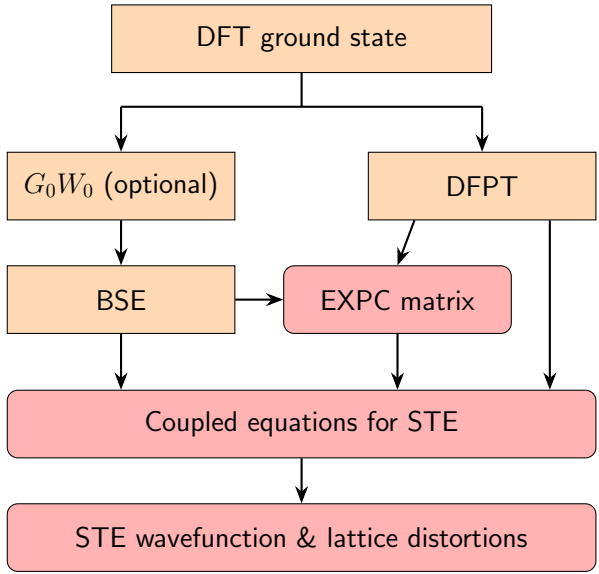
\begin{figure}[h]
    \centering
    \input{figures/tools/workflow_exciton_polaron}
    \caption{Workflow for constructing the \gls{exph} matrix elements and computing \glspl{ste}.}
    \label{fig:workflow_exciton_polaron}
\end{figure}

\gls{exph} connects optical excitations to the lattice degrees of freedom. In some crystals and molecules, the creation of an exciton induces a charge-density redistribution, which can in turn drive a local lattice distortion. This lattice distortion feeds back to the exciton by renormalizing the excitonic wavefunction. The composite \gls{qp} formed by an exciton dressed by its accompanying lattice distortion is commonly referred to as an exciton-polaron, and it is also known as a \acrshort{ste}.

\Glspl{ste} can be obtained by solving a system of coupled equations that describe the interaction between the \gls{ste} wavefunction and the lattice distortion. The theoretical framework and computational workflow shown in~\cref{fig:workflow_exciton_polaron} follow Ref. \cite{yang2022novalApproach}. Similar equations were independently formulated later and applied by other groups \cite{BaiPRLAbInitioSTE,Dai2024ExcitonicPolarons,Dai2024Theory_of_excitonic_polarons}. In this approach, the \gls{exph} interaction matrix \cite{ChenPRL2020}, which is essentially the excited-state force along phonon coordinates, is constructed by combining \gls{dfpt} with the momentum-dependent \gls{bse}. The local lattice distortion is expanded into phonon normal modes, and the \gls{ste} wavefunction is represented using momentum-dependent exciton wavefunctions. Notably, this formulation explicitly avoids modeling of \glspl{ste} by supercells, \ie enabling the entire computational procedure to be carried out within the primitive cell.

We take the CO molecule as a representative example to illustrate how the \gls{exph} framework can be used to predict structural relaxation in the excited state. The C–O bond length increases from its ground-state value of 1.13~\AA\ to 1.22~\AA\ in the $1^{1}\Pi$ excited state. Although CO exhibits only a single vibrational mode, our approach circumvents the conventional multi-step geometry optimization required by the excited-state force method, directly providing the relaxed excited-state structure.

%% file: figures/tools/workflow_exciton_polaron.tex
\begin{tikzpicture}[
  scale=0.75, transform shape,
  every node/.append style={font=\sffamily\Large, align=center},
  arrow/.style={thick,->,>=Stealth},
  link/.style={thick},
  topbox/.style={rectangle, minimum height=12mm, text width=6.5cm, draw=black, fill=orange!30},
  sidebox/.style={rectangle, minimum height=12mm, text width=3.8cm, draw=black, fill=orange!30},
  widebox/.style={rectangle, minimum height=12mm, text width=3.8cm, draw=black, fill=orange!30},
  smallbox/.style={rectangle, rounded corners=4pt, minimum height=12mm, text width=3.8cm, draw=black, fill=red!30},
  resultbox/.style={rectangle, rounded corners=4pt, minimum height=12mm, text width=10.2cm, draw=black, fill=red!30},
  finalbox/.style={rectangle, rounded corners=4pt, minimum height=12mm, text width=10.2cm, draw=black, fill=red!30}
]

\node (dft) [topbox] {DFT ground state};

\node (gw)   [sidebox, below=14mm of dft, xshift=-32mm] {$G_0W_0$ (optional)};
\node (dfpt) [widebox, below=14mm of dft, xshift=32mm] {DFPT};

\node (bse) [sidebox, below=8mm of gw] {BSE};
\node (gmn) [smallbox, below=8mm of dfpt, xshift=-15mm] {EXPC matrix};

\coordinate (midRow3) at ($(bse.south)!0.5!(dfpt.south)$);
\node (cb) [resultbox, below=20mm of midRow3] {Coupled equations for STE
};

\node (psi) [finalbox, below=8mm of cb] {STE wavefunction \& lattice distortions};

\coordinate (busY) at ($(dft.south)+(0,-6mm)$);
\coordinate (busL) at (busY -| gw.north);
\coordinate (busR) at (busY -| dfpt.north);

\draw[link]  (dft.south) -- (busY);
\draw[link]  (busL) -- (busR);
\draw[arrow] (busL) -- (gw.north);
\draw[arrow] (busR) -- (dfpt.north);

\draw[arrow] (gw.south) -- (bse.north);

\coordinate (dfptLeftStart) at ($(dfpt.south)+(-12mm,0)$);
\draw[arrow] (dfptLeftStart) -- (gmn.north);

\coordinate (dfptRightStart) at ($(dfpt.south)+(12mm,0)$);
\coordinate (cbRight) at (dfptRightStart |- cb.north);
\draw[arrow] (dfptRightStart) -- (cbRight);

\draw[arrow] (bse.east) -- (gmn.west);

\coordinate (cbLeft) at (gmn.south |- cb.north);
\draw[arrow] (gmn.south) -- (cbLeft);

\coordinate (cbBse) at (bse.south |- cb.north);
\draw[arrow] (bse.south) -- (cbBse);

\draw[arrow] (cb.south) -- (psi.north);

\end{tikzpicture}

%% file: sections/tddft.tex
\subsection{State of the art}
\label{sec:SOTA_TDDFT}

\gls{tddft} is an extension of \gls{dft} for describing electronic excitations of systems under external time-dependent perturbations. Formally, it is an exact theory as proven by the Runge-Gross theorem \cite{runge1984density}, which establishes a one-to-one correspondence between the time-evolution of the complex many-body wavefunction and the time-dependent density. Compared to methods based on Green functions, \gls{tddft} is less well suited for materials with strong electron-hole interaction, but offers a superior balance between computational cost and accuracy~\cite{Ferre2015}, making it applicable to significantly larger and more complex systems. Due to this compelling advantage, \gls{tddft} has become one of the most widely used methods in quantum chemistry and materials science to study excitations in molecules, nanostructures, and bulk materials~\cite{cui2025spin,jacquemin2009extensive,adamo2013calculations,casida2012progress}. In practical calculations, \gls{tddft} can be applied either in the frequency domain or in the time domain, each of them offering advantages and limitations~\cite{ullrich2025snapshot,pela2021all,pemmaraju2020simulation,maitra2016perspective}. In the former, the time-dependent density is expressed directly in frequency space, typically within the \gls{lr} regime, where density variations are treated as a first-order response to an external perturbative potential. For instance, \gls{lr}-\gls{tddft} has been widely adopted to study optical spectra, electron-energy-loss spectra, low-lying excitation energies, and spin-flip transitions~\cite{cui2025spin, jacquemin2009extensive,maitra2016perspective,ullrich2014brief}. More recently, it has even been applied to describe x-ray Thomson scattering in warm dense matter~\cite{moldabekov2025applying}, magnetic excitations in ferromagnetic materials~\cite{skovhus2022influence}, self-trapped excitons in perovskites~\cite{xu2025polarity}, and excited-state conical-intersection topologies in molecular systems~\cite{Taylor2024}.

Although the initial success of \gls{tddft} can be mostly attributed to \gls{lr}-\gls{tddft}, recent advances in ultra-fast spectroscopy have triggered a crescent interest in the \acrfull{rt} formulation as a theoretical tool for modeling transient spectroscopies, where full dynamical information about the excitation process is required~\cite{tussupbayev2015comparison}. In \gls{rt}-\gls{tddft}, the electron density is propagated explicitly in time in a non-perturbative way, providing direct insight into photoinduced processes in various systems - from molecules to nanostructures. It has been successfully employed to simulate a broad range of phenomena, including non-linear optical processes~\cite{zhao2025hybrid}, high-harmonic generation~\cite{sato2025technical}, attosecond charge injection and photodissociation~\cite{tussupbayev2015comparison}, field-driven spin dynamics~\cite{cui2025spin}, strong-field and thermal ionization~\cite{fuks2015timeresolved}, pump-probe responses~\cite{sato2025technical, buades2021attosecond}, energy transfer mechanisms~\cite{zhao2025hybrid}, optically driven demagnetization~\cite{simoni2025spin}, and plasmonic resonances~\cite{caruso20262025}. Another important capability enabled by \gls{rt}-\gls{tddft} is the coupling of electron dynamics with nuclear motion, allowing for non-adiabatic \gls{md} that can track changes in atomic positions in response to electronic excitations, \eg laser pulses~\cite{fuks2015timeresolved,pela2022ehrenfest}. One recurrent approach is Ehrenfest dynamics, in which the nuclei evolve classically according to Newton's equations, while the forces acting on them are obtained from the time-dependent electronic density computed via \gls{rt}-\gls{tddft}. This strategy has been employed in numerous studies; for example, to model solvated DNA under proton irradiation~\cite{shepard2021simulating}, coherent phonon generation~\cite{shinohara2012nonadiabatic}, charge separation in photovoltaic materials~\cite{andrea2013quantum}, and excited carrier dynamics in carbon and BN nanostructures~\cite{krasheninnikov2007role,andermatt2016combining,long2011ab}. Furthermore, the \gls{md} description can be extended beyond the classical treatment of nuclei to incorporate quantum ionic effects within a multicomponent formalism~\cite{blum2024roadmap, lloydhughes20212021, maitra2016perspective}.

The accuracy of \gls{tddft}, and consequently its agreement with experiments, depends strongly on the choice of the \gls{xc} functional in \gls{rt}-\gls{tddft} or, correspondingly, of the \gls{xc} kernel in \gls{lr}-\gls{tddft}~\cite{ullrich2014brief, kuemmel2017chargetransfer, ullrich2025snapshot, maitra2016perspective}. While considerable effort has been devoted in recent years to improving \gls{xc} functionals and kernels~\cite{maitra2016perspective, rigamonti2015estimating, fuks2015timeresolved, authier2020dynamical, maitra2017charge, manna2018quantitative, refaelyabramson2015solidstate, mahler2022localized, pemmaraju2019valence}, the importance of numerical precision in \gls{tddft} calculations is often overlooked. Without stringent numerical control, it is generally impossible to disentangle errors due to approximations (\gls{xc} functional and kernel) from those introduced by the numerical implementation itself. In this context, like for \gls{dft} and $GW$ calculations, a full-potential all-electron implementation offers a unique advantage. Based on the \gls{lapwlo} basis, \exciting provides efficient \gls{lr}- and \gls{rt}-\gls{tddft} implementations, the latter including the capability of non-adiabatic \gls{md} via Ehrenfest dynamics.

\subsection{Methodology}
\label{sec:Methodology_TDDFT}

The time-evolution of the many-body wavefunction $\mwf(t)$ representing an $N$-electron system subjected to a time-dependent external potential $\vExt(\pos, t)$ is governed by the time-dependent Schrödinger equation. Given the initial state $\mwf(0)$, the Runge-Gross theorem guarantees that there is a unique map between $\mwf(t)$ and the time-dependent electron density $\nKS(\pos,t)$~\cite{runge1984density}. This correspondence allows for replacing the complex object $\mwf(t)$ by $\nKS(\pos,t)$ as the central variable, thereby making the study of non-equilibrium quantum dynamics much more tractable. Similar to static \gls{dft}, in \gls{tddft}, $\nKS(\pos, t)$ is obtained via a set of time-dependent \gls{ks} functions $|\kswf_{n \kk}\rangle$, whose evolution is governed by:
\begin{equation}\label{eq:rt_tddft_evolution}
\frac{\mathrm{d} }{\mathrm{d} t} |\kswf_{n \kk}(t)\rangle= -\im\hKS (t)\, |\kswf_{n \kk}(t)\rangle.
\end{equation}
The time-dependent \gls{ks} Hamiltonian is given by
\begin{equation}
\hKS (t) = \hKS (0) + \Delta \hKS (t) + \vExt(t),
\end{equation}
where $\Delta \hKS (t)$ accounts for the changes in the Hartree and \gls{xc} potentials:
\begin{equation}
\Delta \hKS (t) = 
\left[ \vH(t)-\vH(0) \right] + 
\left[ \vXC(t)-\vXC(0) \right].
\end{equation}
One practical way to obtain spectroscopic properties is to assume that $\vExt$ is small and to apply perturbation-theory within \gls{lr}-\gls{tddft}. One central component is the reducible polarizability, $\reducpola$, also referred to as the density-density response function, which characterizes the response of the  electron density due to the external potential:
\begin{equation}
\reducpola(\pos, \pos', t-t') = \frac{\delta \nKS(\pos,t)}{\delta \vExt(\pos',t')}.
\end{equation}
When transformed to the frequency domain and reciprocal space, $\reducpola$ is given by a Dyson equation
\begin{equation}\label{eq:lr_tddft_polarizability}
\reducpola(\qq, \freq) = \pola(\qq, \freq)+\pola(\qq, \freq)\left[\barcoul(\qq) + \xckernel(\qq, \freq)\right]\reducpola(\qq, \freq) ,
\end{equation}
where $\pola$ denotes the independent-particle polarizability, and $\xckernel$ is the \gls{xc} kernel: 
\begin{equation}
\xckernel(\pos, \pos', t-t') = \frac{\delta \vXC(\pos, t)}{\delta \nKS(\pos', t')}.
\end{equation}
$\xckernel$ measures the first-order response of $\vXC$ to electron-density fluctuations. Analogous to the \gls{xc} functional in static \gls{dft}, the choice of $\xckernel$ is the primary factor determining the accuracy of the calculated spectra: More sophisticated kernels representing higher levels of theory are generally expected to yield improved agreement with experiment~\cite{byun2020time}.

From the polarizability, the microscopic dielectric function $\diel$ is obtained as
\begin{equation}
\diel^{-1}(\qq, \freq) = 1 + \barcoul(\qq)\,\reducpola(\qq, \freq),
\end{equation}
the macroscopic dielectric function follows as
\begin{equation}
\dielmac(\qq, \freq) = 1 / \diel^{-1}_{00}(\qq, \freq).
\end{equation}
$\diel_{\mathrm{mac}}$ is a complex function whose imaginary part is proportional to the absorption spectrum~\cite{OnidaRMP2002}. Furthermore, it gives access to several related frequency-dependent quantities, often termed {\it optical constants}, such as the loss function, $\loss(\qq, \freq)$, and the optical conductivity, $\cond$:
\begin{equation}
\loss(\qq, \freq) = -\imag[\dielmac^{-1}(\qq, \freq)],
\end{equation}
\begin{equation}
\cond(\qq, \freq) = -\im\frac{\freq}{4\pi}[\dielmac(\qq, \freq)-1].
\end{equation}

A different strategy, as adopted in \gls{rt}-\gls{tddft}, is to remain in the time-domain and explicitly propagate the \gls{ks} wavefunctions according to \cref{eq:rt_tddft_evolution}.
This equation is usually solved using a time-evolution operator (propagator) $\timeevol$~\cite{gomez2018propagators}:
\begin{equation}
\timeevol(t+\Delta t, t) = \TimeOrdering\!
\left[ \exp\!\left(
-\im\!\int_t^{t+\Delta t}
\hKS(t')\,\mathrm{d}t'
\right)
\right] .
\end{equation}
The wavefunctions are then evolved in discrete time steps $\Delta t$ as:
\begin{equation}\label{eq:rt_tddft_propagator}
|\kswf_{n \kk}(t+\Delta t)\rangle= \timeevol(t+\Delta t, t) \,|\kswf_{n \kk}(t)\rangle.
\end{equation}
A common approximation in \gls{rt}-\gls{tddft} is the so-called adiabatic approximation, in which $\vXC(t)$ does not contain memory effects. Instead, it depends solely on the instantaneous KS density, and retains the same functional form as in ground-state \gls{dft}, with the ground-state density simply being replaced by its time-dependent counterpart.

While \gls{rt}-\gls{tddft} is usually employed to predict the real-time evolution of a system under a time-dependent external potential, such as a laser pulse, it can be also used to determine frequency-dependent optical constants. By adopting the velocity-gauge, the interaction with the external field $\vExt$ is expressed as:
\begin{equation}
    \vExt(t) = \vExt(0) + \frac{\vecpot(t)^2}{2\speedOfLight^2}-\frac{\im}{\speedOfLight}\vecpot(t)\cdot\nabla.
\end{equation}
In this gauge, the vector potential is $\vecpot(t) = -\speedOfLight\int_0^t \efield(t')\,{\rm d}t'$, where $\efield$ stands for the applied electric field, assumed to be spatially homogeneous within the unit cell. To obtain the optical response functions, the macroscopic current density $\currentDensity(t)$ is evaluated as
\begin{equation}
\currentDensity(t) = \sum_{n\kk} \kptweight{\kk}\,\occ{n}{\kk}\,\langle \kswf_{n\kk}(t)| \shup{\im} \nabla| \kswf_{n\kk}(t)\rangle - \frac{N\vecpot(t)}{\speedOfLight \unitCellVol},
\end{equation}
with $N$ being the number of valence electrons per unit cell. Upon Fourier transformation to the frequency domain, the $i,j$ tensor-components of $\diel_{\mathrm{mac}}$ are given by
\begin{equation}
\dielmac^{ij}(\freq) = \delta_{ij} + \frac{4\pi\im}{\freq}\frac{J_i(\freq)}{E_j(\freq)}
\end{equation}
with $i,j \in {x, y, z}$ denoting Cartesian directions.

\subsection{Linear-response TDDFT in \excitingb}
\label{sec:New_LR_TDDFT}

In \exciting, the matrix elements of the independent-particle polarizability are evaluated as
\begin{equation}
\pola_{\GG \GG'}(\qq,\omega) = \frac{1}{N_\kk \unitCellVol}\sum_{nn'\kk}\frac{\occ{n}{\kk}-\occ{n'}{\kk+\qq}}{\eigval_{n\kk} -\eigval_{n'\kk+\qq}+\freq+\im \eta}\,
M_{nn'\kk}(\GG,\qq)\,
M^*_{nn'\kk}(\GG',\qq)\,,
\end{equation}
where the $\qq$-dependent matrix elements $M_{nn'\kk}(\GG,\qq)$ are defined in \cref{eq:plane_wave_matrix_elements}. $\pola_{\GG \GG'}(\qq,\omega)$ is then inserted into the Dyson equation (\cref{eq:lr_tddft_polarizability}) to obtain the reducible polarizability $\reducpola_{\GG \GG'}(\qq,\omega)$. This formulation includes \glspl{lfe}, which are accounted for by the microscopic $\GG\neq 0$ components of the induced response. In \cref{eq:lr_tddft_polarizability}, the exchange-correlation kernel $\xckernel$ must be specified. In \exciting, the following kernels are implemented:
\begin{itemize}
    \item \gls{rpa}, defined by $\xckernel = 0$;
    \item Adiabatic \gls{lda}, which is derived from the \gls{lda} \gls{xc} potential;
    \item Long-range corrected kernels, including both a static form, $\xckernel = \alpha/\qq^2$~\cite{reining2002excitonic}, and a dynamic version, $\xckernel = (\alpha + \beta\freq^2)/\qq^2$~\cite{botti2005energy};
    \item Family of bootstrap kernels~\cite{rigamonti2015estimating, sharma2011bootstrap};
    \item BSE-derived many-body kernel~\cite{marini2003bound,reining2002excitonic}.
\end{itemize}
In addition, the \gls{ipa} is also available. This corresponds to setting $\reducpola(\qq, \freq) = \reducpola_0^\pI(\qq, \freq)$ in \cref{eq:lr_tddft_polarizability}. In the long-wavelength limit $\qq\rightarrow0$, the matrix elements $M_{nn'\kk}$ are expressed in terms of the momentum matrix elements $\PP_{nn'\kk}$, which allows for computing the full macroscopic dielectric tensor $\dielmac^{ij}(\freq)$. Optionally, the anomalous Hall contribution can be included in the tensor, which enables computing \gls{moke} spectra~\cite{Fan2017}. More information on the implementation of \gls{lr}-\gls{tddft} in \exciting can be found in Refs.~\cite{phdThesisSagmeister,sagmeister2009time,gulans2014exciting}.

\subsection{Real-time \acrshort*{tddft} in \excitingb}
\label{sec:New_RT_TDDFT}

To solve \cref{eq:rt_tddft_propagator}, several propagators are implemented in \exciting, namely the simple exponential, the exponential at midpoint, the approximate enforced time-reversal symmetry scheme, the commutator-free Magnus expansion of fourth order, the exponential using a basis of the Hamiltonian eigenvectors, and the classical fourth-order Runge-Kutta method~\cite{pela2021all}. Using one of these propagators, \cref{eq:rt_tddft_propagator} is solved by expanding each \gls{ks} state in a chosen basis set $|\basis_{n' \kk}\rangle$, with time-dependent coefficients $c_{nn' \kk}(t)$:
\begin{equation}\label{eq:rt_tddft_basis}
|\kswf_{n \kk}(t)\rangle = \sum_{n'} c_{nn' \kk}(t)\, |\basis_{n' \kk}\rangle.
\end{equation}
Two basis sets are implemented:
\begin{itemize}
    \item The default \gls{lapwlo} basis functions, known for achieving the ultimate precision in describing \gls{ks} states~\cite{pela2021all}; 
    \item The \gls{gs} \gls{ks} wavefunctions $|\kswf_{n \kk}(0)\rangle$, which typically offer a more computationally efficient alternative~\cite{pela2024speeding}. In this case, the basis set includes excited states, whose number must be carefully converged.
\end{itemize}

\begin{figure}[ht]
    \centering
    \includegraphics[width=0.8\textwidth]{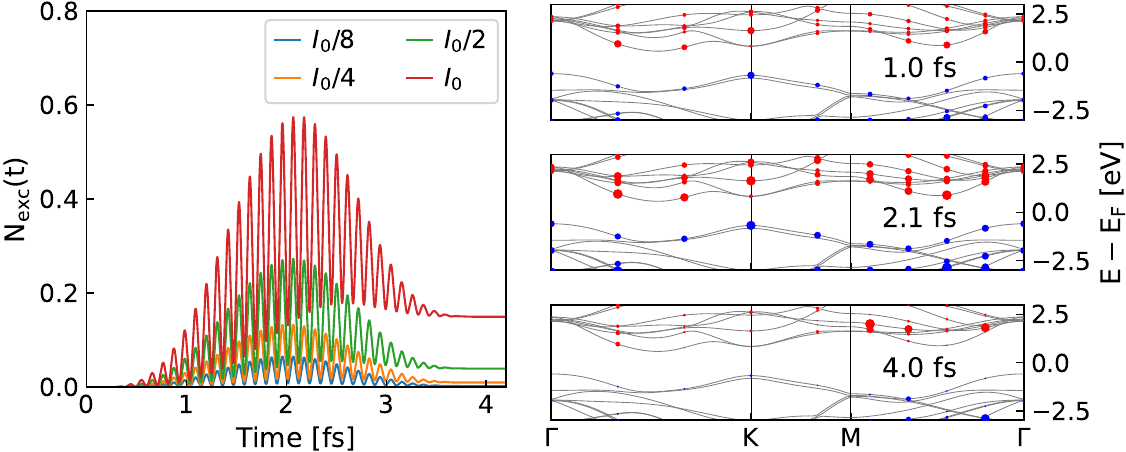}
    \caption{Left: Number of excited electrons per unit cell in WSe$_2$ for various laser intensities ($I_0 = 1.5$ TW/cm$^2$). Right: Bandstructure of WSe$_2$, showing the populations of excited electrons (red) and holes (blue) in the conduction and valence bands, respectively, at times 1.0 (top), 2.1 (middle), and 4.0 fs (bottom).}
    \label{fig:rt-tddft-n_exc_WSe2}
\end{figure}

To monitor the effects of the external electric field, \exciting can output the time-dependent fluctuations in charge density, $\nKS(\pos, t)-\nKS(\pos, 0)$. It can provide insight into how electrons are driven by the laser pulse and how chemical bonds are affected during the excitation process. Furthermore, by projecting the \gls{tdks} states $|\kswf_{n\kk}(t)\rangle$ onto the reference ground-state wavefunctions, \exciting provides the number of excited electrons and holes per unit cell together with their distribution in $\kk$-space~\cite{pela2021all}:
\begin{equation}
\Delta \occ{n}{\kk}(t) = \sum_{n'} \occ{n'}{\kk}(0)\, \left | \langle \kswf_{n\kk}(0)|\kswf_{n'\kk}(t)\rangle \right|^2 - \occ{n}{\kk}(0),
\label{eq:fs_occupation}
\end{equation}
where a positive (negative) value corresponds to the presence of an excited electron (hole). To exemplify this feature, \cref{fig:rt-tddft-n_exc_WSe2} illustrates the number of excited electrons per unit cell in WSe$_2$ subjected to a laser pulse of frequency 1.9~eV and duration 4.0~fs. The intensity of the applied field is varied from $I_0 = 1.5$ TW/cm$^2$ down to $I_0/8$. \cref{fig:rt-tddft-n_exc_WSe2} also shows the distribution of excitations in $\kk$-space at times 1.0 (top), 2.1 (middle), and 4.0~fs (bottom).

For users more interested in methodological details, besides the adiabatic approximation, two other models of $\Delta \hKS(t)$ are implemented: 
\begin{itemize}
    \item $\vXC(t)$ is kept fixed at its ground-state value $\vXC(0)$. In the linear regime, this corresponds to the \gls{rpa}.
    \item Both $\vH(t)$ and $\vXC(t)$ are frozen to their initial values, so that the vector potential $\vecpot(t)$ is the only driver of the time evolution. This represents the \gls{ipa}.
\end{itemize}
Strictly speaking, these latter schemes lie outside the formal \gls{tddft} framework and should instead be regarded as \gls{tdks} approaches. Nevertheless, they can be useful to disentangle the individual contributions of the \gls{xc} and Hartree potentials to the excitation dynamics.

\subsection{Femtosecond dynamics for ultrafast pump-probe spectroscopy}
\label{sec:New_fsDynamics_TDDFT}

\gls{rt}-\gls{tddft} can also provide a first-principles description of the pump step in ultrafast pump-probe experiments. The pump pulse is modeled through a time-dependent vector potential whose parameters, such as duration, frequency, polarization, and fluence, are chosen to match experimental conditions. In the presence of this external driving field, the \gls{ks} states are explicitly propagated in real time according to \cref{eq:rt_tddft_evolution}, yielding the non-equilibrium electronic state induced by the pump. The resulting non-equilibrium electronic state is monitored through projections onto the \gls{gs} \gls{ks} manifold. In practice, we construct the $\kk$- and band-resolved occupations, according to \cref{eq:fs_occupation}, and the photoinduced changes $\Delta\occ{n}{\kk}(t)=\occ{n}{\kk}(t)-\occ{n}{\kk}(0)$. These femtosecond occupations retain the transient redistribution in energy and $\kk$-space induced by the pump pulse. The occupation numbers $\occ{n}{\kk}(t)$ evaluated at selected pump-probe delays are then provided as input to the non-equilibrium \gls{bse} implementation (see \cref{sec:New_PPS2_PP}).

\subsection{Non-adiabatic molecular dynamics}
\label{sec:New_NAMD_TDDFT}

The \gls{bo} approximation is one of the most widely adopted approaches in \textit{ab initio} \gls{md}. Under this approximation, the nuclei are considered point charges moving on a potential energy surface generated by the electrons and other nuclei. The electrons, in turn, are treated quantum mechanically; they respond adiabatically to the nuclear motion, remaining in their eigenstate (typically the \gls{gs}) at each time step~\cite{pela2022ehrenfest, scherrer2017mass, curchod2013trajectorybased, crespootero2018recent, you2021firstprinciples, nelson2020non, tully1990molecular}. Despite its success, the \gls{bo} approximation is inadequate for certain scenarios in modeling the interaction between electromagnetic waves and matter: Processes in which electrons are excited by ultra-fast time-dependent fields and subsequently dissipate energy through coupling to lattice vibrations are inherently non-adiabatic and thus require a treatment beyond the \gls{bo} approximation~\cite{pela2022ehrenfest,crespootero2018recent}. Prominent examples of such phenomena, which are of growing interest, include pump–probe experiments~\cite{caruso20262025} (see \cref{sec:PP}). An efficient approach to non-adiabatic \gls{md} is the Ehrenfest method, which combines a classical treatment of the nuclei with a non-adiabatic evolution of the electronic wavefunctions, usually within the framework of \gls{rt}-\gls{tddft}~\cite{ojanpera2012nonadiabatic, pela2022ehrenfest, wang2011timedependent}. 

Ehrenfest dynamics has been implemented in \exciting, enabling the efficient simulation of complex non-adiabatic processes. Within the velocity-gauge, as assumed in this implementation, the action integral for the total system of nuclei and electrons is~\cite{pela2022ehrenfest}
\begin{equation}
\int_{t_1}^{t_2}\!\mathrm{d}t\left[
\sum_{\atomidx}\frac{M_\atomidx\,\dot{\atomincell}^2_\atomidx}{2} + \frac{Z_\atomidx}{\speedOfLight}\dot{\atomincell}_\atomidx\cdot\vecpot(t)
\right]
+
\int_{t_1}^{t_2}\!\mathrm{d}t
\sum_{n\kk}\occ{n}{\kk}\,\kptweight{\kk }
\left\langle
\kswf_{n\kk}(t)
\left|
\shup{\im}\frac{\partial}{\partial t} - \hKS
\right|
\kswf_{n\kk}(t)
\right\rangle.
\end{equation}
%
It comprises two terms. The first one describes the nuclear motion and its coupling with the external electric field, whereas the second term represents the electronic contributions. Extremizing the action integral with respect to $|\kswf_{n\kk}\rangle$ leads to  \cref{eq:rt_tddft_evolution} as the electronic equation of motion. Since the \gls{lapwlo} basis depends on the atomic positions, $|\basis_{n}\rangle$ in \cref{eq:rt_tddft_basis} change over time following the nuclear motion. Taking this into account gives:
\begin{equation}
\sum_{n'} \langle \phi_{n\kk}|\phi_{n'\kk} \rangle\,\dot{c}_{n'n''}(t) = -\sum_{n'}\left[\im \langle \phi_{n\kk}|\hKS|\phi_{n'\kk} \rangle(t) 
+
\sum_\alpha
\dot{\atomincell}_\alpha 
\cdot
\left\langle \phi_{n\kk}\middle|\frac{\partial \phi_{n'\kk}}{\partial \atomincell_\alpha} \right\rangle 
\right] c_{n'n''}(t). 
\end{equation}
The corresponding equation for the motion of the nuclei is:
\begin{equation}
M_\atomidx\,\ddot{\atomincell}_\atomidx(t) = -\frac{Z_\atomidx}{\speedOfLight}\frac{{\rm d}\vecpot}{{\rm d}t}(t) + \force_{{\ts\mathrm{HF}},\atomidx}(\atomincell_\atomidx,t) + \force_{\mathrm{corr},\atomidx}(\atomincell_\atomidx,t).
\end{equation}
The first term on the right-hand side denotes the external force due to the interaction with the applied electric field; $\force_{{\ts\mathrm{HF}},\atomidx}$ is the Hellmann-Feynman force; and $\force_{\mathrm{corr},\atomidx}$ represents the so-called Pulay corrections. Details on the implementation can be found in Ref.~\cite{pela2022ehrenfest}.

\subsection{Effective Schr\"{o}dinger equation with dynamical Berry-phase field coupling}
\label{sec:New_SE_TDDFT}

While the velocity gauge can be employed in a wide range of problems, it has two major drawbacks. First, at low frequencies of the external field, velocity-gauge calculations may exhibit non-physical divergences due to limitations in numerical accuracy~\cite{aversa_prb_1995, virk_prb_2007}. Another disadvantage is related to the introduction of nonlocal potentials in the time-dependent Hamiltonian, such as scissor-correction operators or nonlocal self-energies. In this case, each operator must be gauge-transformed, resulting in more complicated expressions that can hinder numerical implementations~\cite{bertsch_prb_2000, pemmaraju2018velocity}. Additionally, the velocity operator used to calculate the density current must be modified to incorporate all nonlocal contributions to the Hamiltonian~\cite{delsole_prb_1993, sangalli_prb_2017}. These difficulties are absent when the length gauge is employed. However, a straightforward use of the standard field-coupling term $\efield(t) \cdot \pos$ violates the \gls{pbc} usually employed in calculations with extended systems. An effective Schr\"{o}dinger equation approach that employs the length gauge in a \gls{pbc}-compatible manner via the dynamical Berry phase, was developed in Refs.~\cite{souza_prb_2004, attaccalite_prb_2013} and implemented in the YAMBO package~\cite{ attaccalite_prb_2013}.

With the \gls{rt}-\gls{tddft} module of \exciting, it is possible to use the dynamical Berry-phase approach to numerically integrate the following equations of motion for the spatially periodic parts of the valence \gls{ks} wavefunctions $\ket{v_{m \kk}(t)} = \ee^{- \im \kk \cdot \pos} \ket{\kswf_{m \kk}(t)}$:
\begin{equation}\label{eq:dynamics}
     \im \frac{\drm}{\drm t}  \ket{v_{m \kk}(t)}  = \left\{ \ee^{- \im \kk \cdot \pos'} \left[\hKS (0) + \Delta \hat{h}_{\rm QP} + \Delta \hKS (t) \right] \ee^{ \im \kk \cdot \pos} + \im \efield(t) \cdot \tilde{\partial}_{\kk} \right\}\ket{v_{m \kk}(t)},
\end{equation}
where $\pos'$ and $\pos$ correspond to the coordinates of the bra- and ket-states respectively, $\tilde{\partial}_{\kk}$ denotes gauge-invariant $\kk$-derivative, and $\Delta \hat{h}_{\rm QP}$ is a scissor-correction operator~\cite{baraff_prb_1984, OnidaRMP2002}, which rigidly shifts the \gls{cb} by $\Delta \eigval_{\rm QP}$ to mimic the \gls{qp} gap:
\begin{equation}\label{eq:scissor}
    \Delta \hat{h}_{\rm QP} = \Delta \eigval_{\rm QP} \sum_{m \in {\rm CB}} \ket{\kswf_{m \kk}} \!\bra{\kswf_{m \kk}}.
\end{equation}
We note that \cref{eq:dynamics} lies beyond the scope of standard \gls{dft}, which implies a one-to-one mapping of the electron density and the one-particle \gls{ks} potential. More details on the numerical implementation of the propagation of \cref{eq:dynamics} in \exciting will be published elsewhere~\cite{tumakov_2026}.

After the propagation, the time-dependent macroscopic polarization can be obtained using the modern theory of polarization~\cite{king_prb_1993, resta_revmodphys_1994}:
\begin{equation}\label{eq:polarization}
    \polarization_{\alpha} = \frac{ \mathbf{a}_{\alpha}}{\pi \unitCellVol N_{\kk_{\alpha}^{\perp}}} \sum_{\kk_{\alpha}^{\perp}} \,\imag\!\left[ \ln\prod_{i = 1}^{N_{\kk_{\alpha}} - 1} {\rm det} \Braket{v_{m \kk} | v_{n \kk_{\alpha}^{+}}} \right],
\end{equation}
where $N_{\kk_{\alpha}}$ and $N_{\kk_{\alpha}^{\perp}}$ denote the number of $\kk$ points along cell direction $\alpha$ and orthogonal to it, respectively. The polarization can be used to calculate optical response functions, such as the macroscopic dielectric function:
\begin{equation}\label{eq:diel}
    \dielmac^{ij}(\freq)= \delta_{ij} + 4 \pi P_{i}(\freq) / E_{j}(\freq)\:.
\end{equation}

\begin{figure}[ht]
    \centering
    \includegraphics[scale=0.9]{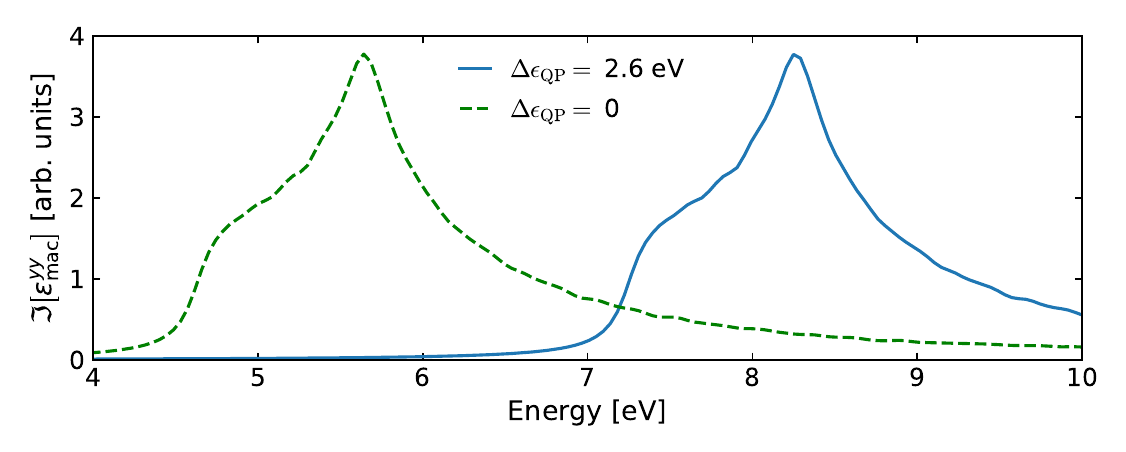}
    \vspace*{-2mm}
    \caption{Imaginary part of the dielectric function of an h-BN monolayer calculated from the time-dependent macroscopic polarization obtained as a response to a delta-shaped pulse via propagating \cref{eq:dynamics}. Solid and dashed lines correspond to a scissors shift of 2.6 eV and 0 eV used in \cref{eq:scissor}, respectively. A Lorentzian broadening of 0.1~eV was applied in the Fourier transform of the postprocessing-step.}
    \label{fig:rt-bp-hbn}
\end{figure}

As an illustration, we calculate the linear-response absorption spectrum of an h-BN monolayer by applying an external field in the form of a Dirac delta function $\efield(t) = \delta(t = 0)\, \mathbf{e}_y$, which is flat in the frequency domain. A scissor shift of $\eigval_{\rm QP} =$ 2.6~eV is used in the calculations to mimic the h-BN \gls{qp} gap of 7.25~eV~\cite{paleari_2dmats_2018}. \cref{fig:rt-bp-hbn} shows the imaginary part of the corresponding component of the dielectric function, $\diel_{\mathrm{mac}}^{yy}(\freq)$. As one can see, simply including the scissor operator in the dynamical equation within the length gauge leads to the desired shift in the linear-response spectrum.

%% file: sections/pumpprobe.tex
\subsection{State of the art}
\label{sec:SOTA_PP}

Pump-probe spectroscopy is a state-of-the-art time-resolved technique for investigating light-matter interaction, particularly non-equilibrium processes like exciton dynamics~\cite{ashoka2022}. Advances in laser technology enable this technique to access a broad energy window, from the infrared and visible \cite{montanaro2020} to the extreme ultraviolet~\cite{rebholz2021} and X-ray domains~\cite{geneaux2019}. In pump-probe spectroscopy experiments, transient features have been widely reported and associated with photoinduced carriers and excitonic effects~\cite{garratt2021ultrafast,garratt2022direct,hillyard2009atomic,palmieri2020mahan,ZuerchNatComm17,GarrattStructuralDynamics24}. Interpreting these experimentally observed transient signals requires theoretical frameworks that can accurately describe the excited-state electronic structure and its evolution under sequential light–matter interactions.

Several theoretical approaches have been developed to simulate pump-probe experiments, including Bloch-equation-based formalisms \cite{picon2019attosecond, CistaroJChThComp23} and model Hamiltonian approaches \cite{malakhov2023exciton, hansen2023theoretical}, which do not provide a truly predictive, first-principles description of real materials. As an alternative, \gls{tddft} has become a popular \textit{ab initio} approach for simulating the pump-probe process \cite{garratt2022direct, ZuerchNatComm17}. However, this mean-field formulation does not explicitly capture many-body interactions. More accurate correlated wavefunction techniques, including a time-dependent configuration-interaction-singles approach \cite{pabst2012} and \gls{cc} theory \cite{SkeidsvollPRA20}, can accurately treat the electron correlation, but the high computational cost limits their application to atoms or molecules. 

Many-body approaches based on the \gls{bse} have proven highly successful for describing both optical absorption and \gls{xas} in solids, capturing electron-hole interactions essential for valence and core excitations (see \cref{sec:BSE}). In standard applications, \gls{bse} calculations treat either optical excitations or core-level excitations starting from the ground state, \ie in a {\it static} manner. Extensions to non-equilibrium implementations remain limited. While \gls{bse}-based methods have been applied to describe optical-pump / optical-probe scenarios involving valence excitons \cite{sangalli2023exciton}, studies of X-ray or extreme-ultraviolet spectra of optically excited materials typically employ approximate treatments of valence excitations and do not adopt a unified many-body description of valence and core excitations \cite{vinson2022advances,KleinJPC23,KleinJACS22}. Overcoming such shortcomings, a fully consistent many-body description of X-ray absorption from excitonic initial states has only recently been achieved by treating both optical and core-excited states within a unified theoretical framework \cite{Farahani2024PRB}. To capture the time evolution of the carrier distribution induced by pump excitation, a theoretical approach based on the non-equilibrium \gls{bse} formalism has recently been developed \cite{Rossi2025,Qiao2025}.

\subsection{Optical pump / X-ray probe spectroscopy}
\label{sec:New_PPS1_PP}

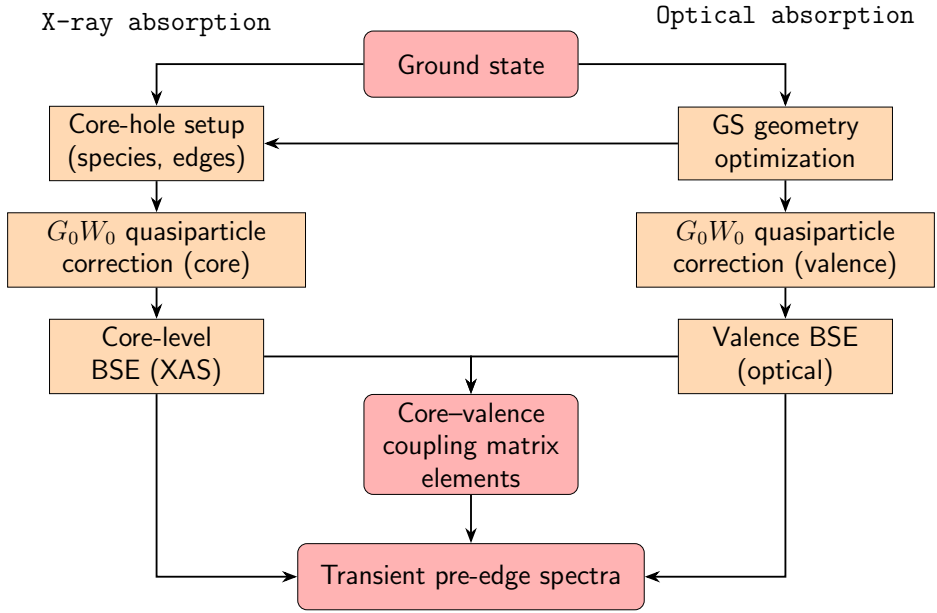
\begin{figure}[h]
    \centering
    \input{figures/tools/workflow_Pump-Probe}
    \vspace*{-8mm}
    \caption{Workflow for simulating optical pump / X-ray probe spectra using the \prexas package based on \exciting data.}
    \label{fig:workflow_Pump-Probe}
\end{figure}
    
A framework based on the \gls{bse} has recently been developed by some of us to describe optical-pump / X-ray-probe experiments in solids~\cite{Farahani2024PRB}. It extends the \gls{bse} formalism to pump–probe scenarios, in which an optical pulse first generates a valence exciton and a subsequent X-ray pulse probes the resulting transient electronic structure by inducing a transition from a core state to vacant states below the Fermi level. This approach enables valence-excited states to be treated as initial states and core-excited states as final states within the same formalism. The key novelty here is the consistent inclusion of electron–hole interactions for both valence and core excitations. This enables a direct connection to be made between transient pre-edge features in X-ray absorption spectra and the microscopic structure of valence excitons. This framework, sketched in Fig. \ref{fig:workflow_Pump-Probe} provides a rigorous basis for modeling and interpreting pump–probe X-ray experiments on photoexcited materials. It is based on second-order time-dependent perturbation theory combined with \gls{bse} calculations. The  implementation in the \prexas package is available on GitHub \cite{git-PreXAS-Exciton}. It provides a first-principles route for modeling polarization- and element-resolved X-ray signatures of excitonic states in solids. 

\prexas requires outputs from two independent \exciting \gls{bse} calculations. One run describes optical valence-to-conduction excitations, while the other one yields core-to-conduction excitations. The code then evaluates transition weights connecting these excitonic states. The execution of \prexas requires only a few input parameters: the excitation energies $\omega_o$ and $\omega_x$ of the optical and X-ray pulses, the numbers $N_{\lambda_v}$ and $N_{\lambda_c}$ of considered valence–conduction and core–conduction excitations, the lifetime broadenings $\eta_I$ and $\eta_F$ of the intermediate and final states, and the polarizations of the pump and probe pulses.

\begin{figure}[th]
    \centering
    \includegraphics[width=0.9\linewidth]{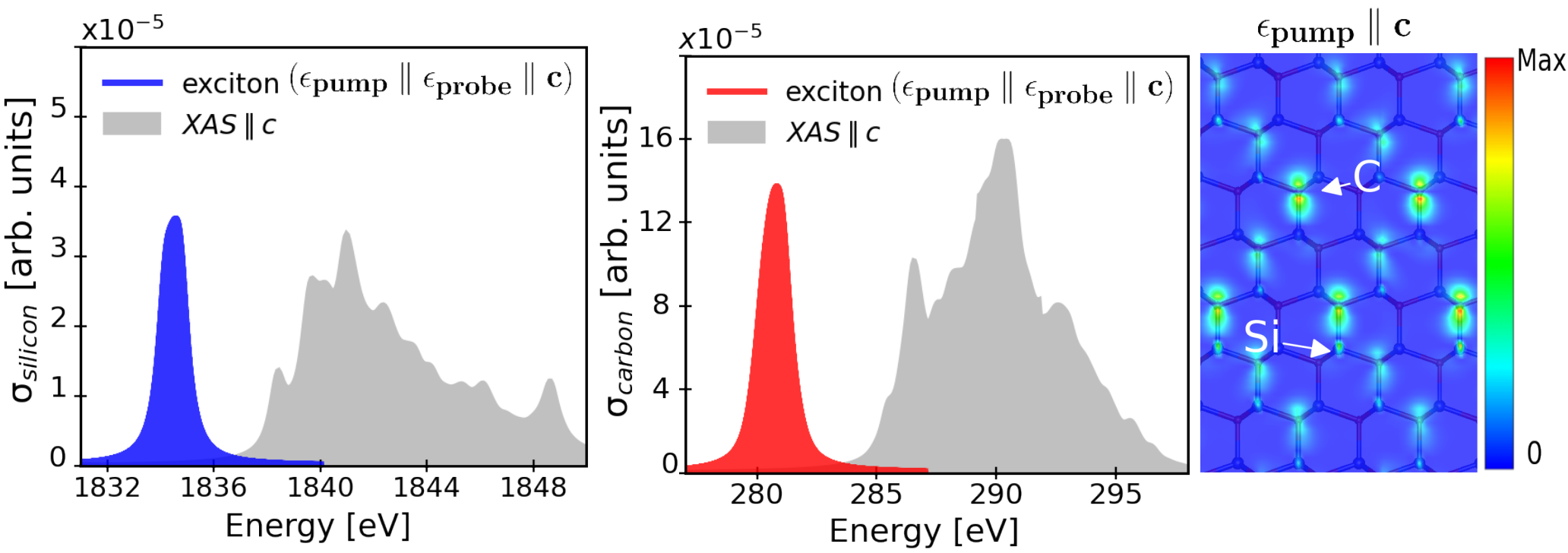}
    \caption{Silicon K-edge \gls{xas} (left) and carbon K-edge spectra (middle panel) in 4H-SiC in terms of the absorption cross section $\sigma$. The blue and red peaks are features that have emerged in the \gls{xas} pre-edge region due to optically excited states for the polarizations of the pump and probe pulses parallel to the crystallographic $\mathbf c$ direction. These peaks are magnified for better visibility. The \gls{gs} \gls{xas} spectra are shown in gray. Right: 2D cuts of the averaged hole distribution of the excitonic states after the action of a pump pulse polarized parallel to $\mathbf c$. Figure adapted from Ref.~\cite{Farahani2024PRB} (copyright American Physical Society).}
    \label{fig:P&P.SiC.png}
\end{figure}

As an example, we show calculations for 4H-SiC in Fig.~\ref{fig:P&P.SiC.png}. We analyze the hole distribution (right panel) obtained for pump-pulse polarizations along the crystallographic $\mathbf c$ direction, together with the corresponding polarization-resolved pre-edge X-ray absorption signals (left and middle panels). The hole distribution cut exhibits predominantly carbon and silicon $p$-type characters oriented along the $\mathbf c$ axis. The associated pre-edge spectra reveals a strong peak caused by an enhanced transition probability from the core level into these valence states compared to the perpendicular direction (not shown here). This behavior reflects the sensitivity of core-level transitions to orbital symmetry and demonstrates a direct correspondence between the spatial orientation of the excitonic hole distribution and the intensity of the pre-edge feature in the X-ray absorption spectrum. For details, we refer to Ref.~\cite{Farahani2024PRB}.

\subsection{Pump-probe spectroscopy based on non-equilibrium BSE}
\label{sec:New_PPS2_PP}
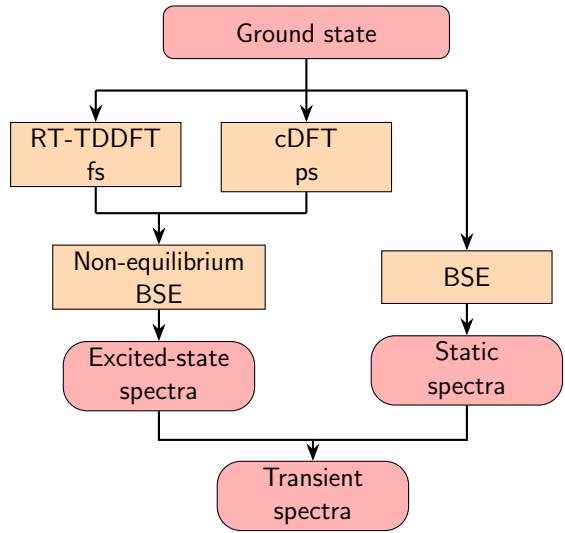
\begin{figure}[H]
    \centering
    \input{figures/tools/workflow_pump-probe_noneq-BSE}
    \caption{Workflow for simulating pump-probe spectra using non-equilibrium BSE.}
    \label{fig:workflow_pump-probe_noneqBSE}
\end{figure}

Some of us have developed a computational approach based on the newly implemented non-equilibrium \gls{bse} (see~\cref{sec:New_NonEq_BSE}), as shown in~\cref{fig:workflow_pump-probe_noneqBSE}. It makes use of \gls{rt}-\gls{tddft} and \gls{cdft}, respectively, to model the pump-induced dynamics across different timescales. At femtosecond time delays, carrier populations are extracted from \gls{rt}-\gls{tddft} simulations (see~\cref{sec:New_fsDynamics_TDDFT}), where Kohn-Sham orbitals evolve under a time-dependent vector potential representing the optical pump. At longer, \ie picosecond, timescales, thermalized carrier occupations are obtained according to the Fermi-Dirac distribution using \gls{cdft} (see~\cref{sec:New_cDFT_GS}). The resulting photoexcited carrier populations are incorporated into the \gls{bse} Hamiltonian, modifying both the transition matrix elements and the screened Coulomb interaction. These modifications account for two essential many-body effects that govern transient optical responses, namely Pauli blocking and Coulomb screening. The \gls{ta} spectrum is given as the difference between the excited-state and static spectra. 

\begin{figure}[H]
    \centering
    \includegraphics[width=0.75\linewidth]{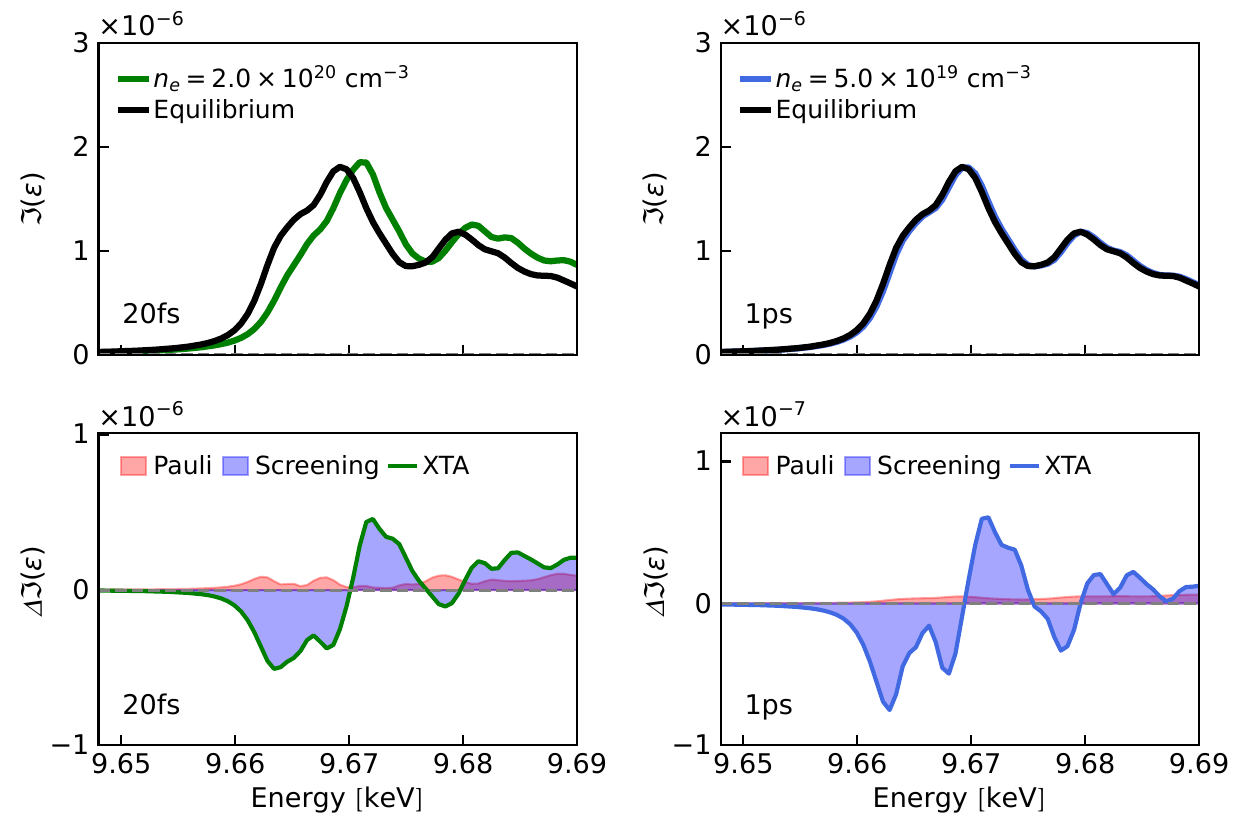}
    \vspace*{-2mm}
    \caption{ Top panels: Static (black) and excited-state \gls{xas} spectra (green/blue) at different carrier densities,  $n_{e}$, on the femtosecond (left) and picosecond (right) timescales for the Zn K-edge of ZnO. Bottom panels: Corresponding \gls{ta} spectra and their decomposition into Coulomb screening (blue) and Pauli blocking (red) effects.}
   \label{fig:ZnO}
\end{figure}

This approach has been successfully applied to a range of semiconductors, including WSe$_2$, CsPbBr$_3$, and ZnO~\cite{Rossi2025,Qiao2025}. In~\cref{fig:ZnO}, we show the pump-induced absorption spectra for the K-edge of Zn in ZnO and its evolution across different timescales. By disentangling the contributions from Pauli blocking and photoinduced Coulomb screening, we identify the latter as the dominant mechanism governing the spectral response~\cite{Rossi2025}. Overall, this framework provides a quantitative method for simulating and interpreting pump-probe experiments. It can capture transient excitations on different time scales, from femtoseconds to picoseconds, and across a broad spectral range, from the visible and near-infrared to the extreme ultraviolet and X-ray regimes.

%% file: figures/tools/workflow_Pump-Probe.tex
\begin{tikzpicture}[node distance=.75cm, every node/.append style={font=\sffamily\large}] 
\centering
\scalebox{0.88}{
\tikzstyle{startstop} = [rectangle, rounded corners, minimum width=2cm, minimum height=1cm, text centered, text width=3cm, draw=black, fill=red!30]
\tikzstyle{process} = [rectangle, minimum width=2cm, minimum height=1cm, text width=3cm, text centered, draw=black, fill=orange!30]
\tikzstyle{processG} = [rectangle, minimum width=4.5cm, minimum height=1cm, text width=4cm, text centered, draw=black, fill=orange!30]
\tikzstyle{decision} = [diamond, minimum width=1.cm, maximum height=1cm, text width=3cm, text centered, draw=black, fill=green!30]
\tikzstyle{arrow} = [thick, ->, >=Stealth]

\centering
\node (DFT) [startstop] {Ground state};
\node (species) [process, left of=DFT, xshift=-4cm, yshift=-1.2cm] {Core-hole setup (species, edges)};
\node (opt) [process, right of=DFT, xshift=4cm, yshift=-1.2cm] {\gls{gs} geometry optimization};
\node (G0c) [processG, below of=species, yshift=-.86cm] {$G_0W_0$ quasiparticle correction (core)};
\node (G0v) [processG, below of=opt, yshift=-.86cm] {$G_0W_0$ quasiparticle correction (valence)};
\node (ex-c) [process, below of=G0c, yshift=-.86cm] {Core-level BSE (\gls{xas})};
\node (ex-v) [process, below of=G0v, yshift=-.86cm] {Valence BSE (optical)};
\node (momentum) [startstop, below of=DFT, yshift=-5.cm] {Core–valence coupling matrix elements};
\node (package) [startstop, below of=DFT, yshift=-7.cm, text width=5cm] {Transient pre-edge spectra};

\draw [arrow] (DFT) -| node[anchor=center, yshift=0.7cm]{\texttt{Optical absorption}}(opt);
\draw [arrow] (DFT) -| node[anchor=center, yshift=0.6cm]{\texttt{X-ray absorption}} (species);
\draw [arrow] (opt) -- (G0v);
\draw [arrow] (species) -- (G0c);
\draw [arrow] (G0v) -- (ex-v);
\draw [arrow] (G0c) -- (ex-c);
\draw [arrow] (opt) -- (species);
\draw [arrow] (ex-v) |- (package);
\draw [arrow] (ex-c) |- (package);
\draw [arrow] (ex-v) -| (momentum);
\draw [arrow] (ex-c) -| (momentum);
\draw [arrow] (momentum) -- (package);
}
\end{tikzpicture}

%% file: figures/tools/workflow_pump-probe_noneq-BSE.tex
\begin{tikzpicture}[
  scale=0.7, transform shape,
  every node/.append style={font=\sffamily\Large, align=center},
  arrow/.style={thick,->,>=Stealth},
  link/.style={thick},
  startstop/.style={rectangle, rounded corners=4pt, minimum height=10mm, text width=5.2cm, draw=black, fill=red!30},
  process/.style={rectangle, minimum height=10mm, text width=3.0cm, draw=black, fill=orange!30},
  bigprocess/.style={rectangle, minimum height=10mm, text width=3.8cm, draw=black, fill=orange!30},
  spectra/.style={rectangle, rounded corners=9pt, minimum height=10mm, text width=3.4cm, draw=black, fill=red!30}
]

\node (gs) [startstop] {Ground state};

\node (rttddft) [process, below=12mm of gs, xshift=-40mm] {RT-TDDFT\\fs};
\node (cdft)    [process, below=12mm of gs]              {cDFT\\ps};

\node (neqbse) [bigprocess, below=10mm of cdft, xshift=-28mm] {Non-equilibrium\\BSE};
\node (bse)    [process,    right=22mm of neqbse]             {BSE};

\node (exspec) [spectra, below=6mm of neqbse] {Excited-state\\spectra};
\node (gsspec) [spectra, below=6mm of bse]    {Static\\spectra};

\coordinate (midSpec) at ($(exspec.south)!0.5!(gsspec.south)$);
\node (trans) [spectra, below=10mm of midSpec] {Transient\\spectra};

\coordinate (busY)  at ($(gs.south)+(0,-6mm)$);
\coordinate (busL)  at (busY -| rttddft.north);
\coordinate (busC)  at (busY -| cdft.north);
\coordinate (busR)  at (busY -| bse.north);

\draw[link]  (gs.south) -- (busC);
\draw[link]  (busL) -- (busR);
\draw[arrow] (busL) -- (rttddft.north);
\draw[arrow] (busC) -- (cdft.north);
\draw[arrow] (busR) -- (bse.north);

\coordinate (mY) at ($(neqbse.north)+(0,6mm)$);

\coordinate (mL) at (mY -| rttddft.south);
\coordinate (mC) at (mY -| cdft.south);
\coordinate (mMid) at (mY -| neqbse.north);
\draw[link]  (rttddft.south) -- (mL);
\draw[link]  (cdft.south)    -- (mC);
\draw[link]  (mL) -- (mC);
\draw[arrow] (mMid) -- (neqbse.north);

\draw[arrow] (neqbse.south) -- (exspec.north);
\draw[arrow] (bse.south)    -- (gsspec.north);

\coordinate (mix) at ($(trans.north)+(0,4mm)$);
\draw[link]  (exspec.south) |- (mix);
\draw[link]  (gsspec.south) |- (mix);
\draw[arrow] (mix) -- (trans.north);

\end{tikzpicture}

%% file: sections/tools.tex
\subsection{State of the art}
\label{sec:SOTA_tools}

In modern computational materials science, the usage of an electronic-structure code can extend far beyond a single calculation. Research practices increasingly rely on automated workflows, large-scale data generation~\cite{Poul2025workflows}, systematic convergence studies~\cite{Bonacci2023workflows}, and advanced post-processing tools that transform raw outputs into physically meaningful observables. At the same time, users expect user-friendly interfaces, reproducible workflows, and seamless integration with external frameworks such as high-throughput platforms, data infrastructures, and analysis environments. As a result, the state of the art in the field has shifted from monolithic, standalone \gls{dft} programs towards rich software ecosystems that combine simulation engines with workflow frameworks, analysis tools, continuous testing, and comprehensive documentation. For a more detailed review on workflows in materials science, we refer to Ref.~\cite{speckhard2025workflows}.

Building on this evolution, modern electronic-structure research is typically supported by a diverse set of auxiliary codes that complement the core simulation engine. These include code-specific and code-agnostic post-processing frameworks for extracting materials properties, scripting interfaces for systematic input generation and output parsing, and workflow layers that orchestrate complex, multi-step simulations. Such components are most commonly implemented in \python. Together, these tools reduce the manual effort required to perform routine tasks, promote reproducibility, and enable scalable computational studies.

\python interfaces to electronic-structure codes have become building blocks for modern automation and reproducibility, as exemplified by \abinit, which is supported by the \python library AbiPy, or \vasp, which is supported by py4vasp. The development of \excitingtools~\cite{excitingtoolsPaper} follows this approach, providing a \python interface layer for input/output handling and workflow automation. Similarly, the \gls{ase} \cite{asePaper} offers a general \python interface to multiple electronic-structure codes through so-called calculators, including \vasp, \quantumespresso~\cite{Giannozzi2017}, \gpaw~\cite{Enkovaara2010}, and many more, serving as a basis for workflow automation and steering of atomistic simulations. \excitingtools has been integrated into \gls{ase}, where it serves as the interface to \exciting.

Workflow libraries for atomistic and electronic-structure calculations are, in principle, largely independent of the underlying code, as they implement abstract concepts such as tasks, dependencies, provenance, and execution backends~\cite{Steensen2025workflows}. In practice, however, the implementation of robust and efficient workflows requires a tight coupling to the specifics of each electronic-structure code, including its input parameters, output formats, error modes, and restart capabilities. Prominent examples include \gls{asr}~\cite{GJERDING2021110731}, which enables the creation of higher-level workflow layers on top of code-specific \gls{ase} calculators, thereby defining reusable workflows for ground-state and materials-property calculations, executed with MyQueue~\cite{Mortensen2020} or TaskBlaster~\cite{Larsen2025}. Similarly, the Materials Project~\cite{Jain2013_MaterialsProject} software stack combines pymatgen~\cite{ONG2013314} for structured input/output handling with workflow engines such as \jobflow~\cite{jobflowPaper} and atomate2~\cite{Atomate22025}, providing automated and provenance-aware workflows primarily targeting codes like \vasp and \quantumespresso. Another widely used framework is AiiDA~\cite{Huber2020}, which offers a code-agnostic workflow engine and provenance model, but still relies on dedicated plugins to interface individual electronic-structure codes. \exciting is situated within this evolving landscape, offering a rich selection of features, including the high-level \python packages \excitingtools and \excitingworkflow, as well as direct interfaces to the library SIRIUS and the simulation codes Cc4s and elphbolt. 

\subsection{Workflow tools in support of \excitingb}
\label{sec:New_WFs_tools}

In this section, we describe various components of the \exciting software ecosystem. The \python library \excitingtools~\cite{excitingtoolsPaper} provides a high-level, extensible interface, designed to simplify both input generation and output analysis of \exciting. It offers robust parsers that convert \exciting output files into well-structured and serializable \python classes and dictionaries, enabling seamless integration with data-analysis frameworks and higher-level workflow managers. It provides object-oriented classes for constructing and manipulating input files in a flexible and programmatic way. The library aims to support the full range of \exciting's capabilities, including \gls{gs} and excited-state calculations such as $GW$, \gls{bse}, and \gls{rt}-\gls{tddft}. The package is distributed with the \exciting code, and is additionally publicly available on PyPI~\cite{excitingtoolsPyPI}.

Within the \exciting ecosystem, the \python package \excitingworkflow builds directly on top of \excitingtools to provide a framework for defining and executing automated high-throughput calculations and complex workflows with \exciting. It employs the \jobflow library as its underlying workflow framework, enabling the modular definition, execution, and chaining of computational tasks. Designed with scalability and reproducibility in mind, the package supports remote execution on compute clusters with queue managers like \slurm through \jobflowremote. \excitingworkflow forms the basis for systematic and reproducible \exciting calculations within automated workflow environments. It currently implements workflows for monitoring the convergence of various input parameters and for the automated optimization of the \gls{lo} basis functions. The convergence workflows are designed to be highly dynamic. In each step, the selected parameter is updated, the corresponding calculation is executed, the convergence criterion is evaluated, and additional steps are generated only when necessary. Robust error-handling mechanisms are included to automatically recover from common issues, such as time-outs or insufficient memory. \excitingworkflow will be released in the near future.

In addition to these more comprehensive workflow tools, we also maintain the \python package \newline \excitingscripts, which provides practical scripts for using \exciting. These scripts give users convenient access to commonly used functions, such as setting up convergence tests in a human readable form, updating parameters in input files, plotting results, and more. Currently, the main purpose of the package is to support the comprehensive suite of \exciting \texttt{tutorials} (see \cref{sec:portfolio_code}). The code is distributed together with the \exciting source code. To allow for more frequent updates, we also publish it as a standalone package in PyPI, available at \url{https://pypi.org/project/excitingscripts/}.

\subsection{Computing elastic constants: \bftt{ElaStic}}
\label{sec:New_Elastic_tools}

The \python package \elastic~\cite{golesorkhtabar2013elastic} allows users to calculate second and third-order elastic coefficients using \exciting~\cite{lion2022elastic} or other electronic-structure codes. Applications in conjunction with Quantum Espresso~\cite{PhysRevB.103.115204,Yu2023} and VASP~\cite{Jin2023} are among recent examples. The package has recently been fully revised using the newest \python standards, and many features have been updated to reduce the computational burden and facilitate integration into high-throughput workflows~\cite{speckhard2026elaStic2}. In order to speed up the selection of Lagrangian vectors, the \matid library is used, which is a robust \python API that makes use of Spglib~\cite{togo2024spglib}. To lessen the computational load, a new set of Lagrangian vectors is introduced. The new code presents an automated procedure to determine the optimal polynomial-fit order and the maximum strain to include in the fitting process. Finally, the code is modularized and configured via a YAML file so that it can be easily integrated into a \jobflow workflow. These new additions allow one to run automated workflows to calculate elastic coefficients efficiently and reliably.

\subsection{Interface to SIRIUS}
\label{sec:New_SIRIUS_tools}

Complex systems, such as large organic crystals, metal-organic frameworks, disordered materials, or surfaces and interface systems are hard to tackle with standard implementations. To overcome the memory and scalability limitations associated with such problems, \exciting has been interfaced with SIRIUS~\cite{sirius_repo}, a domain-specific library for electronic-structure calculations. SIRIUS exposes more parallelization, thus allowing \exciting to efficiently handle much larger systems. As reference cases, calculations of a pyrene-MoS$_2$ interface and BA$_2$PbI$_4$ show that the \texttt{exciting-SIRIUS} interface achieves speedups of about 5$\times$ and 35$\times$, respectively, compared to standard \exciting runs.

\subsection{Coupled-cluster calculations through an interface to Cc4s}
\label{sec:New_CC_tools}

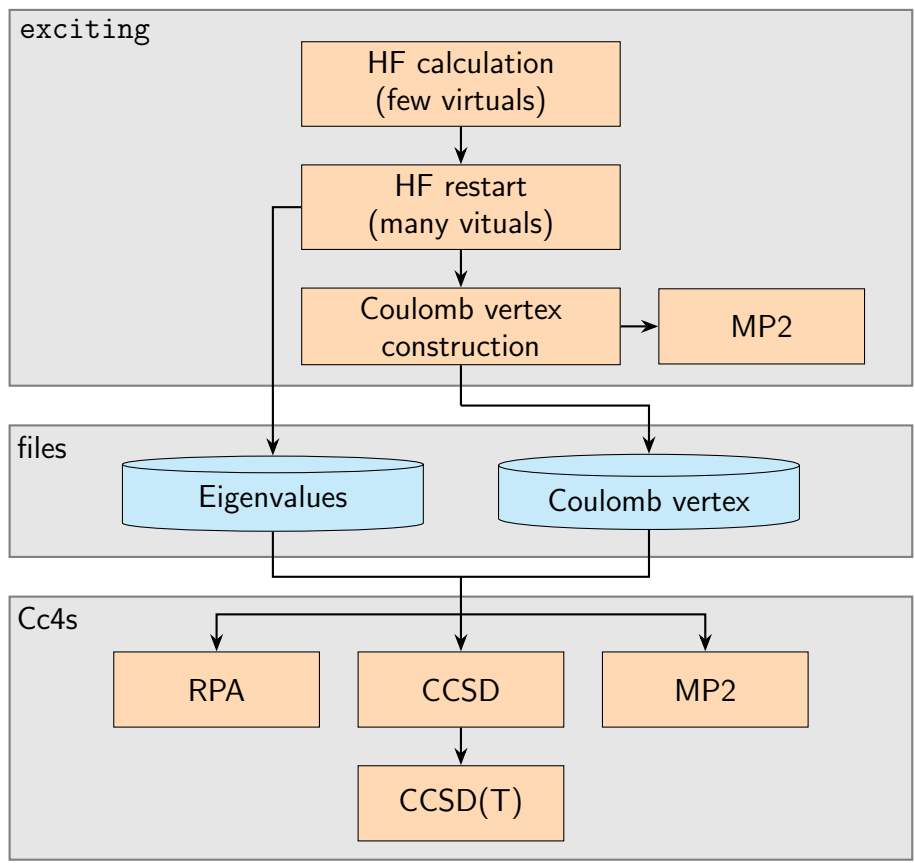
\begin{figure}[H]
    \centering
    \input{figures/tools/workflow_cc4s}
    \caption{Workflow for correlation-energy calculations using Cc4s. 
    The acronyms are explained in the text.}
    \label{fig:workflow_cc4s}
\end{figure}

Cc4s is an open-source code~\cite{Cc4s-code} that implements quantum-chemistry methods~\cite{Moller1934,Bartlett07,Furche2008} for solving the many-electron Schr\"{o}dinger equation. Although the code is designed for periodic systems~\cite{Gruber2018}, it also supports the calculation of atoms and molecules. The workflow for computing the electron correlation energy is shown in \cref{fig:workflow_cc4s} and consists of the following three computational steps: (i) a reference \gls{hf} calculation, (ii) a calculation of the Coulomb vertex and band energies, and (iii) a calculation of the correlation-energy using \gls{rpa}, second-order M{\o}ller-Plesset perturbation theory (MP2), the \gls{cc} method considering single and double excitations (CCSD), or going beyond by considering triple excitations perturbatively in the CCSD(T) method. Steps (i) and (ii) are implemented in \exciting, and using the interface to Cc4s, performs step (iii). This workflow has been used in proof-of-principle calculations reproducing the complete-basis limit of small finite systems~\cite{Grueneis2026}. 

\subsection{Transport calculation through an interface to elphbolt}
\label{sec:New_elphbolt_tools}

The open-source software elphbolt~\cite{Protik2022a} solves the coupled \glspl{bte} with the aid of Wannier-Fourier interpolation. \exciting has the capability of computing \glspl{mlwf}, phonons, and \gls{epc} constants. It thus allows for the export of the localized real-space representations of the electronic Hamiltonian, the dynamical matrices, and \gls{epc} matrix elements, which are then used as an input for elphbolt calculations. This interface complements a direct implementation of transport properties \cite{Caruso2019} based on Ref. \cite{Scheidemantel2003}.

%% file: figures/tools/workflow_cc4s.tex
\begin{tikzpicture}[node distance=2cm, every node/.append style={font=\sffamily\large}] 
\tikzstyle{startstop} = [rectangle, rounded corners, minimum width=2cm, minimum height=1cm,text centered, text width=4cm, draw=black, fill=red!30]
\tikzstyle{startstop_small} = [rectangle, rounded corners, minimum width=2cm, minimum height=1cm,text centered, text width=2.5cm, draw=black, fill=red!30]

\tikzstyle{process} = [rectangle, minimum width=2cm, minimum height=1cm, text width=4cm, text centered, draw=black, fill=orange!30]
\tikzstyle{process_small} = [rectangle, minimum width=2cm, minimum height=1cm, text width=2.5cm, text centered, draw=black, fill=orange!30]

\tikzstyle{io} = [trapezium, trapezium left angle=70, trapezium right angle=110, minimum width=4cm, text centered, draw=black, fill=blue!30]

\tikzstyle{database} = [cylinder, minimum height=1cm, minimum width=4cm, shape border rotate=90,  aspect=0.1, text centered, draw=black, fill=cyan!20]

\tikzstyle{background_box1} = [fill=gray!20, draw=gray, thick, minimum width=12cm, minimum height=5cm, inner sep=0pt]
\tikzstyle{background_box2} = [fill=gray!20, draw=gray, thick, minimum width=12cm, minimum height=1.75cm, inner sep=0pt]
\tikzstyle{background_box3} = [fill=gray!20, draw=gray, thick, minimum width=12cm, minimum height=3.5cm, inner sep=0pt]
\tikzstyle{arrow} = [thick, ->, >=Stealth]
\tikzstyle{link} = [thick, -]
\node (bgbox1) [background_box1, xshift=0cm, yshift = 0cm] {};
\node (bgbox1_text) [xshift=-5cm, yshift = -3mm,align=left,text centered] at (bgbox1.north) {\exciting };
\node (hf1) [process,xshift=0cm, yshift = -1cm]  at (bgbox1.north)  {HF calculation \\(few virtuals)};
\node (hf2) [process, below=5mm of hf1] {HF restart \\(many vituals)};
\node (vertex) [process, below=5mm of hf2] {Coulomb vertex \\construction};
\node (exc_mp2) [process_small, right=5mm of vertex] {MP2};
\draw [arrow] (hf1) -- (hf2);
\draw [arrow] (hf2) -- (vertex);
\draw [arrow] (vertex) -- (exc_mp2);

\node (bgbox2) [background_box2, below=0.5cm of bgbox1, xshift=0cm, yshift = 0cm] {};
\node (bgbox2_text) [xshift=-5.5cm, yshift = -3mm,, align=left] at (bgbox2.north) {files };
\node (eig_file) [database,xshift=-2.5cm, yshift = -1cm]  at (bgbox2.north) {Eigenvalues};
\node (vertex_file) [database,xshift=2.5cm, yshift = -1cm]  at (bgbox2.north) {Coulomb~vertex};
\coordinate (pt_exciting_files) at ($(bgbox2.north)+(0,2.5mm)$);
\draw [link] (vertex) |- (pt_exciting_files);
\draw [arrow] (pt_exciting_files) -| (vertex_file);
\draw [arrow] (hf2) -| (eig_file);
\node (bgbox3) [background_box3, below=0.5cm of bgbox2, xshift=0cm, yshift = 0cm] {};
\node (bgbox2_text) [xshift=-5.5cm, yshift = -3mm, align=left] at (bgbox3.north) {Cc4s};
\node (ccsd) [process_small, yshift = -1.25cm] at (bgbox3.north) {CCSD};
\node (mp2) [process_small, right=5mm of ccsd] {MP2};
\node (rpa) [process_small, left=5mm of ccsd] {RPA};
\node (ccsd-t) [process_small, below=5mm of ccsd] {CCSD(T)};
\coordinate (pt1_files_cc4s) at ($(bgbox3.north)+(0,2.5mm)$);
\draw[link]  (eig_file) |- (pt1_files_cc4s);
\draw[link]  (vertex_file) |- (pt1_files_cc4s);
\coordinate (pt2_files_cc4s) at ($(bgbox3.north)+(0,-2.5mm)$);
\draw[link]  (pt1_files_cc4s) |- (pt2_files_cc4s);
\draw [arrow] (pt2_files_cc4s) -| (mp2);
\draw [arrow] (pt2_files_cc4s) -| (ccsd);
\draw [arrow] (pt2_files_cc4s) -| (rpa);
\draw [arrow] (ccsd) -- (ccsd-t);
\end{tikzpicture}

%% file: sections/data.tex
\subsection{State of the art}
\label{sec:SOTA_data}

Data-centric approaches have become an important part of materials science. This also concerns electronic-structure theory, as summarized in a recent roadmap article~\cite{kulik2022roadmap}. High-quality data are crucial for the success of \gls{ml} and are currently a major bottleneck~\cite{speckhard2025big}. Therefore, open materials databases are invaluable for overcoming this problem. Such databases are often established in the context of high-throughput calculations~\cite{Curtarolo2012_Aflowlib, Jain2013_MaterialsProject,Kirklin2015_OQMD,Schmidt2024_Alexandria}, and their amount of data is growing rapidly. NOMAD \cite{Draxl2018NOMAD, Scheidgen2023NOMAD} has taken a different approach by inviting the electronic-structure community to contribute and share their data. Since it was opened to the public in 2014, about 19 million entries from computational materials science have been published, including contributions from smaller projects and large high-throughput studies. Through continuous development of the NOMAD software~\cite{Scheidgen2023NOMAD}, an increasing number of computational techniques is now supported, including excited-state methods from \gls{mbpt}, \gls{tddft}, \gls{dmft}, and \gls{md}. NOMAD also hosts benchmark data, for example, a comparison of x-ray spectra produced by three different codes, \ie Ocean, \exciting, and XSPECTRA \cite{meng2024}. 

The NOMAD dataspace is often used for \gls{ml}, \eg training a \gls{llm} for crystal-structure prediction \cite{Antunes2024_CrystalLLM}. Combining \gls{ml} with multi-scale modeling techniques, property prediction for disordered alloys has been demonstrated~\cite{Rigamonti2024}. An emerging, most popular application of \gls{ml} is the training of force fields. At the time of writing of this review, new models are published monthly~\cite{Riebesell2025_MatbenchDiscovery}. We emphasize that NOMAD hosts the raw data of the Alexandria database~\cite{Schmidt2024_Alexandria}, which is frequently used as training data for high-performing \gls{ml} force fields. Beyond crystal structure optimization and prediction, \gls{ml} models are used for predicting material properties directly from crystal structures~\cite{Xie2018_CNNmodels, bechtel2025band}, classification tasks~\cite{Ghiringhelli2015_CriticalRole, Stanev2018_TcClassification}, recommendation of novel compounds\cite{Hayashi2022_Recommender, Griesemer2025_Recommender}, or crystal structure generation~\cite{DeBreuck2025_CrystalStructureGeneration}. Due to the large number of recent publications, we refer the interested reader to reviews focused on the topic~\cite{Schmidt2019_MLReview, Choudhary2022_MLReview, Cheng2026_AIDiscoveryReview}. 

We note that there is a severe lack of available training data generated with higher-level methods. In fact, with a few exceptions, \eg Ref.~\cite{Knoesgaard2022_GWML}, most \gls{ml} models are trained on \gls{dft} results obtained with semilocal functionals. Moreover, there is a bottleneck concerning error quantification of computational results, which is needed to assess data quality and how it is impacted by methods and approximations or computational precision. Some work in this direction is provided by studies dedicated to quantifying similarity~\cite{Kuban2022_DOSFingerprints, Kuban2022_QualityAssessment} and error estimates~\cite{Carbogno2022,Speckhard2025}. 

\exciting is well supported by the NOMAD data infrastructure. The output files created by \exciting are recognized by NOMAD parsers, and their content is extracted according to the underlying metadata schema~\cite{Ghiringhelli2017_metadata}. At present, NOMAD contains several tens of thousands \exciting calculations from 73 authors from all over the world.

\subsection{Data-driven error quantification}
\label{sec:New_error_data}

The numerical precision of a calculation depends on the computational parameters, most importantly the \gls{bz} sampling and the basis-set quality (see \cref{sec:SOTA_LAPW}). In order to quantify how computational settings affect a calculation's precision, we can train \gls{ml} models. Here, we choose the basis-set size as an example. This approach can be used to predict, for instance, the total energy of an expensive calculation from a less precise one and even to perform an extrapolation to the \gls{cbs} limit~\cite{kraus2021extrapolating,Speckhard2025}. \gls{cbs} extrapolation facilitates the reuse of data by placing calculations from different sources on a more equal footing. By quantifying the precision of existing calculations, \eg from the NOMAD data infrastructure, the model can help reuse them for a different purpose. Typically, \gls{cbs} extrapolation uses a series of calculations with increasing basis-set size. In recent works, some of us have trained different models to predict the \gls{cbs} energy from a single calculation, which is a more difficult task~\cite{Carbogno2022,Speckhard2025}.

\begin{figure}[h]
    \centering
    \includegraphics[width=0.8\linewidth]{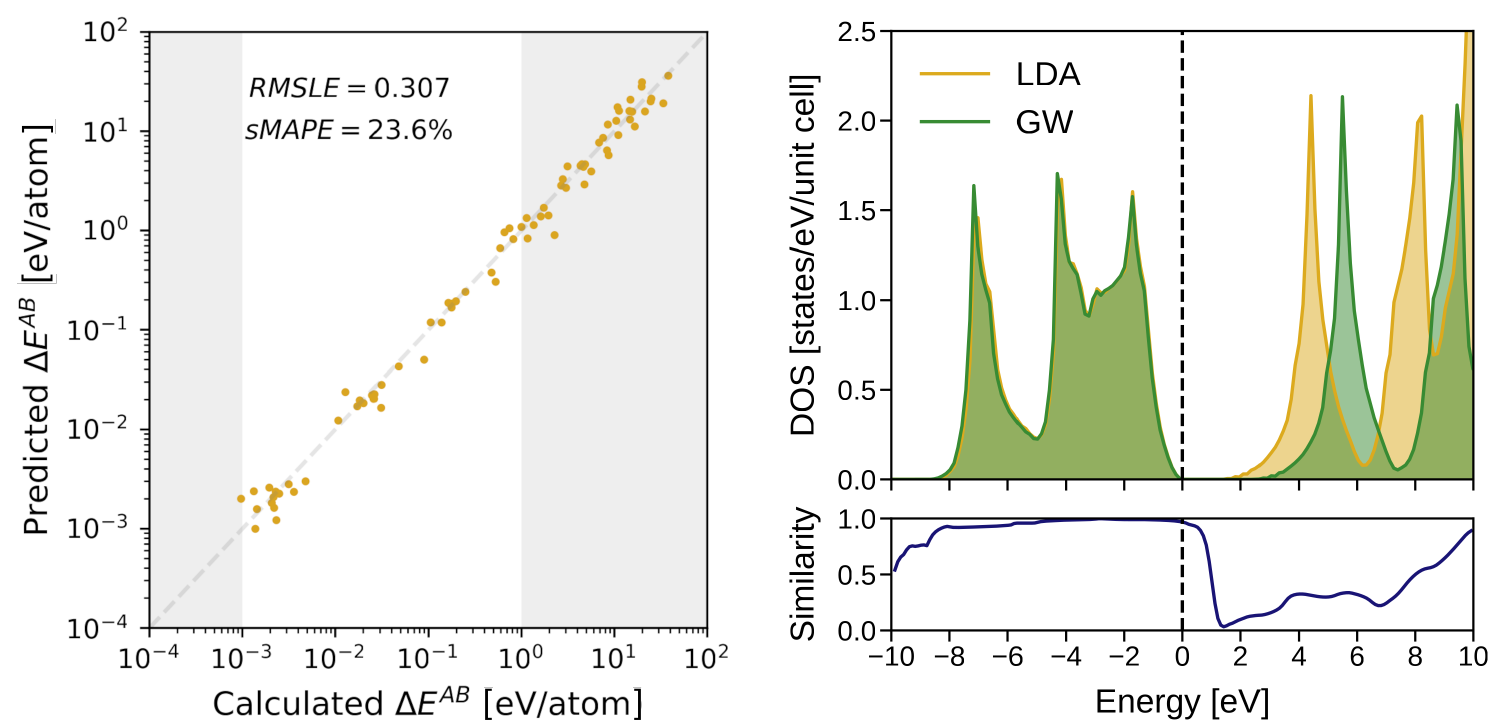}
    \caption{Examples of data-driven error quantification. Left: Results of a SISSO model to predict $\Delta E^{AB}$, \ie the difference in the total energy from a calculation done with basis-set size $N_{\rm b}$ and the \gls{cbs} value~\cite{Speckhard2025}. Note the log-log scale to help visualize smaller $\Delta E^{AB}$ values. Values between 1 meV/atom and 1 eV/atom (white background) are of particular interest to DFT practitioners. Right: \gls{dos} of SiC~\cite{Pardini2018_NOMADData}, computed with \gls{lda} (yellow) and the $GW$ approximation (green), aligned at the valence band maximum. Bottom right: Similarity between the \gls{lda} and the $GW$ result as a function of energy.}
    \label{fig:data_quality_assessment}
\end{figure}

We want to highlight one model here, which was trained using the symbolic regression method SISSO~\cite{ouyang2018sisso, purcell2023recent}. Symbolic regression techniques have the benefit that the resulting model can be expressed as an arithmetic formula of the input features \cite{Ghiringhelli2015_CriticalRole}. This makes the resulting models highly interpretable. In the left panel of \cref{fig:data_quality_assessment}, we showcase the performance of such a model on a benchmark dataset from Ref.~\cite{Carbogno2022} consisting of diverse binary materials, computed with a systematic variation in the basis-set size from artificially low values to very large values. As a result, the energy differences between these calculations, $\Delta E^{AB}$, span several orders of magnitude. This makes the ML task more difficult~\cite{Speckhard2025}. By optimizing for the root-mean-squared-logarithmic-error (RMSLE) of $\Delta E^{AB}$, the SISSO model performs well across the target's wide range of values for the diverse set of materials in the test set~\cite{Speckhard2025}. It can be used to recommend, for a given material and a targeted tolerance in the resulting property, the optimal basis-set size for a computation. Recent work has expanded this focus to quantify the uncertainty in more complex properties such as the lattice parameters, band gaps, and the density of states with respect to the basis-set size and the $\kk$-point mesh~\cite{Speckhard2026thesis}.

The need for data-quality assessment concerns not only scalar quantities like energies, but also spectral properties. Typically, the assessment of spectra is performed qualitatively by visual inspection to analyze, for example results from different levels of theory. Applied to big data, \eg for large scale benchmark tests, this approach is not feasible. One way of approaching this challenge is the use of similarity metrics~\cite{Isayev2015_MaterialsCartograpgy, Kuban2022_DOSFingerprints, Kuban2022_QualityAssessment}, which can be defined for any type of data and allow to quantify how (dis)similar results are. In the top right panel of \cref{fig:data_quality_assessment}, we show an example for this approach by presenting the \gls{dos} of SiC~\cite{Pardini2018_NOMADData} computed by two different approaches, \ie at the \gls{dft} level, using the \gls{lda} functional, and at the $GW$ level, using the $G_0W_0$ approach on top of the \gls{lda} result. Comparing them, we observe the typical blue-shift of the conduction bands, since the $GW$ result exhibits a larger band gap. We quantify this difference using the method described in Ref.~\cite{Kuban2024_MADAS}: We use fingerprints~\cite{Kuban2022_DOSFingerprints} to encode the \glspl{dos} of both calculations in a small energy window around a selected reference energy. We then compare these fingerprints for a range of different reference energies, obtaining an energy-resolved quantitative measure of the similarity of the two \glspl{dos}, which is shown in the bottom right panel of \cref{fig:data_quality_assessment}. As a similarity metric, we use the Tanimoto coefficient~\cite{Willett1998_Similarity}. In the valence bands, the two results overlap almost perfectly, resulting in a similarity value of almost $1$. Around the Fermi level at $0$ eV, the similarity drops sharply, due to the larger band-gap of the $GW$ result. Identifying and quantifying the effect of different approximations on the results of electronic structure calculations is key to understanding, which material classes require which level of theory for an accurate description. Such tools are especially important for large-scale benchmarks and scans of the whole materials space, as they allow to quickly assess the impact of different levels of theory to the computational results.

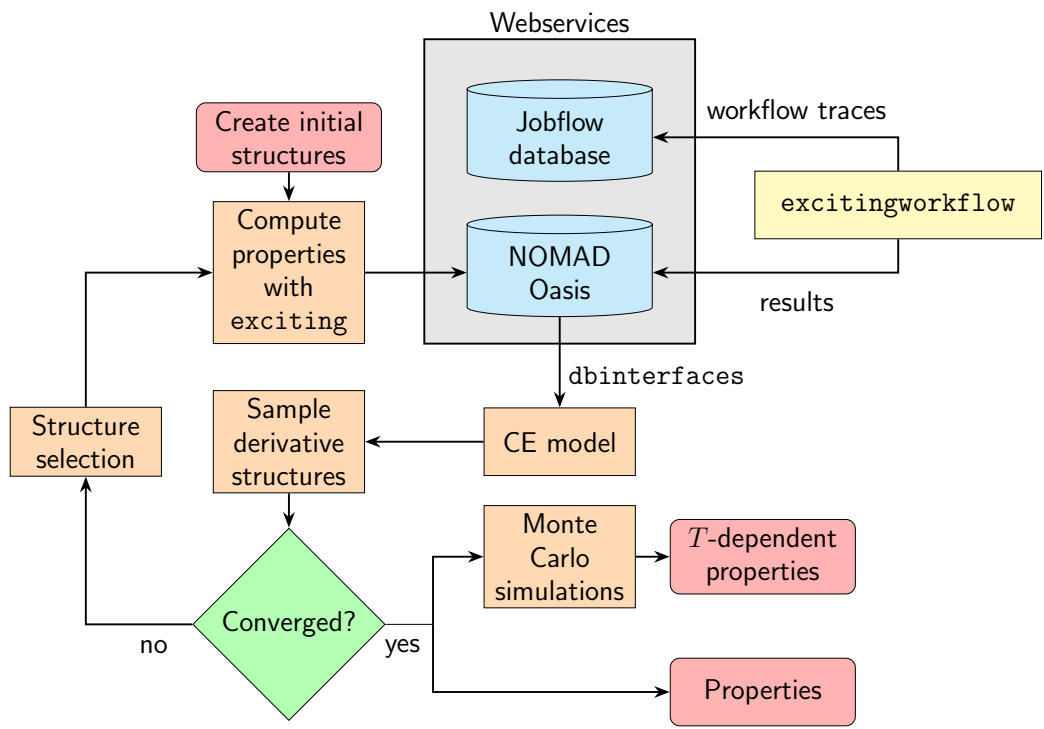
\begin{figure}[h]
    \centering
    \input{figures/data/workflow-ce-nomad}
    \vspace*{-6.5mm}
    \caption{Integration of NOMAD infrastructure for different applications. Both a NOMAD Oasis and a Jobflow database (blue) are running as webservices (gray box). The interaction with workflows using the \excitingworkflow framework is indicated by the yellow box in the top right. One the left, we see the workflow for building \gls{ce} models with \exciting data using an interface to the NOMAD Oasis. Individual workflow tasks are shown in orange, data are shown in red, and decision points are shown in green.}
    \label{fig:workflow-ce-nomad}
\end{figure}

\subsection{NOMAD Oasis for \excitingb data}
\label{sec:New_Oasis_data} 

The NOMAD software (see \cref{sec:SOTA_data}) can be installed locally for the organization of a research group's data or for collaborations between groups. Such an installation, which can be customized to individual needs, is called NOMAD Oasis. The authors of this review manage an Oasis at the Humboldt-Universit\"at zu Berlin, which is specifically configured for a research group focused on electronic-structure calculations. Starting with practicalities, we have increased, for example, the limit of unpublished uploads that a user can have. This allows group members to upload and share unfinished or even failed calculations for discussions with their colleagues. Increasing the allowed upload size makes it possible to share raw (binary) data that are too large for the central NOMAD instance. The Oasis is also accessible via the public internet, which allows for uploads directly from HPC centers. Obviously, the access to our Oasis is restricted to group members and collaborators, since it contains unpublished research.

Overall, we make use of the flexible setup of NOMAD Oasis: Individual components of the system, such as the database and web interface, are running in Docker (\url{https://www.docker.com/}) containers. We have extended the Oasis infrastructure by hosting a MongoDB (\url{https://www.mongodb.com/}) database in a container. This is shown in \cref{fig:workflow-ce-nomad}, where the gray box indicates the setup described here: A Jobflow database is running as a webservice next to the NOMAD Oasis. This configuration allows group members to upload and store workflow traces and other data produced by workflows implemented in \excitingworkflow (see \cref{sec:New_WFs_tools}), or any other framework based on \jobflowremote. At the same time, the input and output files are uploaded as raw data to the Oasis, where all relevant information is extracted from the raw files. This setup is specifically useful for running high-throughput workflows, since uploading and sharing can be automatized, making new data immediately available to collaborators.

Our setup also has benefits from an infrastructure point of view. From the user's perspective, it simplifies the usage of different tools, since they can access different services under a single URL. For the administrator of the Oasis and the Jobflow database, the setup and maintenance of the services is simplified, since everything can be managed with a single configuration file. We provide the template for this setup at \url{https://github.com/exciting/sol-oasis}, including documentation for adopting it for other groups.

\subsection{Building \acrshort*{ce} models with \excitingb, \cellb, and NOMAD}

\begin{figure}[h]
    \centering
    \includegraphics[width=1.0\linewidth]{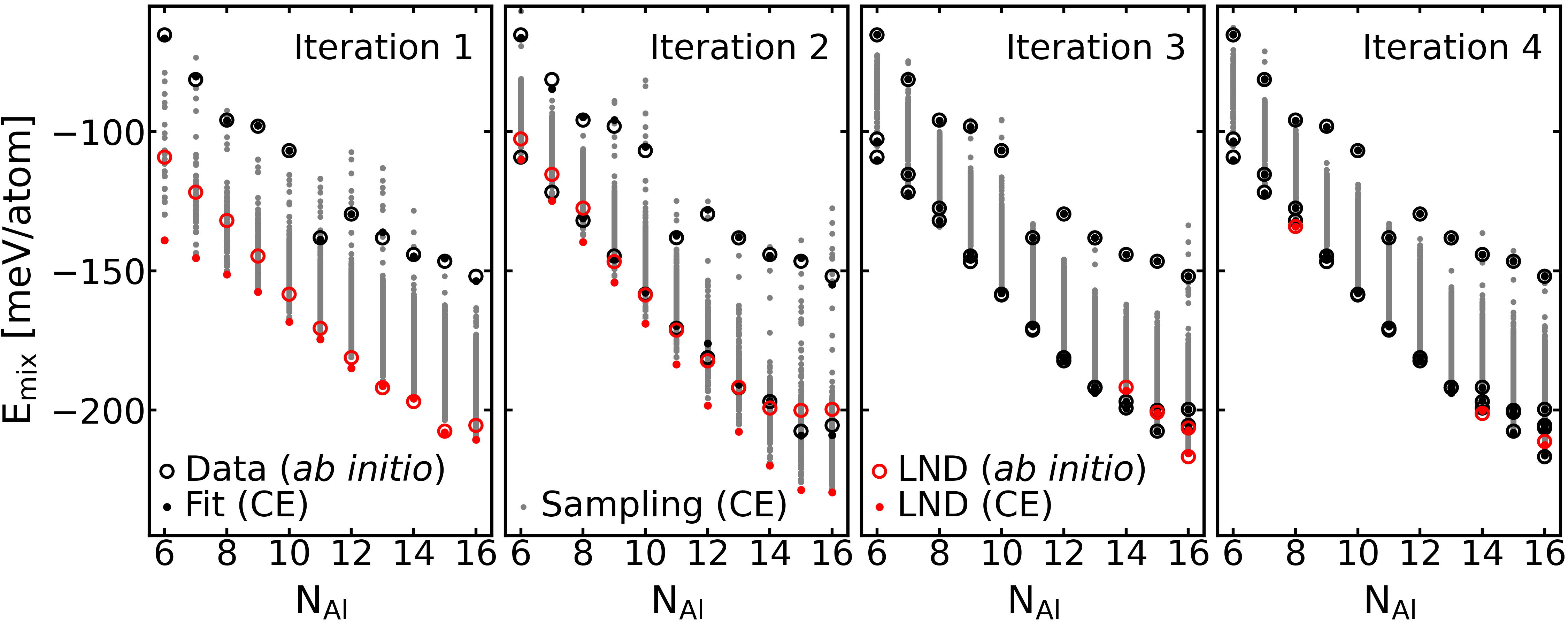}
    \caption{Construction of a \gls{ce} model for the energy of mixing of the clathrate compound Ba$_8$Al$_{x}$Si$_{46-x}$ with \exciting, \cell, and NOMAD using the workflow of \cref{fig:workflow-ce-nomad}. N$_\mathrm{Al}$ is the number of Al substituents in the clathrate.}
    \label{fig:workflow-clathrate}
\end{figure}

Access to thermodynamic properties of materials with \exciting, such as the ground states of alloys and order-disorder transitions, is made possible by building \acrfull{ce} \cite{Sanchez1984,Connolly1983} models based on training data computed with \exciting. For this purpose, a workflow integrating \exciting with \cell \cite{Rigamonti2024} and NOMAD has been implemented, which allows for creating training structures, computing them with \exciting, uploading the data to the NOMAD Oasis (see ~\cref{sec:New_Oasis_data}), and finally creating accurate \gls{ce} models for accessing various properties in a numerically efficient manner. The workflow is sketched in \cref{fig:workflow-ce-nomad}. First, an initial set of structures is created, as indicated by the red box in the top left. For all structures, \gls{dft} calculations are performed with \exciting and the input and output files are uploaded to the NOMAD Oasis, where they are parsed and the computed properties are extracted. Using the \texttt{dbinterfaces} module of \cell, these results are then acquired from the Oasis and used as target values for the construction of a \gls{ce} model with \cell. Then, derivative structures are sampled, \eg by generating all possible structures up to a certain supercell size, and their properties are predicted using the \gls{ce} model. We then check for convergence. This typically involves evaluating if the model's cross validation score is below a target threshold (green box). Other criteria, as for instance verifying that no new \acrlong{gs} structures are found among the previously sampled structures (in the case that the total energy is the modeled property), may be relevant to determine convergence. If convergence is not achieved, we select additional derivative structures to increase the diversity and size of the training set. For this, we use different approaches that are discussed in more detail in Ref.~\cite{Rigamonti2024}. For the new selected configurations, properties are again computed with \exciting and uploaded to the Oasis. With the extended training set, we build a new \gls{ce} model, and the process is repeated until the model is converged. (We note that the same procedure also works with the public NOMAD instance.) 

By modeling the total energy per particle, one can then perform statistical thermodynamics simulations on large supercells, \eg employing Metropolis \gls{mc} simulations with \cell \cite{Troppenz2023}. This enables the prediction of temperature-dependent properties, such as the specific heat $C_p$. By modeling additional properties, as for instance the lattice parameters, one can calculate their \gls{mc} average at different concentrations and temperatures. This workflow has been used to find the ground-state structures of the intermetallic clathrate alloy Ba$_8$Al$_{x}$Si$_{46-x}$, which is a promising material for thermoelectric applications. The so obtained models were used to describe its phase diagram \cite{Rigamonti2024}. \cref{fig:workflow-clathrate} shows the results of the workflow's iterative process. In the panel labeled "Iteration 1", starting from a set of random structures (black empty circles), a \gls{ce} model is built. This model is used to sample derivative structures using simulated annealing (gray dots), yielding low-energy structures with predicted energies indicated by red dots. The latter are selected to enlarge the training set by computing their \textit{ab initio} energies (red empty circles). In iteration 1, a significant disagreement between the computed and predicted energies indicates that convergence has not been achieved. Therefore, following the workflow, the process is repeated in iterations 2 to 4, augmenting the training set in each iteration by adding \gls{lnd} structures, resulting in a \gls{ce} model with an accuracy of about 1 meV/atom. 

This workflow also allows for leveraging advanced \gls{ml} techniques for the prediction of nonlinear properties, as demonstrated in Ref.~\cite{Stroth2025}, where non-linear \gls{ce} models were used to predict, \eg the \gls{ks} band gap of clathrates.

\subsection{Visualizing Fermi surfaces: \fsvisualb}
\label{sec:New_FSvisual_data}

\exciting can be used to calculate \glspl{fs}. The resulting data, \ie electronic band energies evaluated on a very dense \textbf{k}-grid in the \gls{bz}, are stored in the human-readable bxsf file format established by the XCrySDen software \cite{Kokalj1999XCrySDen}. For visualization, we here introduce \fsvisual, a \python-based, modular framework that offers interactive 3D viewing, allowing for rotation and providing zoom functionality. Unlike other stand-alone tools, such as FermiSurfer \cite{FermiSurfer} or XCrySDen, \fsvisual generates \gls{html} files for its visualizations by utilizing the capabilities of the Plotly~\cite{plotly} package. The resulting \gls{fs} plots can be accessed easily, either directly or through any web browser, without the need for specialized software, and can be shared and embedded in websites as interactive widgets. As such, we plan to integrate \fsvisual into NOMAD (see~\cref{sec:SOTA_data}), to allow for interactive exploration of the uploaded data.

\begin{figure}[h]
    \centering
    \includegraphics[width=0.75\linewidth]{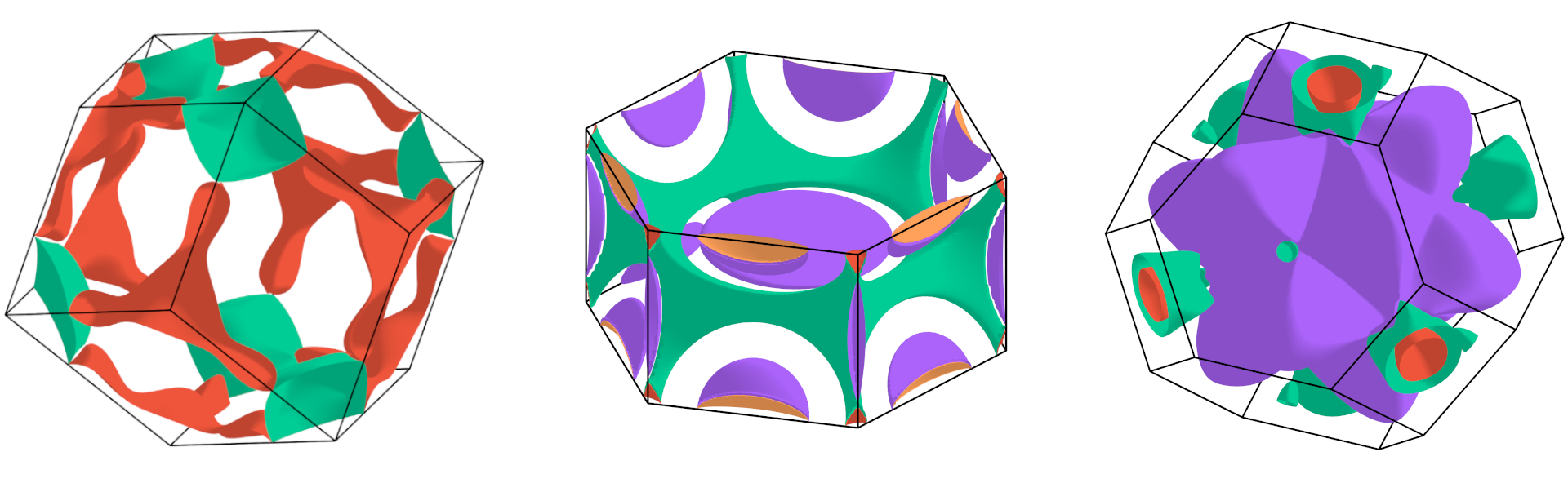}
    \vspace*{-1mm}
    \caption{Fermi surfaces of Ba, Mg, and Ru (left to right), calculated using the PBEsol \gls{xc} functional, inside the \gls{bz} (black lines) of the bcc, hcp, and fcc structures, respectively. The individual sheets of the \gls{fs} can be distinguished by their color.}
    \label{fig:Fermi_surfaces}
\end{figure}

Figure \ref{fig:Fermi_surfaces} showcases the \glspl{fs} of Ba, Mg, and Ru inside the respective \glspl{bz}. Ba exhibits two open sheets (colored in red and green), while Mg shows four open sheets (colored in red, green, purple, and orange). Ru displays two open sheets (colored in red and green) alongside a closed sheet (colored in purple), as well as a closed sheet, which is hidden inside the purple one. Using \fsvisual, each branch, and also the \gls{bz}, can be viewed individually. We used \fsvisual to create interactive visualizations of the \glspl{fs} of 37 elemental metals, which were computed using \exciting. They can be explored at the \exciting webpage, see \url{https://exciting-code.org/fermi_surfaces}. More details of the code and a detailed analysis of the data will be published elsewhere~\cite{Stutz2026_FSvisual}. The source code is already available at \url{https://github.com/exciting/FSvisual}.

%% file: figures/data/workflow-ce-nomad.tex
\begin{tikzpicture}[node distance=2cm, every node/.append style={font=\sffamily\large}] 
\scalebox{0.9}{
\tikzstyle{startstop} = [rectangle, rounded corners, minimum width=2cm, minimum height=1cm,text centered, text width=2.5cm, draw=black, fill=red!30]
\tikzstyle{io} = [trapezium, trapezium left angle=70, trapezium right angle=110, minimum width=2cm, minimum height=1cm, text width=3cm, text centered, draw=black, fill=blue!30]
\tikzstyle{process} = [rectangle, minimum width=2cm, minimum height=1cm, text width=2cm, text centered, draw=black, fill=orange!30]
\tikzstyle{interface} = [rectangle, minimum width=2cm, minimum height=1cm, text width=2cm, text centered, draw=black, fill=yellow!30]
\tikzstyle{decision} = [diamond, minimum width=2cm, minimum height=1cm, text width=2cm, text centered, draw=black, fill=green!30]
\tikzstyle{arrow} = [thick, ->, >=Stealth]
\tikzstyle{background box} = [fill=gray!20, draw=black, thick, minimum width=4cm, minimum height=4.5cm, inner sep=0pt]

\tikzstyle{database} = [cylinder, minimum height=1.5cm, minimum width=1.5cm, shape border rotate=90,  aspect=0.1, text centered, draw=black, fill=cyan!20]

  \node (start) [startstop] {Create initial structures};
  \node (pro1) [process, below of=start] {Compute properties with \exciting};
  \node (bgbox) [background box, right of=pro1, xshift=2cm, yshift = 1.2cm] {};
  \node (bgbox_text) [above of=bgbox, yshift = 5mm] {Webservices};
 \node (db) [database, right of=pro1, xshift=2cm, text width=2.5cm] {NOMAD Oasis};
  \node (jfdb) [database, above of=db, text width=2.5cm] {Jobflow database};
  \node (excwf) [interface, right of=bgbox, xshift=30mm, yshift = -2mm, text width = 4cm] {\texttt{excitingworkflow}};
  \node (pro2) [process, below of=db, yshift=-5mm]{CE model};
  \node (pro3) [process, below of=pro1, yshift=-5mm]{Sample derivative structures};
  \node (dec1) [decision, below of=pro3, yshift=-7mm] {Converged?};
  \node (str) [process, left of=pro3, yshift=-0.0cm, xshift=-1cm] {Structure selection};
  \node (pro5) [process, right of=dec1, xshift=2cm, yshift=1cm] {Monte Carlo simulations};
  \node (stop) [startstop, right of=pro5, xshift=1cm] {$T$-dependent properties};
  \node (stop2) [startstop, below of=stop] {Properties}; 

  \draw [arrow] (start) -- (pro1);
  \draw [arrow] (pro1) -- (db);
  \draw [arrow] (db) -- node[anchor=west, yshift=-2mm] {\texttt{dbinterfaces}} (pro2); 
  \draw [arrow] (excwf) |- node[anchor=north, yshift=0.7cm, xshift=-1.5cm] {workflow traces} (jfdb); 
  \draw [arrow] (excwf) |- node[anchor=south, yshift=-0.7cm, xshift=-1.5cm] {results} (db); 
  \draw [arrow] (pro2) -- (pro3);
  \draw [arrow] (pro3) -- (dec1);
  \draw [arrow] (dec1)  -| node[anchor=north, xshift = 1cm, yshift=-1mm] {no} (str);
  \draw [arrow] (str) |- (pro1);
  \coordinate  (inv) at ([xshift=0.7cm]dec1.east);
  \draw  (dec1) -- node[anchor=north, xshift = -1mm, yshift=-1mm] {yes} (inv); 
  \draw [arrow] (inv) |- (pro5);
  \draw [arrow] (pro5) -- (stop);
  \draw [arrow] (inv) |- (stop2);
}
\end{tikzpicture}

%% file: sections/outlook.tex
\exciting is always under active development, steadily extending its methodological scope and computational capabilities across ground state, excited state, and non-equilibrium simulations. Here, we list some topics that are currently under development and will be deployed in subsequent releases, or are on our wish list: For ground-state calculations, we are currently extending the hybrid-functional implementation to the spin-degrees of freedom, including consistent treatment of \gls{soc}. Moreover, dielectric-dependent screening will be part of then next release. Also atomic forces and geometry optimization from hybrid functionals are in our focus. The implementation of several \gls{vdw} functionals \cite{Kim2020} will be available soon as well as our band-structure-unfolding code.

Current developments in the $GW$ module include core-level $GW$. Moreover, the effect of $GW$ corrections to electron-phonon matrix elements will be evaluated. Future work will include the incorporation of vertex corrections. Within \gls{tddft}, developments in the \gls{rt}-\gls{tddft} module focuses on incorporating nonlocal, \gls{mbpt}-derived self-energy operators and hybrid functionals to better capture excitonic effects. Real-time simulations of pump-probe experiments involving core electrons will be implemented, along with an extension of real-time propagation to include \gls{soc}. Furthermore, the high accuracy of our full-potential all-electron \gls{lapwlo}-based \gls{tddft} implementations will be leveraged to generate reliable benchmark-quality reference data. Continued developments of the \gls{bse} formalism will allow for extensions of the description of \glspl{ste} in solids, including finite-temperature regimes and real-time exciton-phonon dynamics. Our tools for first and second-order Raman scattering and infrared spectroscopy will be linked to the \gls{dfpt} implementation and made more user-friendly. Pump-probe spectroscopy will be extended towards photoexcited electron populations above the Fermi level, including Pauli-blocking effects on near-edge X-ray absorption. Furthermore, we aim at integrating a non-equilibrium BSE-based pump-probe simulation framework with nonadiabatic molecular dynamics.

Finally, complementary efforts focus on improving and extending our workflows, enhancing  usability of the code's functionality. Python interfaces and analysis tools will continue to be extended to support the latest developments. Verifying and understanding the convergence behavior of \exciting will be supported by the creation and analysis of curated benchmark datasets, which will allow us to suggest input parameters for \exciting calculations. With a focus on high-throughput calculations, we plan to further strengthen the integration of \excitingtools and \excitingworkflow with the NOMAD infrastructure. We will make use of \texttt{FSvisual} to display Fermi surfaces within the NOMAD framework.
All developments are ideally accompanied by HPC-friendliness, making them highly parallelizable and computationally efficient on various platforms, including CPUs and GPUs. 

Collectively, these ongoing and planned developments reaffirm \exciting's position as the benchmark all-electron code, continuously expanding its portfolio of state-of-the-art \gls{gs} and theoretical spectroscopy features.